\documentclass[namedreferences,hyperref,optionalrh]{spr-sola}

\usepackage{graphicx}        
\usepackage{multirow}  
\usepackage{color}           

\newcommand{\new}[1]{#1}
\newcommand{\old}[1]{}

\definecolor{violet}{rgb}{0.5,0,1}

\chardef\us=`\_

\usepackage{ulem}
\definecolor{dblue}{RGB}{0, 0, 100}

\begin{document}

\begin{article}

\begin{opening}

\title{Polarimeter to Unify the Corona and Heliosphere (PUNCH)}

\author[addressref={aff-swri},corref,email={deforest@boulder.swri.edu}]{\inits{C.E.}\fnm{Craig~E.}\ \lnm{DeForest}\orcid{0000-0002-7164-2786}}
\author[addressref={aff-ucar},email={sgibson@ucar.edu},]{\inits{S.E.}\fnm{Sarah~E.}\ \lnm{Gibson}\orcid{0000-0001-9831-2640}}
\author[addressref={aff-swri},email={rkillough@swri.org}]{\inits{R.L.}\fnm{Ronnie}\ \lnm{Killough}}
\author[addressref={aff-ral}]{\inits{N.R.}\fnm{Nick~R.}\ \lnm{Waltham}\orcid{0000-0002-4170-7104}}
\author[addressref={aff-swri}]{\inits{M.N.}\fnm{Matt~N.}\ \lnm{Beasley}\orcid{0009-0004-0646-0523}}
\author[addressref={aff-nrl}]{\inits{R.C.}\fnm{Robin~C.}\ \lnm{Colaninno}\orcid{0000-0002-3253-4205}}
\author[addressref={aff-swri}]{\inits{G.T.}\fnm{Glenn~T.}\ \lnm{Laurent}\orcid{0009-0000-6578-326X}}
\author[addressref={aff-swri}]{\inits{D.B.}\fnm{Daniel~B.}\ \lnm{Seaton}\orcid{0000-0002-0494-2025}}
\author[addressref={aff-swri}]{\inits{J.M.}\fnm{J.~Marcus}\ \lnm{Hughes}\orcid{0000-0003-3410-7650}}
\author[addressref={aff-hq}]{\inits{M.}\fnm{Madhulika}\ \lnm{Guhathakurta}\orcid{0000-0001-5357-4452}}
\author[addressref={aff-gsfc}]{\inits{N.M.}\fnm{Nicholeen~M.}\ \lnm{Viall}\orcid{0000-0003-1692-1704}}
\author[addressref={aff-gmu}]{\inits{R.}\fnm{Raphael}\ \lnm{Atti\'e}\orcid{0000-0003-4312-6298}}
\author[addressref={aff-iiap}]{\inits{D.}\fnm{Dipankar}\ \lnm{Banerjee}\orcid{0000-0003-4653-6823}}
\author[addressref={aff-reading}]{\inits{L.}\fnm{Luke}\ \lnm{Barnard}\orcid{0000-0001-9876-4612}}
\author[addressref={aff-noaa}]{\inits{D.A.}\fnm{Doug~A.}\ \lnm{Biesecker}}
\author[addressref={aff-ral}]{\inits{M.M.}\fnm{Mario~M.}\ \lnm{Bisi}\orcid{0000-0001-6821-9576}}
\author[addressref={aff-goettingen}]{\inits{V.}~\fnm{Volker}\ \lnm{Bothmer}\orcid{0000-0001-7489-5968}}
\author[addressref={aff-swri}]{\inits{A.}\fnm{Antonina}\ \lnm{Brody}}
\author[addressref={aff-ucar}]{\inits{J.}\fnm{Joan}\ \lnm{Burkepile}\orcid{0000-0002-9959-6048}}
\author[addressref={aff-usyd}]{\inits{I.H.}\fnm{Iver~H.}\ \lnm{Cairns}\orcid{0000-0001-6978-9765}}
\author[addressref={aff-swri}]{\inits{J.L.}\fnm{Jennifer~L.}\ \lnm{Campbell}}
\author[addressref={aff-swri}]{\inits{T.R.}\fnm{Traci~R.}\ \lnm{Case}}
\author[addressref={aff-swri}]{\inits{A.}\fnm{Amir}\ \lnm{Caspi}\orcid{0000-0001-8702-8273}}
\author[addressref={aff-hq}]{\inits{D.}\fnm{David}\ \lnm{Cheney}}
\author[addressref={aff-udel}]{\inits{R.}\fnm{Rohit}\ \lnm{Chhiber}\orcid{0000-0002-7174-6948}}
\author[addressref={aff-ral}]{\inits{M.J.}\fnm{Matthew~J.}\ \lnm{Clapp}}
\author[addressref={aff-cu}]{\inits{S.R.}\fnm{Steven~R.}\ \lnm{Cranmer}\orcid{0000-0002-3699-3134}}
\author[addressref={aff-ral}]{\inits{J.A.}\fnm{Jackie~A.}\ \lnm{Davies}\orcid{0000-0001-9865-9281}}
\author[addressref={aff-cires}]{\inits{C.A.}\fnm{Curt~A.}\ \lnm{de~Koning}\orcid{0000-0002-9577-1400}}
\author[addressref={aff-swri}]{\inits{M.I.}\fnm{Mihir~I.}\ \lnm{Desai}\orcid{0000-0002-7318-6008}}
\author[addressref={aff-swri}]{\inits{H.A.}\fnm{Heather~A.}\ \lnm{Elliott}\orcid{0000-0003-2297-3922}}
\author[addressref={aff-ucar}]{\inits{S.}\fnm{Samaiyah}\ \lnm{Farid}\orcid{0000-0003-0223-7004}}
\author[addressref={aff-cua}]{\inits{B.}\fnm{Bea}\ \lnm{Gallardo-Lacourt}\orcid{0000-0003-3690-7547}}
\author[addressref={aff-swri}]{\inits{C.}\fnm{Chris}\ \lnm{Gilly}\orcid{0000-0003-0021-9056}} 
\author[addressref={aff-swri}]{\inits{C.}\fnm{Caden}\ \lnm{Gobat}\orcid{0000-0003-1268-8845}}
\author[addressref={aff-cu}]{\inits{M.H.}\fnm{Mary~H.}\ \lnm{Hanson}\orcid{0009-0005-8304-4037}}
\author[addressref={aff-ral}]{\inits{R.A.}\fnm{Richard~A.}\ \lnm{Harrison}\orcid{0000-0002-0843-8045}}
\author[addressref={aff-swri}]{\inits{D.M.}\fnm{Donald~M.}\ \lnm{Hassler}\orcid{0000-0001-8830-1200}}
\author[addressref={aff-swri}]{\inits{C.}\fnm{Chase}\ \lnm{Henley}}
\author[addressref={aff-swri}]{\inits{A.M.}\fnm{Alan~M.}\ \lnm{Henry}}
\author[addressref={aff-apl}]{\inits{R.A.}\fnm{Russell~A.}\ \lnm{Howard}\orcid{0000-0001-9027-8249}}
\author[addressref={aff-ucsd}]{\inits{B.V.}\fnm{Bernard~V.}\ \lnm{Jackson}\orcid{0000-0001-7664-5858}}
\author[addressref={aff-stellar}]{\inits{S.}\fnm{Samuel}\ \lnm{Jones}}
\author[addressref={aff-ucar}]{\inits{D.}\fnm{Don}\ \lnm{Kolinski}}
\author[addressref={aff-swri}]{\inits{D.A.}\fnm{Derek~A.}\ \lnm{Lamb}\orcid{0000-0002-6061-6443}}
\author[addressref={aff-swri}]{\inits{F.}\fnm{Florine}\ \lnm{Lehtinen}}
\author[addressref={aff-swri}]{\inits{C.}\fnm{Chris}\ \lnm{Lowder}\orcid{0000-0001-8318-8229}}
\author[addressref={aff-ucar}]{\inits{A.}\fnm{Anna}\ \lnm{Malanushenko}\orcid{0000-0001-5066-8509}}
\author[addressref={aff-udel}]{\inits{W.H.}\fnm{William~H.}\ \lnm{Matthaeus}\orcid{0000-0001-7224-6024}}
\author[addressref={aff-pu}]{\inits{D.J.}\fnm{David~J.}\ \lnm{McComas}\orcid{0000-0001-6160-1158}}
\author[addressref={aff-swri}]{\inits{J.}\fnm{Jacob}\ \lnm{McGee}}
\author[addressref={aff-aber}]{\inits{H.}\fnm{Huw}\ \lnm{Morgan}\orcid{0000-0002-6547-5838}}
\author[addressref={aff-ncra}]{\inits{D.}\fnm{Divya}\ \lnm{Oberoi}\orcid{0000-0002-4768-9058}}
\author[addressref={aff-gmu}]{\inits{D.}\fnm{Dusan}\ \lnm{Odstrcil}\orcid{0000-0001-5114-9911}}
\author[addressref={aff-ral}]{\inits{C.}\fnm{Chris}\ \lnm{Parmenter}}
\author[addressref={aff-swri}]{\inits{R.}\fnm{Ritesh}\ \lnm{Patel}\orcid{0000-0001-8504-2725}}
\author[addressref={aff-udel}]{\inits{F.}\fnm{Francesco}\ \lnm{Pecora}\orcid{0000-0003-4168-590X}}
\author[addressref={aff-swri}]{\inits{S.}\fnm{Steve}\ \lnm{Persyn}}
\author[addressref={aff-noaa}]{\inits{V.J.}\fnm{Victor~J.}\ \lnm{Pizzo}\orcid{0000-0002-7110-2325}}
\author[addressref={aff-hq}]{\inits{S.P.}\fnm{Simon~P.}\ \lnm{Plunkett}}
\author[addressref={aff-apl}]{\inits{E.}\fnm{Elena}\ \lnm{Provornikova}\orcid{0000-0001-8875-7478}}
\author[addressref={aff-apl}]{\inits{N.E.}\fnm{Nour~Eddine}\ \lnm{Raouafi}\orcid{0000-0003-2409-3742}}
\author[addressref={aff-swri}]{\inits{J.A.}\fnm{Jillian~A.}\ \lnm{Redfern}\orcid{0000-0002-4507-3372}}
\author[addressref={aff-irap}]{\inits{A.P.}\fnm{Alexis~P.}\ \lnm{Rouillard}\orcid{0000-0003-4039-5767}}
\author[addressref={aff-swri}]{\inits{K.D.}\fnm{Kelly~D.}\ \lnm{Smith}}
\author[addressref={aff-swri}]{\inits{K.B.}\fnm{Keith~B.}\ \lnm{Smith}}
\author[addressref={aff-swri}]{\inits{Z.S.}\fnm{Zachary~S.}\ \lnm{Talpas}}
\author[addressref={aff-ral}]{\inits{S.J.}\fnm{S.~James}\ \lnm{Tappin}\orcid{0000-0002-1878-5243}}
\author[addressref={aff-nrl}]{\inits{A.}\fnm{Arnaud}\ \lnm{Thernisien}\orcid{0009-0006-8066-5548}}
\author[addressref={aff-gsfc}]{\inits{B.J.}\fnm{Barbara~J.}\ \lnm{Thompson}\orcid{0000-0001-6952-7343}}
\author[addressref={aff-swri}]{\inits{S.}\fnm{Samuel}\ \lnm{Van~Kooten}\orcid{0000-0002-4472-8517}}
\author[addressref={aff-swri}]{\inits{K.J.}\fnm{Kevin~J.}\ \lnm{Walsh}\orcid{0000-0002-0906-1761}}
\author[addressref={aff-bc}]{\inits{D.F.}\fnm{David~F.}\ \lnm{Webb}}
\author[addressref={aff-swri}]{\inits{W.L.}\fnm{William~L.}\ \lnm{Wells}}
\author[addressref={aff-esa}]{\inits{M.J.}\fnm{Matthew~J.}\ \lnm{West}\orcid{0000-0002-0631-2393}}
\author[addressref={aff-swri}]{\inits{Z.}\fnm{Zachary}\ \lnm{Wiens}}
\author[addressref={aff-udel}]{\inits{Y.}\fnm{Yan}\ \lnm{Yang}}
\author[addressref={aff-rob}]{\inits{A.}\fnm{Andrei}\ \lnm{Zhukov}\orcid{0000-0002-2542-9810}}


\address[id=aff-swri]{Southwest Research Institute, 1050 Walnut Street Suite 300, Boulder, CO 80302 U.S.A.}
\address[id=aff-ucar]{National Center for Atmospheric Research / High Altitude Observatory, Boulder, CO, U.S.A.}
\address[id=aff-ral]{United Kingdom Research \& Innovation / Science \& Technology Facilities Council / Rutherford Appleton Laboratory, Harwell Campus, Oxfordshire, U.K.}
\address[id=aff-nrl]{Naval Research Laboratory, Washington, DC, U.S.A.}
\address[id=aff-hq]{NASA Headquarters, Washington, DC, U.S.A.}
\address[id=aff-gsfc]{NASA Goddard Space Flight Center, Greenbelt, MD, U.S.A.}
\address[id=aff-gmu]{George Mason University, Innovation Park, VA, U.S.A.}
\address[id=aff-iiap]{Indian Institute of Astrophysics, Bangalore, India}
\address[id=aff-reading]{University of Reading, Reading, U.K.}
\address[id=aff-noaa]{NOAA Space Weather Prediction Center, Boulder, CO, U.S.A.}
\address[id=aff-goettingen]{University of G\"ottingen, Germany}
\address[id=aff-usyd]{University of Sydney, Australia}
\address[id=aff-udel]{University of Delaware, Newark, DE, U.S.A.}
\address[id=aff-cu]{University of Colorado, Boulder, CO, U.S.A.}
\address[id=aff-cires]{Cooperative Institute for Research in Environmental Sciences, Boulder, CO, U.S.A.}
\address[id=aff-cua]{Catholic University of America, Washington, DC, U.S.A.}
\address[id=aff-apl]{Johns Hopkins University Applied Physics Laboratory, Columbia, MD, U.S.A.}
\address[id=aff-ucsd]{University of California, San Diego, CA, U.S.A.}
\address[id=aff-stellar]{Stellar Solutions Inc., Los Altos, CA, U.S.A.}
\address[id=aff-pu]{Princeton University, Princeton, NJ, U.S.A.}
\address[id=aff-aber]{University of Aberystwyth, U.K.}
\address[id=aff-ncra]{National Center for Radio Astrophysics, India}
\address[id=aff-irap]{Institut de Recherche en Astrophysique et Plan\'etologie, Toulouse, France}
\address[id=aff-bc]{Boston College, Boston, MA, U.S.A.}
\address[id=aff-esa]{European Space Agency, Noordwijk, Netherlands}
\address[id=aff-rob]{Royal Observatory of Belgium, Brussels, Belgium}

\runningauthor{DeForest et al.}
\runningtitle{Polarimeter to Unify the Corona and Heliosphere [submitted to \textit{Solar Physics}]}

\begin{motto}
\textit{
Look deep into nature, and then you will understand everything better. -- A. Einstein
}
\end{motto}

\begin{abstract}

The Polarimeter to Unify the Corona and Heliosphere (PUNCH) mission is a 
NASA Small Explorer to determine the cross-scale processes that unify the solar
corona and heliosphere. 
PUNCH has two science objectives: (1) understand how coronal structures become the
ambient solar wind, and (2) understand the dynamic evolution of transient structures, such as
coronal mass ejections, in the young solar wind.  
To address these objectives, PUNCH uses a constellation of four small spacecraft in Sun-synchronous 
low Earth orbit, to collect linearly polarized images of the K corona and young
solar wind.  The four spacecraft each carry one visible-light imager in a 1+3 configuration: a
single Narrow Field Imager solar coronagraph captures images of the outer corona at all 
position angles, and at solar elongations from 1.5° (6 R$_\odot$) to 8° (32 R$_\odot$); 
and three separate  Wide Field Imager heliospheric imagers together capture views of the entire inner 
solar system,
at solar elongations from 3° (12 R$_\odot$) to 45° (180 R$_\odot$) from the Sun. PUNCH images include
linear-polarization data, to enable inferring the three-dimensional structure of visible features 
without stereoscopy.  The instruments are matched in wavelength passband, 
support overlapping 
instantaneous fields of view, and are operated synchronously, to act as a single 
``virtual instrument'' with a 90$^\circ$ wide field of view, centered on the Sun.  PUNCH 
launched in March of 2025 and began science operations in June of 2025. 
PUNCH has an open data policy with no proprietary period, and PUNCH Science Team Meetings
are open to all.

\end{abstract}
\end{opening}
\pagebreak
\section{Introduction}
     \label{S-Introduction} 

The Polarimeter to UNify the Corona and Heliosphere (PUNCH) mission is a NASA Small Explorer to determine 
the cross-scale processes that unify the solar corona and heliosphere. 
PUNCH has two science objectives: (1) to understand how coronal structures become the
ambient solar wind, and (2) to understand the dynamic evolution of transient structures, such as
coronal mass ejections, in the young solar wind. 

\begin{figure}[tbh]
    \centering
    \includegraphics[width=4.5in]{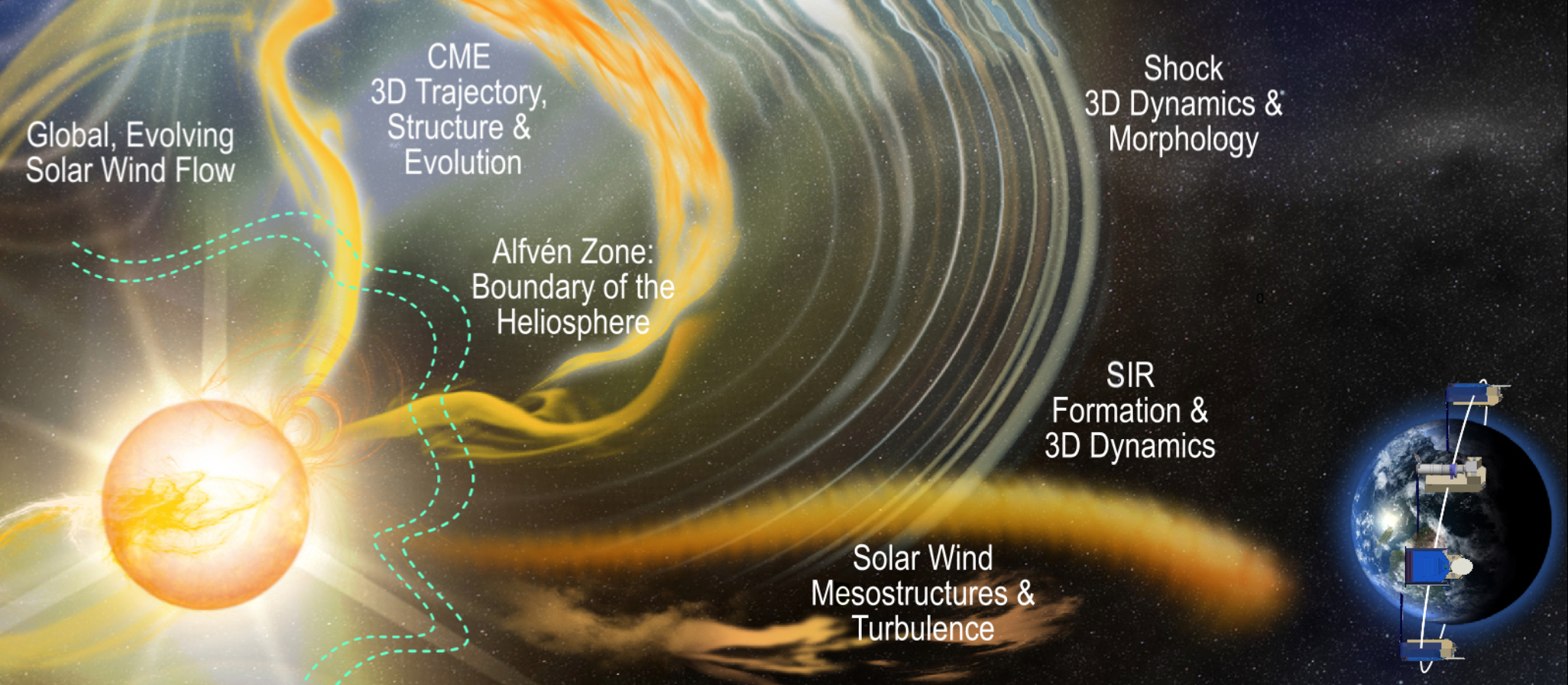}
    \caption{Artist's rendering shows the six major PUNCH scientific topics, each addressed by a science question.  These are organized
    into two major science objectives (understanding the ambient and the dynamic solar wind) in support of a single science goal: to determine the cross-scale processes that unite the solar corona and heliosphere.}
    \label{fig:beauty}
\end{figure}

The two PUNCH science objectives are encapsulated in three science questions each, rendered in Figure \ref{fig:beauty}.  They are:

\begin{enumerate}
    \item Understand how coronal structures become the ambient solar wind.
    \begin{enumerate}
        \item How does the young solar wind flow and evolve on global scales?
        \item Where and how do mesoscale structures and turbulence form in the solar wind?
        \item What are the evolving physical properties of the Alfv\'en surface?
    \end{enumerate}
    \item Understand the dynamic evolution of transient structures in the solar wind.
    \begin{enumerate}
        \item How do coronal mass ejections (CMEs) propagate and evolve in the solar wind, in three dimensions?
        \item How do quasi-stationary stream interaction regions (SIRs) form and evolve?
        \item How do shocks form and interact with the solar wind across spatial scales?
    \end{enumerate}
\end{enumerate}

These objectives are addressed with deep-field, 
polarimetric image sequences of the ``K corona'', consisting of sunlight that is 
Thomson-scattered by electrons in the coronal and solar wind plasma. 
Four separate Observatories, each comprising a small spacecraft 
and a single visible-light imager, orbit the Earth in a Sun-synchronous twilight (6am/6pm) 650~km orbit, maintain widely spaced orbital phases, and act as a single ``virtual coronagraph'' 
with a 90$^\circ$ field of view, to record
the entire inner solar system at solar elongation angles between 1.5$^\circ$ and 45$^\circ$.

PUNCH exists within the context of constantly improving solar and heliospheric studies.  Historically,
solar physics and heliospheric physics have been separate disciplines, divided by the technologies used
to study each portion of the Sun-heliosphere system.  The solar corona has been studied primarily by 
remote sensing, for example with visible light
scattered by free electrons \citep{thomson1903conduction} to form the K corona
\citep[e.g.,][]{billings1966book} although many other remote-sensing technologies also apply.  Remote
sensing generally provides morphology and large scale macrophysics, with some indirect
insight into 
unresolvable microphysics.  Contrariwise, the solar wind has been studied 
primarily by in-situ sampling of the material in interplanetary space
\citep{neugebauer1966}; these technologies yield
detailed diagnosis of plasma conditions at a single moving point in the plasma, with indirect insight 
into large-scale macrophysics.  

Joint studies of the solar corona and solar wind have generally required spanning the two radically different experimental
disciplines, but have borne fruit in connecting the Sun and near-Earth environments 
\citep{Zirker_1977,sheeley_etal_1985,Domingo_1995}. 
ESA's Solar Orbiter mission \citep{muller_etal_2020} carries a hybrid instrument suite specifically designed to enable joint remote/in-situ studies of solar wind features and streams, by associating particular remotely sensed coronal features with the plasma detected in-situ by the spacecraft.
Recently, the Parker Solar Probe \citep{fox_etal_2016} has carried in-situ sampling devices into
the outer reaches of the solar corona itself \citep{kasper_etal_2021}, allowing direct measurement of the microphysics 
in the outer corona with simultaneous remote sensing of the relevant large-scale structures surrounding the system under study. 

Multiple instruments, including the Coriolis Solar Mass Ejection Imager  \citep[Coriolis/SMEI;][]{eyles_etal_2003}, the Solar-Terrestrial Relations Observatory's Sun-Earth Connection Coronal and Heliospheric Investigation \citep[STEREO/SECCHI;][]{howard_etal_2008}, the Solar Orbiter Heliospheric Imager \citep[SOLO/SOLOHI;][]{howard_etal_2020}, and the Parker Solar Probe Wide-field Imager for Solar Probe \citep[PSP/WISPR;][]{Vourlidas_etal_2016}, 
have been built to image large and bright features in the solar wind.  
With the advent of imaging from off the Sun-Earth line, via STEREO/SECCHI, direct, quantitative, simultaneous 
imaging and in-situ studies of Earth-bound events became possible \citep[e.g.,][]{davies_etal_2009, deforest_etal_2013a}, and 
the presence of multiple deep-space missions with intersecting fields of view and on-board in-situ 
instruments has greatly
increased the number of such studies.  PUNCH brings routine global imaging of the solar wind and its
structures into 
the mix of available observations, explicitly enabling a unified treatment of the corona and 
solar wind as a single system.

PUNCH polarimetry is designed to allow
location of features in three dimensions (3D) from a single vantage point:  
both mesoscale solar wind structure \citep[e.g.,][]{viall_deforest_kepko_2021} and coronal mass 
ejections \citep{dekoning_etal_2009,mierla_etal_2009,deforest_etal_2017}.  This is possible because
the K corona consists of sunlight that is Thomson scattered by free electrons.
Thomson scattering polarizes the outbound beam according to the scattering angle \citep{billings1966book,tappin_howard_2009,deforest_etal_2013}.  Each coronal feature is therefore
polarized to a degree that depends on its three-dimensional location.  The basic theory has been known for decades, but was only reduced to practice once image quality and analysis methods improved enough to reveal the polarimetric signal with sufficient fidelity \citep{mierla_etal_2010,deforest_etal_2017}. Many authors have more recently developed the theory and demonstrated the practice, as reviewed by \citet{gibson_etal_2025}. PUNCH is specifically designed to measure the 3D trajectory and structure of small ejecta, CMEs, and large solar-wind features such as stream interaction regions (SIRs).

\begin{figure}
    \centering
    \includegraphics[width=3in]{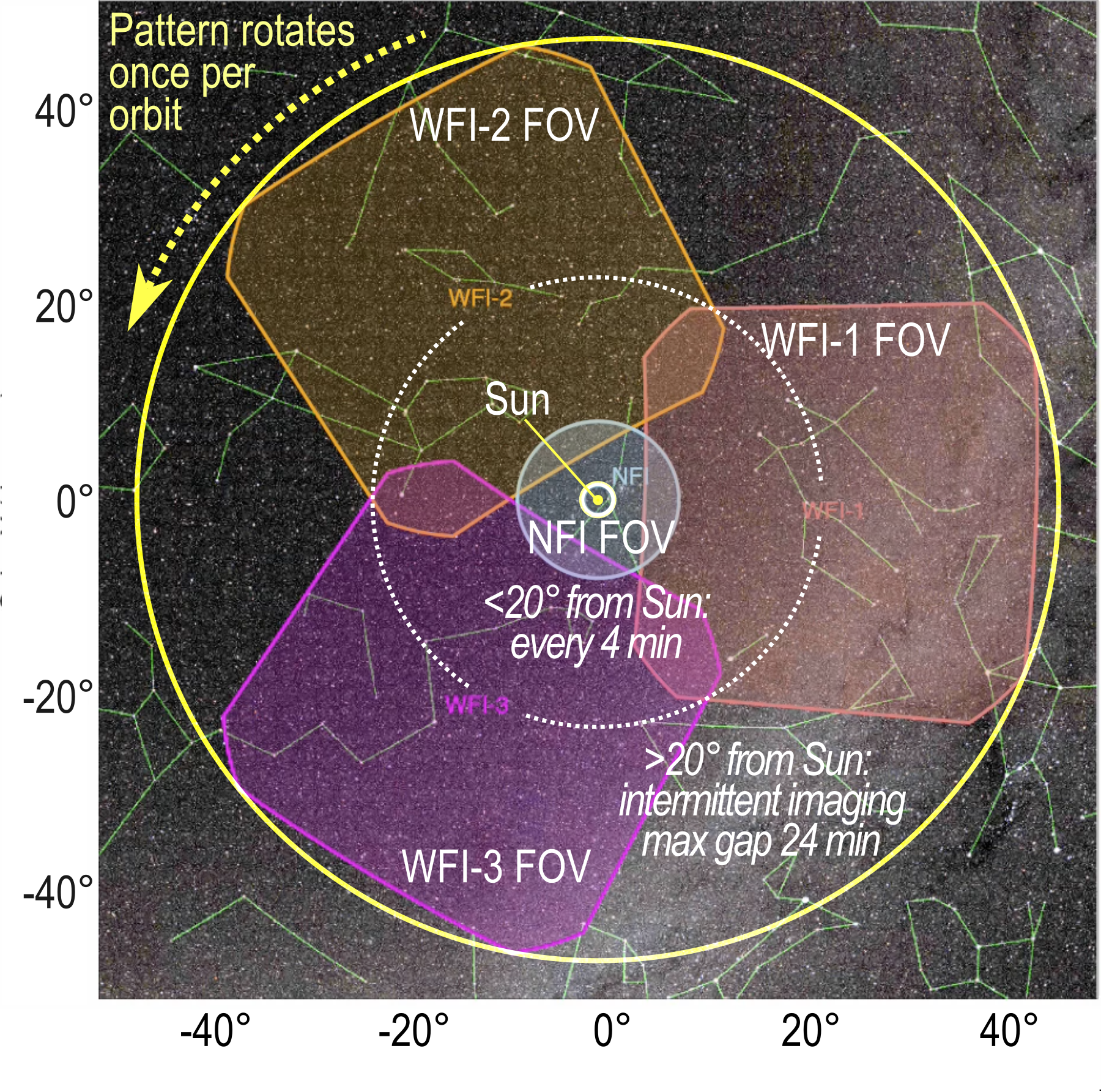}
    \caption{Four PUNCH spacecraft view a 90$^\circ$ wide field of view on the celestial sphere, centered on the Sun.  One Narrow Field Imager (NFI) captures the outer solar corona to an elongation of 32 R$_\odot$.  Three Wide Field Imagers (WFIs) together capture a trefoil instantaneous field of view (IFOV) on the sky.  The trefoil figure rotates once per orbit, capturing the complete field of view (FOV; yellow circle) approximately once every 30 minutes. The fields of view overlap at elongations below 80 R$_\odot$, providing 4 minute cadence in the inner part of the FOV.} 
    \label{fig:trefoil}
\end{figure}

PUNCH launched into space at 2025 March 12 03:10 UTC (2025 March 11 20:10 PDT) from Vandenberg Space Force Base.  The four $\sim$50~kg smallsat Observatories deployed jointly into Sun-synchronous low Earth orbit 
above the day/night terminator line.  The Observatories dispersed over 120 days into a ``1+3'' constellation comprising a single NFI Observatory and, separately, the  
three WFIs mutually separated by 120$^\circ$ in orbital anomaly\footnote{Readers are reminded that orbital ``anomaly'' is an angular parameter that indicates the position of a body in a Keplerian orbit.  Variants are ``true anomaly'', measured relative to the prime focus of the orbital ellipse; ``eccentric anomaly'', measured relative to the center of the orbital ellipse; and ``mean anomaly'', a temporal phase angle relative to the orbital period.  For a circular orbit such the PUNCH nominal orbit, all three variants have the same value.}.  A single Observatory (designated ``PUNCH-NFI00'' or ``NFI-0'')
has a ``Narrow Field Imager'' (NFI, or PUNCH/NFI) coronagraph \citep{colaninno_2025} that images the K corona 
at all position 
angles $\alpha$, at elongations $\varepsilon$ from 1.5$^\circ$ (6 R$_\odot$)\footnote{PUNCH uses ``R$_\odot$'' 
to denote exactly 0.25$^\circ$ on the celestial sphere. In project documents, ``r$_\odot$''  refers to the solar radius of $6.96\times10^8$ m.} to 8$^\circ$ (32 R$_\odot$) from the Sun.  
The remaining three 
Observatories (designated ``PUNCH-WFI01'' etc. or ``WFI-1'' etc.) each have a single ``Wide Field Imager'' (PUNCH/WFI) heliospheric imager \citep{laurent_2025} with an approximately 
40$^\circ$ square field of view, to image a portion of the sky with $\varepsilon$ ranging from 3$^\circ$ to 45$^\circ$ (12 
R$_\odot$
to 180 R$_\odot$).  Three WFIs are needed to ensure sufficient coverage of the outer field of view.  The 
instruments are operated synchronously and have matched passbands, to form a single ``virtual 
instrument'' that can produce one full polarimetric image sequence each 4 minutes, throughout the 2-year nominal
mission.  Figure \ref{fig:trefoil} shows how the individual instrument FOVs fit together to form the overall 
PUNCH FOV.  Every imager's FOV overlaps with the FOV of each of the other three, to ensure cross-calibration and seamless merging of the final data. Photometric sensitivity varies across the FOV but is finer than 1\% of the 
expected Thomson scattering brightness from CME fronts, out to 45$^\circ$ from the Sun.

An additional instrument, the Student Thermal Energetic Activity Monitor 
\citep[STEAM;][]{hanson_2024}, rides on PUNCH-NFI00 and was designed to report the time-dependent 
soft X-ray 
spectrum of the 
Sun, to probe the physics of coronal heating and solar flares.  STEAM is physically 
on board PUNCH-NFI00, but is disabled due to an anomaly discovered during Observatory 
integration and testing, in late 2024.

The following sections are overviews of: the science motivation for PUNCH (Section~\ref{sec:Science}),
observing requirements that are driven by the science (Section ~\ref{sec:Requirements}), 
the PUNCH overall mission design (Section~\ref{sec:Mission}), the instruments 
(Section~\ref{sec:Instruments}),
 the primary science data products (Section~\ref{sec:SOC}), and the expected impact of PUNCH on topics outside the primary science (Section~\ref{sec:additional-science}), followed by a 
brief conclusion (Section~\ref{sec:conclusions}). Details of each major topic
are further expanded in relevant articles in 
this Topical Collection, cited in each section.  
Appendix \ref{sec:appendix-nomenclature} contains a suggested nomenclature, adopted throughout the PUNCH
project,
for the various angles and quantities
needed for PUNCH data analysis.  Appendix \ref{sec:appendix-polarization} provides the basic theory 
of 3-D imaging via polarization.

\section{PUNCH Science}
    \label{sec:Science}

Bridging the measurement gap between solar physics and heliospheric physics is critical to 
understanding the Sun and heliosphere as a single system, and to understanding the 
origin and nature of Earth-impacting structures in the solar wind.  These structures
range from small-scale inhomogeneities in the solar corona and solar wind to large-scale
features such as coronal mass ejections (CMEs), stream interaction regions (SIRs)\footnote{We use `stream interaction region'' (SIR) rather than ``corotating interaction region'' (CIR), because SIR is the more general term.  CIRs are a subset of SIRs that persist for at least a solar rotation \citep{pizzo_78}.}, and 
associated interplanetary shocks.  Although these features' ultimate source is the solar corona, both evolution and
interactions occur en route to Earth, and substructures may form and change via plasma 
instabilities or turbulence.  The plasma encompassing the outer corona and inner heliosphere before significant interplanetary processing
(the ``young solar wind'') is a fundamental missing piece of the Sun-Earth connection.

Despite decades of study \citep[e.g.,][]{neugebauer1966,hundhausen_1972, lopez_87,marsch_93,sheeley_97,zurbuchen_98,bruno_01,borovsky_08,suess_09,cranmer_17,viall_borovsky_2020,viall_deforest_kepko_2021}, quantitative insight into the origins of solar-wind structure and variability has been limited by a lack of direct observations. As a result, scientific studies have remained heavily reliant on modeling to connect the remotely observed corona with heliospheric in-situ data \citep{odstrcil_09, Provornikova_etal2024}. PUNCH is specifically designed to fill in this gap by imaging the evolving plasma with sufficient resolution and sensitivity to address the PUNCH science objectives, described in the subsections below.

\subsection{Science Objective 1: The Ambient Solar Wind}
    \label{S-Objective-1}
    
The solar wind fills and defines the heliosphere \citep{parker_1958,hundhausen_1972}.  Even in its ambient state, 
it is highly variable in both time and space. Ejecta, waves, and turbulence form a fluctuating background that 
is sensitive to local physical characteristics, yet also preserves information about the Sun. Streams of fast 
wind intersperse with slower flows, and the domain boundaries separating these streams evolve in response to 
coronal changes and to solar-wind hydrodynamic and magnetohydrodynamic (MHD) instabilities. Variations also occur in boundaries between 
physical regimes, e.g., the $\beta=1$ and Alfv\'{e}n surfaces. The region between the middle 
corona \citep{west2023defining} and Earth undergoes variability on all observable scales.  PUNCH 
is designed to exploit this variability to measure flow itself (Section \ref{S-WG1A}), the onset of large-scale turbulence (Section \ref{S-WG1B}), and
the important, but largely-unexplored, Alfv\'en surface (or zone) (Section \ref{S-WG1C}).

\subsubsection{Science Topic 1A: Solar Wind Flow}
    \label{S-WG1A}
    
\begin{figure}[tbh]
    \centering
    \includegraphics[width=2.3in]{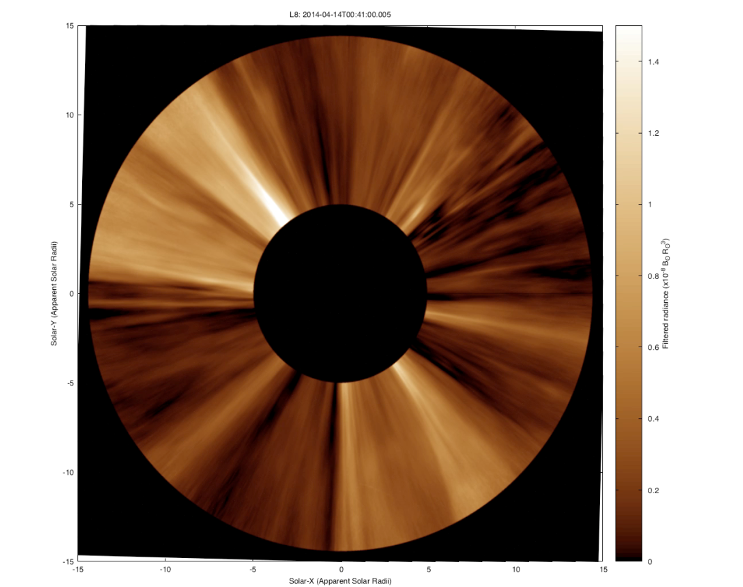}
      \includegraphics[width=2.3in]{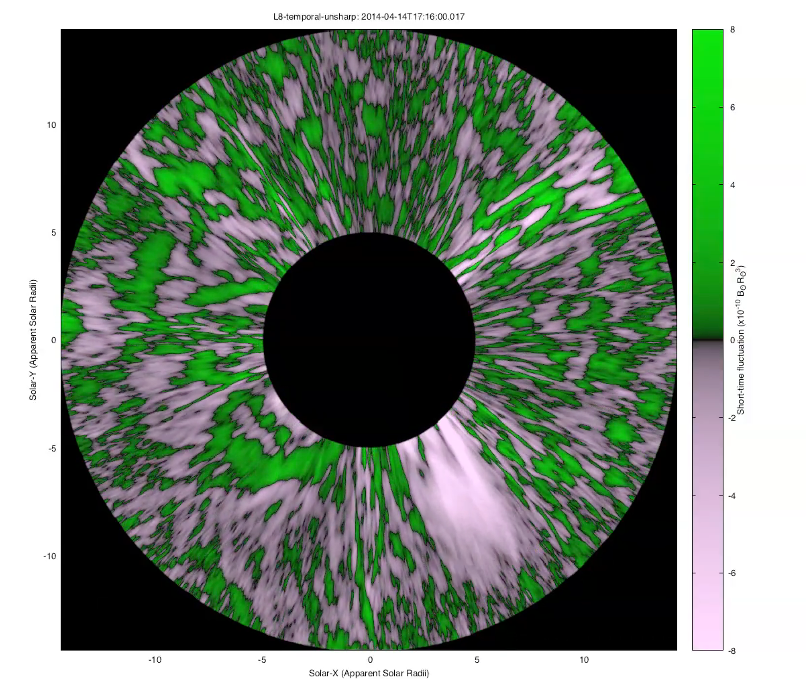}
    \caption{STEREO COR2 white light coronagraph observations from a deep-field campaign \citep{deforest_etal_2018}, with 72 hours of long exposures at 5-minute cadence, reveal outer coronal structure. Left image has had smooth background and stars removed, showing the outer corona dominated by fine ``woodgrain'' structure. Outflow is visible everywhere because of small moving features, and further processing (right image) shows a riotous torrent of blobs and variable streams, tracing flow of the young solar wind.}
    \label{fig:wood-sauron}
\end{figure}

The Ulysses mission \citep{mccomas_00,vonsteiger_00,mccomas_02} revolutionized understanding of the heliosphere when its measurements revealed the solar wind's latitudinal structure. Being an in-situ mission, Ulysses sampled only a single narrow locus along each orbit, and required years to do so. In contrast, imagers reveal instantaneous global structure. Many authors have analyzed solar wind scceleration in the low corona \citep[e.g.,][]{sheeley_97,wang_00}. Deep-field imaging of the mid to outer corona \citep{deforest_etal_2016,deforest_etal_2018} reveals faint moving structures and inhomogeneities in the outer corona at all latitudes and solar wind speeds (Figure~\ref{fig:wood-sauron}), demonstrating the feasibility of tracing solar wind flow and its variability throughout the transition from coronal flow into the separated regime of the young solar wind. These measurements are necessary to constrain current incomplete understanding of solar wind acceleration flow, \citep[e.g.,][]{Cranmer_winebarger_2019}.

\begin{figure}[tbh]
    \centering
    \includegraphics[width=3.in]{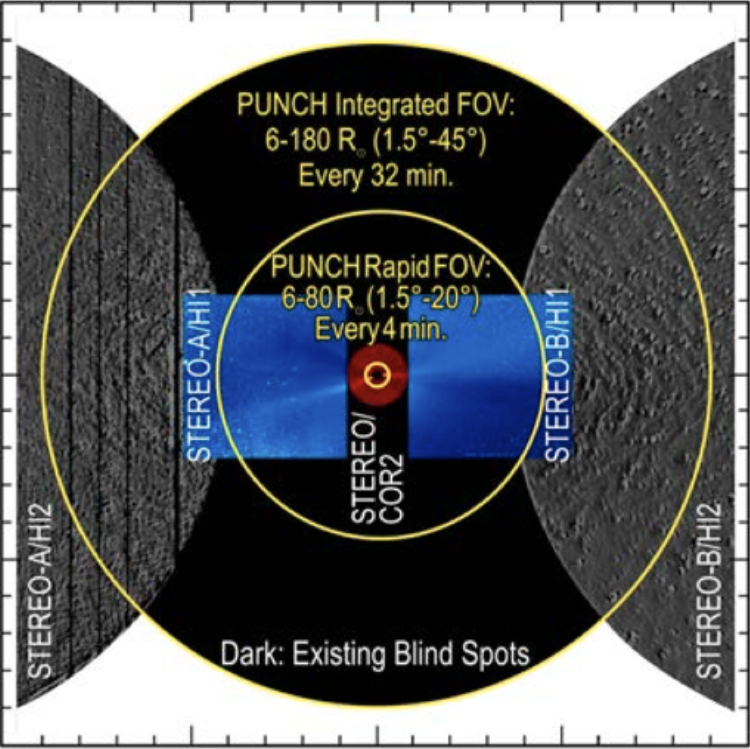}
      \caption{PUNCH eliminates observational gaps between coronal and heliospheric imagers and at the poles, continuously observing the young solar wind with a global field of view (FOV). Shown are STEREO COR2 and HI FOVs, within the full PUNCH FOV.}
    \label{fig:blind}
\end{figure}
    
PUNCH's global field of view (Figure~\ref{fig:blind}) enables continuous tracing of inhomogeneities over all latitudes via deep-field images 10 to 30 times more sensitive than previous coronagraph/heliospheric images and collected every few minutes. Velocity maps are extracted through flow tracking techniques including routine correlation tracking of these inhomogeneities \citep{Attie_etal_2025}.

These observations enable analysis of solar wind acceleration from the outer corona to the inner heliosphere, the changing boundaries between fast and slow solar wind, and the global conditions through which transient structures propagate. They also provide observational ground truth for comparison with theoretical and simulation predictions of evolving global flow, validating and/or falsifying models of solar wind acceleration.

\subsubsection{Science Topic 1B: Mesostructures and Turbulence}
    \label{S-WG1B}

The solar wind near Earth is dominated by fluctuations of unknown origin. Mesoscale structures
(those between the small-scale kinetic regime and large structures like coronal mass ejections or 
stream interaction regions) may form mainly from
solar wind turbulence \citep{elliott_10,matthaeus_11,deforest_etal_2015}, from solar-wind release
\citep{sheeley_97,sheeley_09,borovsky_08,rouillard_etal_2010,viall_etal_2010,neugebauer_12,hardwick_13,kepko_16}, or from a combination of both processes. The dichotomy of mechanisms, and open questions surrounding these mesostructures,\footnote{The term ``microstructure'' has been used since at least the 1990s by the solar imaging community \citep{sheeley_etal_1997}.  We use ``mesostructures'' instead, because the ``micro-'' prefix, used in the context of solar wind science, implies plasma kinetic 
scales -- which are much smaller than the resolution of current imagers including PUNCH.}
were recently reviewed by \citet{viall_deforest_kepko_2021}.  Mesostructures are small in heliospheric imagers, but large compared to Earth's magnetosphere; they are therefore important daily drivers of geospace dynamics and space weather \citep[e.g.,][]{viall_etal2009,dimatteo_etal2022,kurien_etal2024}.  

The characteristics and formation locations of mesostructures are important constraints on models of solar-wind origin and acceleration. In particular, they distinguish models that invoke steady flow along the edges of coronal holes \citep{suess_98,vasquez_03,wang_07} from those that are intrinsically intermittent through interchange reconnection or multiple ejections of small plasmoids \citep{suess_96,fisk_03,endeve_04,rappazzo_05,chen_09,antiochos_etal2011,uritsky_etal2021,raouafi_etal2023}.

Some mesostructures are present in the outer corona and visible in Thomson-scattered light. Two classes of mesoscale density features in the solar wind have been identified out to 0.3 AU; ``puffs'' (including quasiperiodic density features) that originate in the corona \citep{viall_etal_2010,viall_15}, and ``flocculae,'' that become apparent at distances between 0.1 and 0.3 AU \citep{deforest_etal_2016}. Both kinds of variability are faintly discernible with STEREO/SECCHI, but in-situ observations \citep{viall_08,kepko_etal2024} suggest a zoo of similar mesostructure at scales just below the sensitivity limit of that suite. The location and distribution of such features probes the $\beta=1$ surface, where the Sun's magnetic field becomes too weak to stabilize flows against hydrodynamic instabilities in the solar wind.

\begin{figure}[tbh]
    \centering
    \includegraphics[width=3.in]{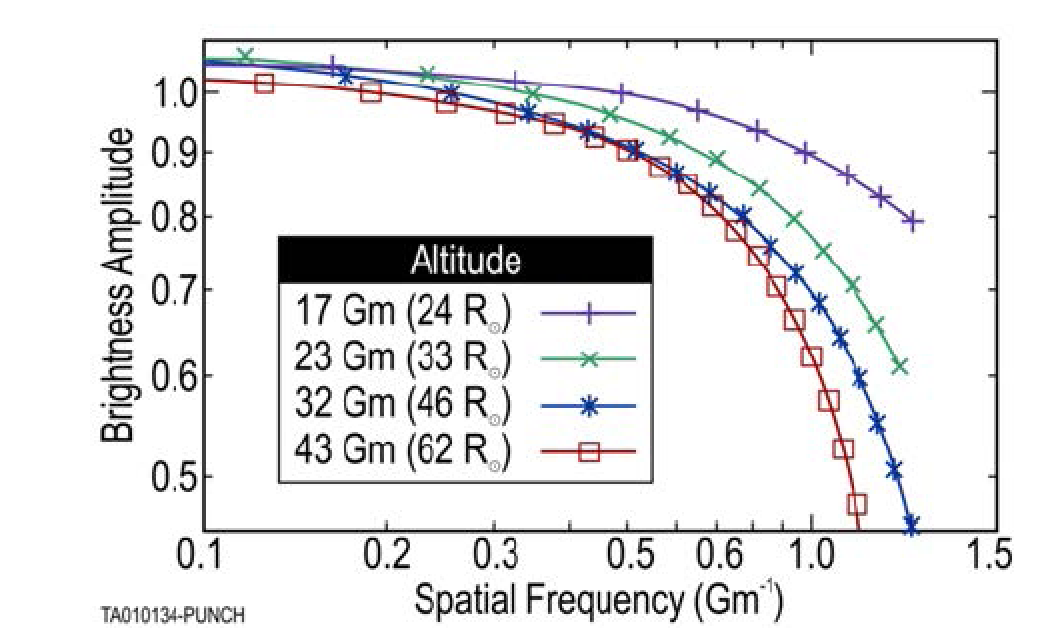}
      \caption{PUNCH is designed to reveal the development of turbulence in the solar wind. Observations of white-light brightness from STEREO/SECCHI reveal spectral steepening with altitude, suggesting onset of turbulence. With 10 times higher sensitivity, PUNCH is designed to resolve the hinted inertial range at spatial frequencies up to 5 Gm$^{-1}$ (200 Mm scale) and to altitudes of 150 Gm.}
    \label{fig:spectrum}
\end{figure}

Evolution of solar wind mesostructures is also a strong probe into the development of large-scale turbulence in the young solar wind. Turbulence may be the key to solving the ``solar-wind heating problem,'' that the solar wind does not cool as expected from adiabatic expansion. For example, at 1 AU the temperatures are 100 times higher than expected from adiabatic propagation alone. It has been proposed that kinetic energy, entrained in large-scale turbulence, cascades downward to kinetic scales in the solar wind, heating it via plasma dissipative effects \citep{coleman_68,leamon_98}. In-situ measurements have revealed a turbulent cascade with a a well-established inertial range at 1 AU \citep{marsch_93,goldstein_99}. Coronal observations suggest the existence of turbulence there as well \citep{coles_89,chae_98,kohl_06,tomczyk_09,liu_14,cranmer_20}, but how and where a turbulent cascade develops remains unclear \citep{zank_92,chhiber_18}. Analysis of images from STEREO/HI-1 \citep{deforest_etal_2016} has shown that the visible radial structures (``striae'') of the solar corona break up via currently unresolved processes at apparent distances between 10$^\circ$ and 20$^\circ$ (40 R$_\odot$ to 80 R$_\odot$) from the Sun. The striae exhibit small lateral motions and spectral evolution, seen close to the HI-1 noise floor, that may indicate the development of a turbulent cascade (Figure~\ref{fig:spectrum}). 

By identifying small faint features, including puffs and perturbations of coronal
striae, and locating them in 3D through polarization analysis (Appendix \ref{sec:appendix-polarization}), PUNCH is intended to explore the origin of solar-wind 
variability (intermittence of the solar source, unstable propagation dynamics, or a combination of the two) and observe the possible development of an
inertial range of turbulent scales in the young solar wind. A wide range of 
mesostructures can be tracked as they form, evolve, and propagate into the heliosphere, 
determining their relationship to each other and to the hypothesized turbulent evolution
of the solar wind. Recent studies with Parker Solar Probe have shown that the turbulent characteristics of the young solar wind within and just above the corona itself, i.e. at solar elongation angles of roughly 10-40 $R_\odot$ from Earth, are quite different from those higher in the inner heliosphere \citep[e.g.,][]
{zhao_etal_2022,zank_etal_2022,bandyopadhyay_etal_2022}.  By separating variation in time from variation in space, PUNCH extends these analyses to probe the onset, development, and potential isotropization of a turbulent cascade in the young solar wind.

Analyzing turbulence with image data gives rise to unique challenges of interpretation, arising from the line-of-sight integration and the angular
(rather than Cartesian) sampling of those lines of sight \citep[e.g.,][their Appendix A]{deforest_etal_2016}. \citet{pecora_etal_2024} discuss at some length the nuances of interpreting PUNCH data to analyze turbulence in the corona and solar wind. \citet{wang_etal_2024} cover analysis of the possible energy-bearing $f^{-1}$ spectrum of density fluctutions in the corona.

\subsubsection{Science Topic 1C: The Alfv\'{e}n Surface}
    \label{S-WG1C}
    
The coronal plasma is in a state of constant expansion into interplanetary space, where it forms the solar wind.
The boundary where the radial flow speed exceeds the fast-mode magnetosonic wave speed is a critical surface beyond which information 
cannot propagate back down to the Sun, at least in the form of MHD waves \citep{weberdavis_67,hundhausen_1972}.  
This ``Alfv\'{e}n surface'', so called because the relevant fast-mode speed is the Alfv\'{e}n speed,  is a natural outer limit of the corona itself, marking a natural border between the extended solar corona and the disconnected solar wind.  It is the true source surface of the interplanetary magnetic field, and its shape and location 
are strong discriminators among solar wind acceleration models. It also plays a role in stellar rotational spin-down \citep{garraffo_15}, which is relevant to the Sun's past and future evolution \citep{metcalfe_16}.  

Models \citep[e.g.,][]{cranmer_07,verdini_09, zhao_10,goelzer_14,tasnim_16} place the location of 
the Alfv\'{e}n surface between about 10 and 30 R$_\odot$, with a complexity that evolves and changes with the solar magnetic cycle.
Fourier filtering of images has revealed the existence of inbound features out
to about 15 R$_\odot$ \citep{deforest_14}, but there remain difficulties
matching the observed speeds and accelerations with solar wind models
\citep{verdini_09,tenerani_16}.
It may be the case that these inflowing parcels are associated with
nonlinear MHD structures such as shocks, jets, or shear instabilities \citep{cranmer_21}, 
complementing expected outflowing parcels from reconnection in the outer corona and heliosphere 
\citep{gosling_05,khabarova_etal_2016}.

In addition, the results of the STEREO deep-field campaign (Figure~\ref{fig:wood-sauron})
suggest that density inhomogeneities may be so great as to disrupt a well-formed 
Alfv\'{e}n surface. In particular, electron density may vary by up to a factor of 10 on
the smallest resolved outer coronal scales of a few tens of Mm
\citep[see also][]{raymond_14}, affecting the speed of MHD waves (slower in dense 
regions; faster in evacuated regions).  For this reason, it may be more appropriate to
think in terms of a fractal or frothy Alfv\'{e}n zone, with profound implications for
the nature of the interplanetary magnetic field boundary \citep{chhiber_etal_2022}.

Recent in-situ studies with Parker Solar Probe have explicitly identified coronal regions where inbound
waves are present, and have therefore identified the Alfv\'{e}n surface and loci within it 
\citep{kasper_etal_2021,ran_etal_2024,zesen_etal_2024}.  Although these in-situ measurements are 
groundbreaking and extremely important, they cannot 
in principle determine the global structure of the surface.

PUNCH studies the location of the Alfv\'{e}n zone through analysis to separate inbound 
and outbound features in the solar wind, and determine the region above which inbound 
motions cease.  Across time, the shape, complexity, and evolution of this zone, and the 
Alfv\'{e}n speed itself, are determined through direct measurement of feature motions. 
Although particular moving features may be due to bulk motion of mesostructures or to waves 
\citep{tenerani_16}, in either case the Alfv\'{e}n speed of the surrounding medium limits 
the relative speed of the bulk flow and the visible feature \citep{cranmer_21}, enabling 
PUNCH's visual-tracking approach to identifying the Alfv\'{e}n surface (or zone).  
PUNCH science on this topic is described in more detail by \citet{Cranmer_etal_2023}.
    
\subsection{Science Objective 2: The Dynamic Solar Wind}
    \label{S-Objective-2}

The impact of large, dynamic structures such as coronal mass ejections (CMEs), stream interaction regions (SIRs), 
and shocks on the solar wind and geospace is well documented \citep[e.g.,][]{schwenn_06,pulkkinen_07}. A deeper 
understanding of the physical processes controlling these space-weather drivers holds the key to improvements in 
predictive capability. However, tracking, monitoring, and diagnosing even the most basic physical properties of 
transients (speed, density, morphology) has remained limited and challenging. This motivates PUNCH's improvements 
in resolution, FOV, and photometry, and also its polarimetric capabilities, which can improve predictions of CME and SIR 
location and arrival time \citep{howard_etal_2013}. 3-D localization using polarimetry improves on unpolarized stereoscopy 
alone, by eliminating multiple-viewing-angle confusion effects. Large-scale structures may be located in several ways: 
constraint of kinematic envelope models \citep{howard_etal_2013}, direct analytic location of 
substructures \citep{deforest_etal_2017} and kinematically constrained tomographic inversion \citep{jackson_06,bisi_08,jackson_etal_2011,morgan_15}, as well as forward modeling \citep[e.g.,][]{Talpeanu2020}.

\subsubsection{Science Topic 2A: Coronal Mass Ejections}
    \label{S-WG2A}
    
Coronal mass ejections (CMEs) are the largest perturbations of the solar wind, and the source of the strongest space
weather events at Earth.  They have been objects of study since the 1970s. However, many aspects of CME physics, particularly how they propagate 
and evolve through the solar wind medium, have not been well constrained by measurement.
Earlier imagers could track only the largest or brightest components of CME structure beyond 
20 R$_\odot$, yielding limited information about how CME sub-structures interact with each other or with the solar wind, 
despite a handful of joint tracking/in-situ studies of particular events 
\citep[e.g.,][]{howard_deforest_2012b,deforest_etal_2013a,palmerio_etal2024}. PUNCH is designed to enable analysis of 
not only bulk CME bodies, including the front trajectory and possible bulk deflection, but also
the interior features that may be remnants of the source structure 
in the lower corona.

The structure and orientation of CME magnetic fields at the Earth largely determine space-weather impact,
representing a central challenge for heliophysics. Current understanding of CME magnetic structure is based
on in-situ sampling \citep{gosling_87,lepping_90}, in combination with near-Sun measurements of CMEs 
\citep{dere_99} and their precursors \citep{gibson_15}. These observations generally support a model of
CMEs as magnetic flux ropes \citep{vourlidas_13}. However, CME magnetic structures are known to change in
the solar wind due to distortion \citep{howard_deforest_2012}, deflection and rotation \citep{isavnin_14},
CME-CME interactions \citep{gopalswamy_01}, and magnetic reconnection/erosion \citep{gosling_05,ruffenach_15}.
Assuming the axis of the CME flux rope can be obtained through model or bulk 3D measurement, its chirality
follows from the direction of circulation about the axis \citep{deforest_etal_2017}.

PUNCH science on the topic of CME propagation and evolution is described in more detail, including a 
thorough review of recent work, by \citet{malanushenko_etal_2025}.

\begin{figure}[tbh]
    \centering
    \includegraphics[width=3.in]{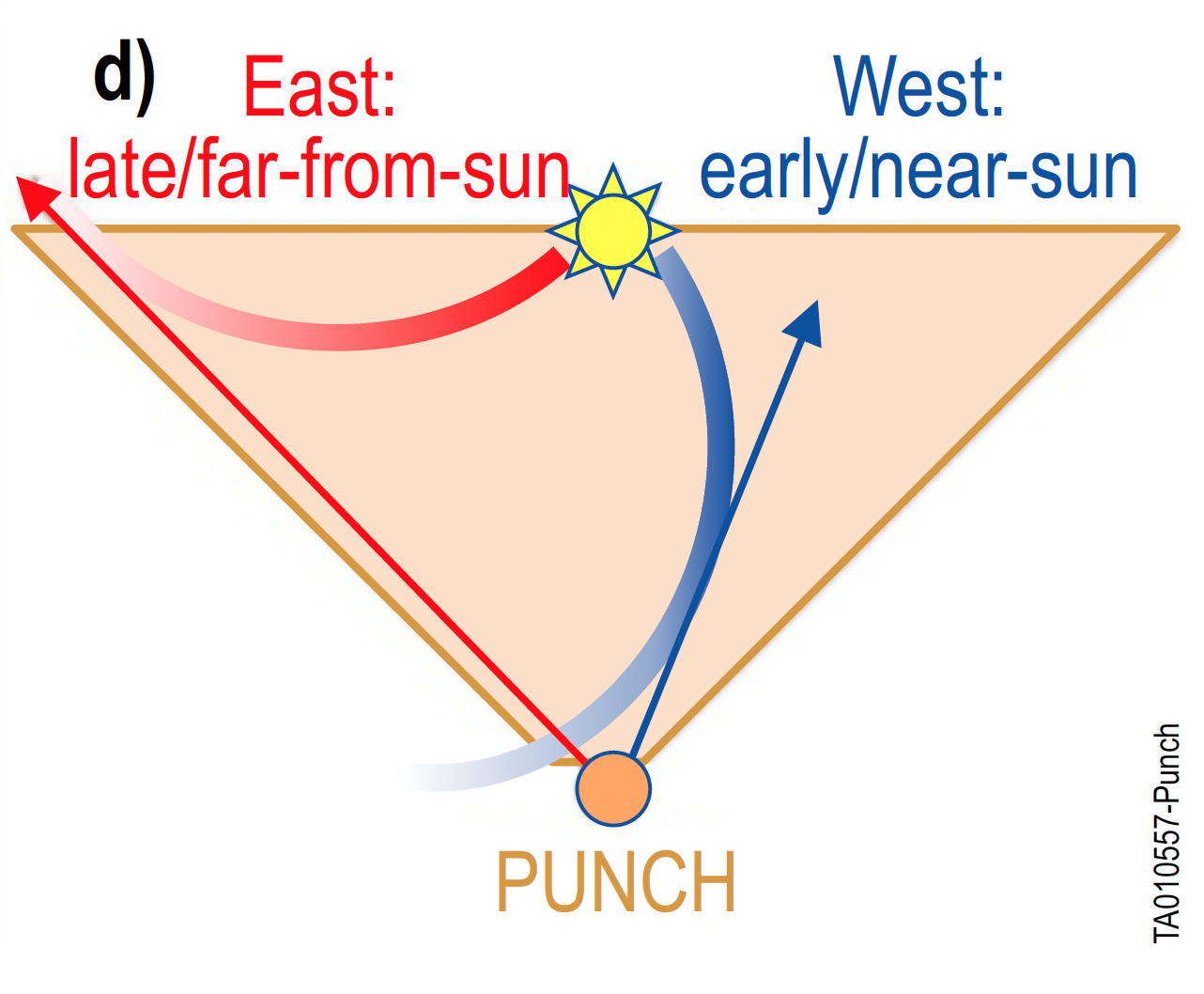}
      \caption{PUNCH observes both early and late manifestations of SIRs, enabling analysis of their birth and evolution.  }
    \label{fig:CIR}
\end{figure}

\subsubsection{Science Topic 2B: Stream Interaction Regions}
    \label{S-WG2B}

Stream interaction regions occur at the interface between radial fast and slow solar-wind streams whose sources rotate with the Sun \citep{smith_76, pizzo_78, pizzo_80, pizzo_82}. Although CMEs are the focus of much current space-weather related work, the quasi-stationary SIRs (including their long-lived subset, co-rotating interaction regions or CIRs), 
are also a significant source of space weather, particularly in the Earth's radiation belts \citep{miyoshi_05}.

Prior to STEREO/HI, SIR research mainly depended on radio interplanetary scintillation (IPS) observations \citep{breen_98, bisi_10, dorrian_10} or used in-situ measurements that typically sample a single track through the feature \citep{tappin_84, pizzo_94, moebius_02, bucik_09}. From the wide-angle view on SIRs provided by heliospheric imagers, we now know SIRs manifest very differently depending on whether one observes from ahead looking east, or from behind looking west \citep{wood_10}. In particular, the east view observes the advancing SIR structure along its developed part, far from the Sun, while the west view observes along a much closer, earlier phase of evolution during the short interval as the SIR sweeps past (Figure~\ref{fig:CIR}). 

In addition to differences expected from simple geometry \citep{sheeley_etal_2008,tappin_howard_2009} we see other distinctions between the two views such as several ray-like features that appear to run into the SIR in the East view \citep{rouillard_etal_2010, williams_11,conlon_15}. Meanwhile, the west view of the SIR structure \citep{sheeley_etal_2008, rouillard_09, rouillard_etal_2010} exhibits distortions and wavelike transients that are mostly short-lived and difficult to track far from the Sun. These are the likely signatures of the early formation of the SIR at the solar-wind regime interface, but their relationship with the later-forming portion of the SIR, the solar wind, and embedded mesoscale structures remains unknown.

PUNCH's high spatial and temporal resolution, polarimetry, and wide FOV encompassing both 
eastward and 
westward directed views are designed to enable a 3D understanding of interplanetary SIR fronts and to 
track evolution of developing turbulence and other substructure in the fronts themselves, to better
understand these sparsely-studied, yet very large, elements of the inner heliosphere.  The mesoscale structure within SIRs, that PUNCH will image and track, is of particular importance in understanding the space weather impact of SIRs on geospace \citep[e.g.,][]{kilpua_etal2015, turner_etal2019, kepko_viall2019, katsavrias_etal2025}.

PUNCH science
on this topic is described in more detail by \citet{dekoning_etal_2025}.

\subsubsection{Science Topic 2C: Shocks}
    \label{S-WG2C}

The basic geometric theory of interplanetary shocks is well established \citep{hundhausen_1985}, but large-scale shock 
structure is rarely simple. Shock morphology depends on the surrounding medium \citep{odstrcil_09,tappin_97}. Many shocks 
and CME sheaths appear to have an asymmetric structure \citep{pizzo_11, feng_12}. Further, some shocks 
appear to form on the CME flanks rather than, as expected, at its nose where density pileup is greatest
\citep{ontiveros_09,wood_09,wood_10b}.

Changes in hydrodynamic shock behavior from turbulence are well documented by aerodynamics studies 
\citep{ribner_53,crichton_69} and laboratory experiments \citep{plotkin_72,williams_73,ribner_86}. 
Magnetohydrodynamic shock-turbulence interactions are just as strong \citep{zank_03} and turbulence 
significantly strengthens shock magnetic fields \citep{giacalone_07}. Simulations suggest that CMEs are 
strongly affected by turbulent instabilities across their shocks \citep{odstrcil_11}, and some 
in-situ shock measurements suggest a corrugated shape \citep{szabo_95,kasper_05} indicating their 
turbulent instability or fluctuations associated with solar-wind variability may be distorting 
shock fronts. Such shock corrugations may be responsible for the acceleration of solar energetic 
particles (SEPs) and type II radio bursts, but are currently not resolvable \citep{cairns_etal_2003,knock_03}.

Prior generations of coronagraphs and heliospheric imagers have been used to explore shocks, as well as the
interaction of CMEs with each other and with other large-scale structures such as SIRs 
\citep{lugaz_12,liu_12,temmer_12,harrison_12,moestl_12,wood_12}. 3D analyses using the Solar Mass Ejection
Imager (SMEI) data \citep{jackson_09,jackson_etal_2011} show that shocks appear ``spotty,'' reminiscent of
expectations for turbulent instabilities. However, such studies have been severely hampered by limitations
associated with instrument sensitivity and motion blur.

PUNCH is intended to observe the spatial irregularities and brightness variations of solar-wind CME 
interactions on scales 10 to 30 times below the scale of the structures' gross envelope. This should
provide a cross-scale picture of shock formation and shock turbulence interactions, and CME/SIR interactions, 
allowing the exploration of solar-wind variability on interplanetary shock behavior.   The PUNCH science topic
of shock physics is described in more detail by \citet{aldayeh_etal_2025}.

\section{Science Requirements}\label{sec:Requirements}
 
\new{Engineering requirements are central to development
of any mission, because large projects in general and space missions in particular
represent trades between capability and cost.  Because the process to develop
mission-level requirements is not as familiar in the scientific literature as reporting of 
the resulting instrument 
performance, we describe here both the PUNCH top-level observing requirements and 
an overview of the methodology used to develop them.}

\new{Each topic in Section \ref{sec:Science} levies specific scientific observing 
requirements on the PUNCH mission: aspects of observation that are needed to resolve the
corresponding science question.  \new{These requirements, such as spatial/angular 
resolution and photometric sensitivity, are specific to each topic and were derived from 
review of the scientific literature at the start of mission definition. Distilling these
scientific requirements into testable engineering requirements meant establishing a complete
minimum set of mission parameters to constrain, rendering all the individual constraints on those parameters
into a form that is 
directly comparable across questions, and identifying a
notional set of ``Baseline Science Requirements'' (BSRs) that would satisfy all of the science questions. 
These requirements 
became the mission definition.  We executed the full derivation process twice, using two different criteria.  We developed the 
BSRs using best-estimate observing requirements to fully address each particular scientific question, 
with an identified analysis path to closure of the question.  We also developed a separate set of 
``Threshold Science Requirements'' (TSRs) based on best-estimate observing requirements to significantly advance
each scientific question.}  The BSRs were used by the instrument and engineering teams to derive a series of 
lower level requirements which, in turn, drove each aspect of the mission design.  The TSRs provided understanding 
of how flexible each requirement might be, enabling development trades or compromises in case they were needed.}

\old{During mission definition, we developed a set of mission-wide requirements that encompasses all these 
specific observing 
requirements. 
Here we describe the seven primary mission-wide observing requirements (``Baseline Science Mission Requirements,'' or BSRs)
and the questions that drive them.  These requirements flow down to the mission definition, and formed the basis of the 
engineering requirements structure for all aspects of PUNCH. }

\new{Developing mission top-level requirements itself requires compromise between three motivations: first, the desire to
fully constrain the capability of a new instrument suite, which determines the minimum set of requirements; second, 
the need for each requirement
to fully verifiable during development, which forces each requirement to refer to testable properties of the mission (without
reference to the science target); third, the need for requirements to be feasible, which forces simple structure 
and requirements that can be analyzed
and tested straightforwardly.}

\new{PUNCH science is broad enough that several requirements were challenging to specify.  The needed 
image resolution varies across
the field of view and is therefore specified at two representative solar elongations.  Observing cadence varies for different
targets and drives overall mission design, so the requirement is more complex to avoid overspecification. Photometric
sensitivity is coupled to image resolution, and therefore required a novel formulation.  These nuances are more fully
described in the subsections below.}
\old{In many cases, an observing requirement varies across the field of view.  For example, spatial resolution can be 
coarser far from the Sun than close to it, because of the known overall expansion of streams and structures in the solar wind.  In these cases, values are specified at two representative radial 
positions: 15 R$_{\odot}$ (3.75°) from Sun center, to capture the outer corona in the vicinity of the Alfv\'{e}n zone and within the region visually dominated by radial striae, and 
80 R$_{\odot}$ (20°) from Sun center,
to capture the young solar wind above the transition from anisotropic to isotropic structure \citep{deforest_etal_2016}.}

Table \ref{tab:reqs} lists the seven BSR requirements; their rationales are in the sections below.

\begin{table}[tbh]
    \begin{center}
    \begin{tabular}{l|c|c|c|c}
        \hline
         \multirow{2}*{\textbf{Title}} & \multirow{2}*{\textbf{Requirement}} & \textbf{Performance} & \textbf{Driver} & \textbf{Sec-}\\
         & & \footnotesize{design/\textit{measured}} & \textbf{(topic)}&\textbf{tion}\\
         \hline
         1: FOV position angle $\alpha$ & $\ge $270$^\circ$ & \textit{360$^\circ$} & all & \ref{SS-fov}\\
         \cline{1-5}
         2: FOV elongation $\varepsilon$    & 2$^\circ$$-$40$^\circ$ & \textit{1.5$^\circ$$-$45$^\circ$} & 2A & \ref{SS-fov}\\
         \cline{1-5}
         \multirow{2}{*}{3: Angular Resolution} & $\varepsilon=3.75^\circ$: $\le $3'  & \textit{1.4'} & {1B} & \multirow{2}{*}{\ref{SS-res}}\\
            & $\varepsilon=20^\circ$: $\le$ 5'  & \textit{1.8'} & {2C} & \\
        \cline{1-5}
        4: \footnotesize{Photometric Sensitivity} & $\varepsilon=3.75^\circ$: $\le$ 320 & 200 &
        \multirow{2}{*}{1B} & \multirow{2}{*}{\ref{SS-phot}}\\
        \footnotesize{($\times10^{-17} B_{\odot}$: $1^{\circ}$, 4 min)} & $\varepsilon=20^\circ$: $\le $10  & 3 & &  \\
        \cline{1-5}
        \multirow{1}{*}{5: Observing Cadence} & \multirow{1}{*}{Per Table \ref{tab:st-cov}} & \textit{Exceeds} & \multirow{1}{*}{all} & \multirow{1}{*}{\ref{SS-res}}\\
        \cline{1-5}
        6: \footnotesize{Polarimetric Sensitivity} & $\varepsilon=3.75^\circ$: $\le$640 & 300 & 1B & \multirow{2}{*}{\ref{SS-pol}}   \\
            \footnotesize{($\times10^{-17} B_{\odot}$: $1^{\circ}$, $4$ min)}           & $\varepsilon=20^\circ$: $\le$20 & 5 & 2A & \\
        \cline{1-5}
        7: Mission Duration & $>$15 months & 24 months & 2A & \ref{SS-fov} \\
        \hline

    \end{tabular}
    \end{center}
    \caption{The seven PUNCH Baseline Science Mission Requirements (BSRs) define the observing characteristics of the mission,
    and drive the engineering requirements structure. Each requirement is explained in the corresponding section below. Requirements 3, 4, and 6 are specified at two separate representative elongation angles $\varepsilon$.
    Some BSRs were fully verified in-flight before article submission; they are marked in \textit{italics}. }
    \label{tab:reqs}
\end{table}

\begin{table}[tbh]
    \begin{center}
    \begin{tabular}{p{1cm}|p{1.15cm}|p{4cm}|p{2cm}|p{1cm}}
        \hline
        \textbf{FOV} & \textbf{$\varepsilon$ range} & \textbf{Requirement} & \textbf{Performance} & \textbf{Driver} \\
        \hline
        \vfill Inner \vfill & \vfill 2.5°-30° \vfill & Image pairs: $\le$8 min. cadence; $\le$24 min. gap between pairs & 4 min. cadence, $<$16 min. gaps & \vfill 1A,2C \vfill \\
        \hline
        Middle & 5°-27.5°  & 20 min. cadence & 4-12 min. & 1C,2B \\
        \hline
        Outer & 2°-40° & 40 min. cadence & 4-30 min. & 1B,2A \\
        \hline
    \end{tabular}
    \end{center}
    \caption{PUNCH cadence requirements vary with elongation angle $\varepsilon$ across the field of view, based on each science question.}
    \label{tab:st-cov}
\end{table}

\subsection{Field of View and Duration}
    \label{SS-fov}

The PUNCH overall field of view (FOV) is required to subtend a minimum of 270° in position angle around the Sun.  This angle 
ensures that, regardless of orientation, the field covers both polar and equatorial regions of the Sun.  The mission achieves 360°
coverage in the typical case, although lunar interference was calculated, pre-flight, to ``wash out'' up to $45^\circ$ of position angle for a few days each month.

PUNCH images are 
required to extend inward to below 8 R$_{\odot}$ (solar elongation $\varepsilon\le2^\circ$), to capture coronal dynamics below 
the pre-flight best estimated
altitude of the Alfv\'{e}n zone; this altitude is driven by all three of the Objective 1 topics.  The field extends 
outward to $45^\circ$ to 
track CMEs well out into the heliosphere and identify internal evolution and/or possible deflection.

The mission duration is set by the need to observe ten well-presented large CMEs over the course of the 
mission lifetime (Section \ref{S-WG2A}).  Based on a launch into the active phases of the solar cycle,
existing counts 
in the CaCTUS catalog \citep{robbrecht_etal_2009}, and a selection factor of 2.5\% to account for 
CME size, presentation, direction of propagation, overlap, and lunar interference, we estimated 0.025 
well-presented, well separated, front-side ``PUNCH large CME events'' per day, driving a mission duration
requirement of 15 months and (with margin) the overall mission design duration of 24 months.

\subsection{Photometry}
    \label{SS-phot}
    
PUNCH photometric sensitivity is specified in ``normalized B$_\odot$'' units, defined below.  
``Sensitivity'' refers to the
instrument's ability to distinguish a feature of the specified radiance (SI units: W m$^{-2}$ sr$^{-1}$)
against the measurement noise floor.  Radiance is specified in units of B$_\odot$, the mean solar
photospheric radiance ($2\times10^7$ W m$^{-2}$ sr$^{-1}$), because the Sun itself illuminates the 
corona and solar wind \citep[e.g.,][]{billings1966book,howard_deforest_2012} and these units are 
therefore both convenient and closely related to the optical depth of observed solar wind features.  

Noise level (and therefore photometric sensitivity) in the PUNCH data is independent 
of the signal under study, because the main signal-dependent element of photometric noise 
(photon shot noise from the Thomson-scattered K coronal signal itself) is small compared to other sources.
Significant contributors to the PUNCH photometric noise are photon shot 
noise from the F coronal background and starfield, read noise from the camera, and 
(in the far field) residual errors in the background model used to 
remove the starfield from the data (Section \ref{sec:SOC}); all of these
elements are fully or approximately independent of the signal from the solar wind, and we therefore treat them as a 
fixed ``noise floor'' against which the desired signal is compared to derive a 
signal-to-noise ratio (SNR). For specification, we consider sensitivity as bare detection, 
i.e. SNR=1 in the 
target feature.

Specifying a radiance sensitivity relevant to all the science topics is problematic because, 
even in PUNCH's signal-independent noise regime, photometric uncertainty depends on 
the apparent size and required time resolution for each feature under study.
In particular, the 
number of independent samples of the noise field is proportional to the solid angle (total number 
of pixels) occupied by a
feature and to the length of time used to accumulate an image for analysis.  These samples all add in
quadrature so that in a feature of size $\Omega_f$, observed for a time $t_{obs}$, the noise $N_f$
in the average brightness (radiance) of a feature scales as:
\begin{equation}
    N_f = {N_0}\sqrt{\Omega_0 t_0 / \Omega_f t_{obs}}\label{eq:noise-omega}
\end{equation}
compared to a reference feature of apparent size $\Omega_0$ with a photometric
noise level of $N_0$ 
in the same data set, observed for a reference interval $t_0$.  In other words, given two
different-sized features with similar brightness and background noise field,
the larger feature will yield less average-brightness noise because its radiance is 
averaged across more pixels; and, likewise, averaging over more time will yield less noise,
because the inferred overall radiance is an average over more exposures.
Moreover, because $\Omega_f \propto \zeta_f^2$ for similarly shaped features with linear image-plane scale $\zeta_f$, Equation \ref{eq:noise-omega} reduces to:
\begin{equation}
    N_f = {N_0}\frac{\zeta_0}{\zeta_f}\sqrt{\frac{t_0}{t_{obs}}}\label{eq:noise-r}
\end{equation}
where $N_0$, $t_0$, and $\zeta_0$ are constants of the observation (i.e. independent
of the feature under study), and we use the greek $\zeta$ to remind ourselves that 
image-plane scale is a measure of angle, not distance.

For convenience, we choose $\zeta_0$ as $1^\circ$ of apparent size ($\Omega_0 = 1$~deg$^2$), and normalize photometric sensitivity 
to that spatial scale; and likewise specify $t_0$ as four minutes, to match the punch
observing cadence described in Section \ref{sec:conops} below, and normalize photometric 
sensitivity to that temporal scale.  
Using these normalized radiance units allows comparison between the sensitivities required 
for the six science topics,
independently of the spatial and temporal resolution required for each. 

\begin{figure}[tbh]
    \centering
    \includegraphics[width=3.5in]{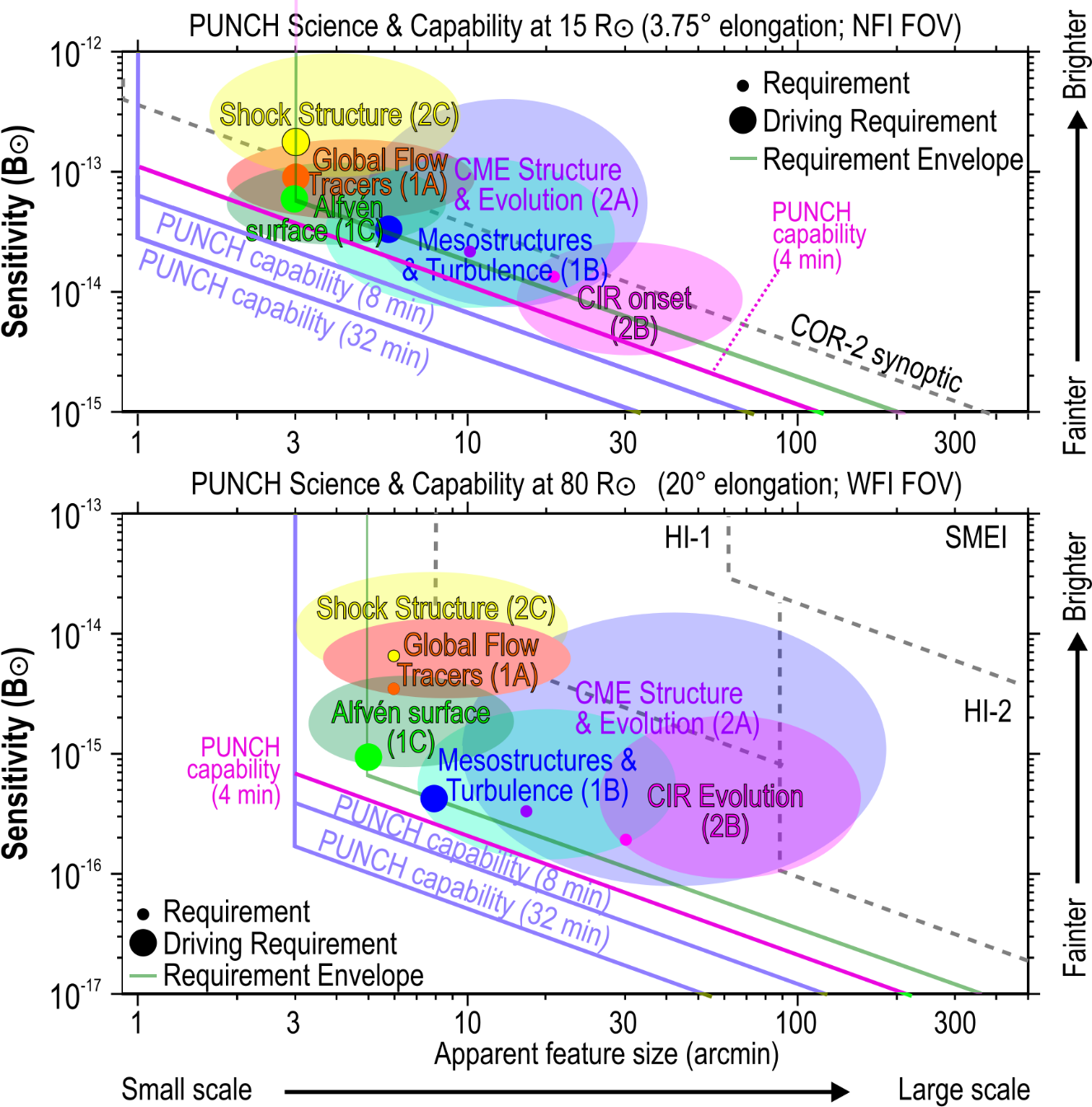}
    \caption{PUNCH resolution and sensitivity requirements are driven by the
    individual science topics, and
    are plotted  relative both to known structures and existing instrument
    capability. Ovals represent particular science topics; green solid line
    represents driving requirements; dashed lines show the observing envelopes
    of prior missions.  Diagonal sensitivity lines mark constant values of $N_0$ and reflect the 
    dependence of photometry spatial scale. In
    both the corona (TOP) and young solar wind
    (BOTTOM), topic 1C drives the resolution requirement and topic 1B drives
    the sensitivity requirement.  Averaging PUNCH 4-minute-cadence images across
    time yields higher sensitivity as shown by the ``8-min'' and ``32-min'' lines at
    the bottom of each plot.}
    \label{fig:photometric_requirements}
\end{figure}

Figure \ref{fig:photometric_requirements} shows the 
interplay between driving resolution and photometric requirements, 
prior instruments, and classes of feature in the corona and young solar wind. 
Loci of constant $N_0$ form diagonal lines on that plot, reflecting the result (illustrated
in Equation \ref{eq:noise-r}) that radiance sensitivity varies with spatial scale.

\subsection{Polarization}
    \label{SS-pol}

Polarization sensitivity is driven by the need to resolve 3D locations of 
ejecta and CMEs (Sections \ref{S-WG1B} and \ref{S-WG2A}).
The basic theory of 3D imaging through polarimetry is in Appendix
\ref{sec:appendix-polarization}. 
Crude linearization of the relationship between the polarization ratio $pB/B$ 
and scattering angle $\chi$ (Appendix \ref{sec:appendix-polarization})
yields a slope
$\Delta\chi/\Delta(pB/B)$ of roughly 0.6$^\circ$ per percent 
polarization, over 
most relevant values of $\chi$.  This ratio connects
precision of 3D spatial position to polarimetric precision for PUNCH 
observations. PUNCH's polarimetric requirements 
are specified as photometric requirements on both $B$ and the excess 
polarized brightness $pB$, based on the known brightness of target 
structures and the required precision of 3D location.

The relationship between photometry of individual camera images
through a polarizing filter, and polarimetry using multiple
images through a trio of such filters, is nontrivial but tractable 
via algebraic analysis.  \citet{deforest_etal2022b} recently revisited
that relationship via derivation from basic polarization physics.  As a 
rule of thumb: noise 
levels in PUNCH unpolarized
$B$ data products (derived from triplets of images through polarizing filters) 
scale approximately as would
those in sums of unpolarized 
exposures; while noise levels in $pB$ data products from PUNCH are worse by
a factor of at most
$\sqrt 3$, compared to the corresponding $B$ data product. 

\subsection{Spatial and Temporal Resolution}
    \label{SS-res}
    
Studying the interplay of turbulent and injected features (Section
\ref{S-WG1B}) drives PUNCH's spatial resolution in the inner field.
The size scale of Sheeley blobs \citep{sheeley_etal_1997} observed with SOHO/LASCO is 
roughly 10$'$; determining whether there is significant substructure requires 
3x over-resolution of the known features \citep{deforest_2007}, leading to the 
requirement of 3$'$. Studying shock evolution (Section
\ref{S-WG2C}) drives spatial resolution in the outer field, based on preliminary 
simulations showing possible shock ``crinkling'' via the Rayleigh-Taylor instability.

The temporal resolution requirement is different in different parts of the PUNCH
field of view.  The fundamental cadence is driven by flow tracking and shock front
analysis (Sections \ref{S-WG1A} and \ref{S-WG2C}) in the inner portion of the field
of view.  An 8 minute requirement derives from the need to avoid the ``mistaken 
identity problem'' \citep{deforest_etal_2007} in feature tracking, in which feature identity can erroneously shift from one visible feature to another if the observing cadence is too sparse.  In the outer
portion of the field of view, 40 minutes is required for coherent tracking of CMEs
across the inner solar system.  

PUNCH operates at a fundamental cadence of 4 minutes,
providing margin against the 8 minute requirement, and in its nominal configuration
the mission achieves 4 minute cadence at all elongations from the inner edge of 
the FOV to approximately 20° from the Sun (marked in Figure \ref{fig:trefoil}).

\section{PUNCH Mission Design \label{sec:Mission}}

The PUNCH mission design was set by physical constraints on dynamic range and data rate.  Two separate 
instrument types (the NFI and WFI, described in Section \ref{sec:Instruments}, are required to 
capture the seven orders of magnitude in dynamic range across the field of view (Figure \ref{fig:brightness}).
An initial cost trade showed that over a two-year mission, ground station contacts would drive cost for a deep-space probe,
and proved too expensive for a Small Explorer program.
Operating from LEO reduced ground costs significantly, but also led to obstruction of the field by the Earth itself. 
Fortunately, the ready availablity of commercial subsystems reduced the cost of a spacecraft enough that
a four-spacecraft constellation could be built and operated within NASA's Small Explorer cost cap.  
PUNCH is therefore a ``1+3'' constellation, with one major instrument mounted on each of four interchangeable spacecraft,
and an instrument complement comprising one NFI and three WFIs. The flight assets thus comprise four Observatories, each
integrating a spacecraft and an instrument.  We chose to fly NFI on its own spacecraft, rather than co-hosting it with a WFI, to maximize the cost benefit of
uniform spacecraft design across the entire constellation.

\subsection{Orbit, launch sequence, and concept of operations\label{sec:conops}}

PUNCH launched into in circular low Earth orbit, initially 650km altitude (in the ionosphere topside F-region), in a Sun-synchronous configuration with 
6am time of ascending node and 97.9° inclination.  This orbit was chosen to ensure that
the missions were always close to the ideal observing geometry relative to Sun and
Earth.  

The spacecraft coordinate system
is defined such that $+X$ (roll axis) is normal to the solar panels at the front of the 
spacecraft, $+Z$ (yaw axis) is normal to the instrument deck,
and $+Y$ (pitch axis) forms a right-handed coordinate system.  During science operations,
each spacecraft maintains $+X$ pointing at the Sun (via uplinked solar ephemeris), and 
rolls so that the zenith is close to the XZ plane.  

\begin{figure}
    \centering
    \includegraphics[width=0.85\linewidth]{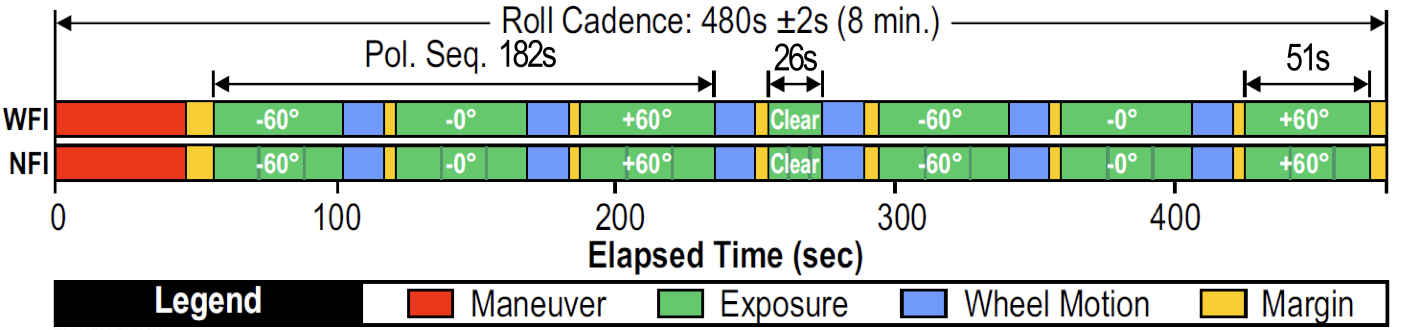}
    \caption{PUNCH observing sequence timeline begins each 8-minute observing window with a roll maneuver, then holds pointing for two three-exposure polarized image sequences and one clear (unpolarized) exposure.  The polarized sequences form a uniform, 4-minute cadence data set.  The observing sequence is repeated for the entire science mission, aside from minimal interruptions for calibration and maneuvering. }
    \label{fig:observing-sequence}
\end{figure}

PUNCH collects primary data on a 4 minute cadence, broken into 8-minute 
observing windows. During each observing window, all four spacecraft 
roll to place $+Z$ ahead of the local zenith, then hold fixed Sun-pointing 
with inertial (celestial) roll constraint, while acquiring image data for 
approximately seven minutes.
The acquisition times are synchronized to within $\pm$1 second across the 
constellation, and each camera acquires photons with an acquisition duty cycle above 50\%
(Figure \ref{fig:observing-sequence}). 
At 650 km altitude, the terrestrial horizon dip is 24.8$^\circ$. The orbital period at the
start of the mission is 98 
minutes, gradually decreasing to 94.5 minutes shortly before re-entry.  Therefore, over 
a single 8 minute observing window all four 
spacecraft traverse 29.4$^\circ$ of orbital anomaly, gradually increasing to 30.4$^\circ$ over the
orbital lifetime of the mission.  This pointing sequence 
maintains the Earth's horizon a minimum of 9.5$^\circ$ below (-Z) the 
spacecraft XY plane at all times, compared to a requirement of 4$^\circ$ for 
WFI.  This allows stray light control for WFI and passive cooling of the 
camera for all four 
instruments.  All operations are scheduled around this regular, constellation-wide, 
sequence of 8-minute observing windows and the 4-minute regular cadence of data acquisition.

The observing sequence is diagrammed in Figure \ref{fig:observing-sequence}, which 
highlights the repetitive nature of the observations.  Each polarized image 
sequence occupies approximately 3 minutes and comprises three image acquisition
cycles through three polarizing filters oriented 60$^\circ$ apart in angle.  In
normal operations, each
8-minute observing window contains two 
such sequences, punctuated at the start by a pause to roll the spacecraft and
at the center by a single clear (unpolarized) exposure from the instrument.  Thus
each observatory acquires seven images per 8-minute observing window.

The four-minute cadence of PUNCH observation, twice the required rate described
in Table \ref{tab:st-cov}, was driven by the need to mitigate
the ``high altitude aurora'' seen by the SMEI heliospheric imager \citep{eyles_etal_2003} above 800km altitude \citep{mizuno_etal_2005}. The 
high altitude aurora is interesting in its own right but impacted SMEI's solar wind imaging.  Collecting a full
polarized sequence every four minutes provides resilience against the aurora, because of the 
parallax from PUNCH's orbital motion: if a particular camera frame is affected by the 
high-altitude aurora, the frame four
minutes earlier or later is unlikely to be affected by the same feature 
in the same image location.  The 8-minute clear sequence was chosen both as a
backup for ease of 
post-processing and to double-check the analysis of the polarized sequences.

The approximately 30° roll sequence permits sky imaging even during lunar phases in which the Moon passes 
through the WFI field of view: even if certain WFI exposure frames are washed out by
moonlight, frames from adjacent roll intervals are unaffected as the Moon is below the baffle plane.  
NFI sees the Moon during the new-Moon phase, but based on experience with earlier LEO 
coronagraphs \citep[e.g., Solwind][]{Sheeley_etal1980} the NFI data were predicted to 
not be strongly affected; and ``first light'' imagery from 2025 May indeed shows that the 
new Moon does not impact NFI observations.

The 4-minute cadence sequence is planned to continue for the entire mission, interrupted
at long intervals by calibration exposure sequences or orbital-trim maneuvers.  Ground 
communications are via low-gain antenna and do not interrupt science. Orbital trim maneuvers,
comprising a slew to rotate the -Z face to the prograde or retrograde direction, a few-second
rocket thrust event, and a slew back to nominal science mode, are each fully contained within
a single 8-minute observing window, to minimize impact to the overall observing cadence.

The four spacecraft launched together on a single Falcon 9 rocket,
and spring-deployed at approximately 1 m/s from the launch vehicle.  WFI-2 
deployed in the forward orbital direction, NFI-0
in the reverse orbital direction, WFI-3 60° from the forward orbital direction, and 
WFI-1 60° from the reverse orbital direction.
This caused the positions to spread gradually over the first 120 days of operation
which were also used for spacecraft
and instrument commissioning.  A small rocket motor on each Observatory provides 
a few m/s of $\Delta v$ both to arrest the 
relative orbital drift and to maintain the three WFI Observatories approximately
120$^\circ$ apart in orbit. 

All four observatories compress and cache image data and then downlink stored images during ground passes. The three WFI observatories execute approximately one ground pass per day, and the NFI observatory executes approximately six ground passes per day.  

\subsection{Spacecraft design\label{sec:spacecraft}}

The PUNCH spacecraft are designed to be as similar as possible, \old{with interchangeable decks}\new{with interchangeable mounting structures and electrical interfaces to support} the lone prime-mission instrument on \new{the} top (+Z face) of each.  The
spacecraft body is a ``bathtub'' design with interior walls, fabricated from a single block of
aluminum.  The interior volume is closed out by a +Z deck lid that \old{is also a mounting
surface for the primary instrument}\new{accommodates support brackets to mount the instrument}.  The -Z surface holds a Motorized Light Band (MLB) 
separation system used to interface each spacecraft to the launch vehicle.

Command and data handling (C\&DH) are via a SwRI-supplied C\&DH box built around a
Centaur single-board computer, with ancillary cards for power handling, analog control, 
and digital control.  The C\&DH system manages spacecraft sequencing, regulation, and fault handling.  It issues commands to commercially procured spacecraft subsystems and operates the instrument directly. The C\&DH system collates and manages separate telemetry streams for engineering/housekeeping data and for science.  These are coded, buffered, and transmitted as separate virtual channels, using the standard Consultative Committee for Space Data Systems (CCSDS) format.

Telemetry, commanding, and tracking are via a dual-band two-way radio, built by Tethers 
Unlimited Inc., that offers S-band uplink and both S-band and X-band downlink via low-gain 
antennas on the top and bottom decks of the spacecraft. The S-band link offers 5 Mbps data 
rate to the ground; the X-band link offers just under 25 Mbps. The receivers are programmed to 
respond to 
spacecraft-unique idle codes transmitted from the ground, affording specificity of contact even
during the early phases of the mission when the four spacecraft were close together. The S-band
is used for commanding and for engineering telemetry; and the X-band is used for downlink of 
science data. For resilience, either band can be used to downlink any type of data from the
spacecraft. The X-band radio on board NFI-0 suffered a failure in 
observatory-level integrated thermal vacuum testing, and NFI-0 is
therefore operated solely with S-band downlink, at an average rate of roughly one ground pass 
every four hours.  The radios were intended to be used simultaneously, but a common manufacturing 
flaw in the radio units limits usage to one transmitter at a time.

\begin{figure}
    \centering{}
    \includegraphics[width=0.9\linewidth]{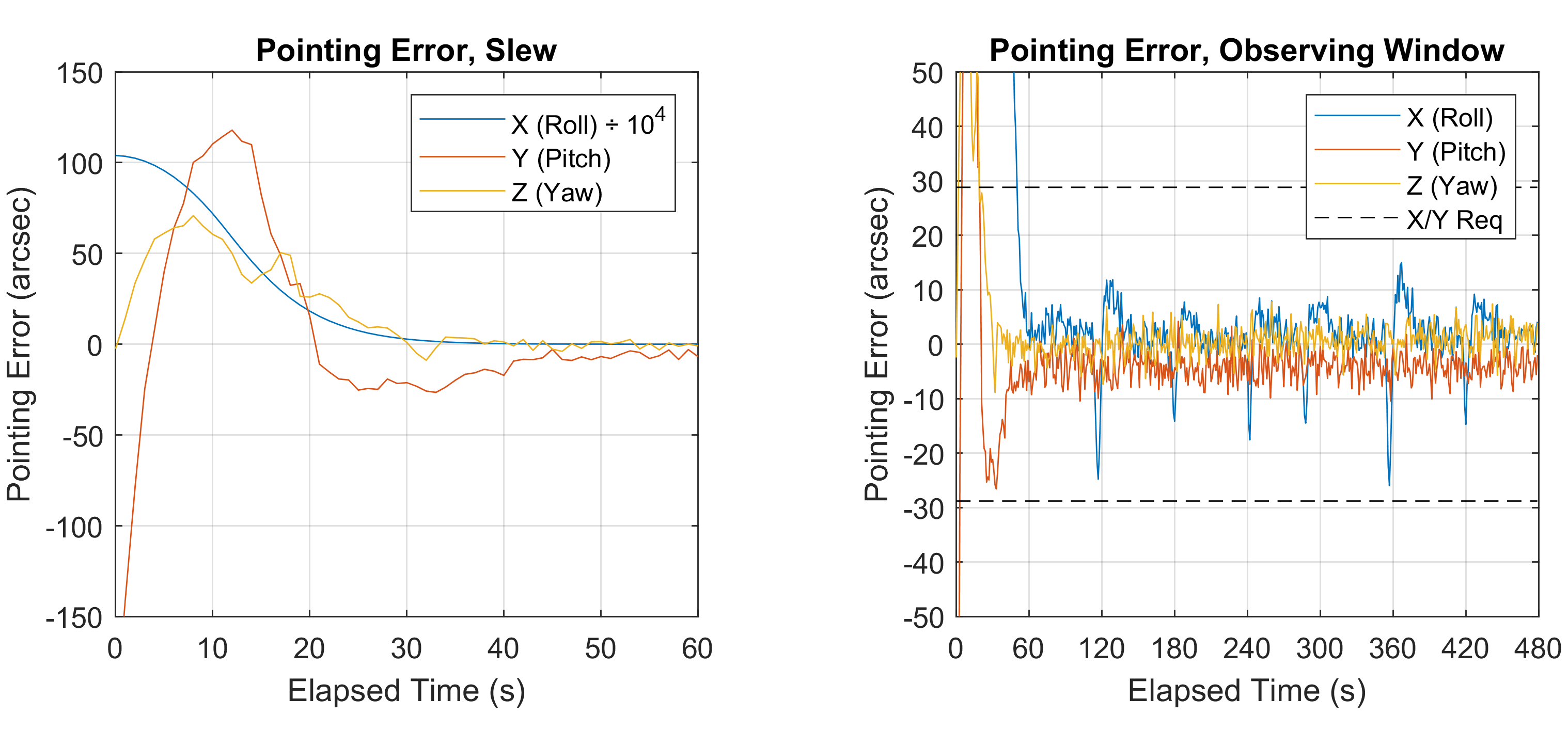}
    \caption{\new{The PUNCH ADCS (Blue Canyon XACT) system exceeds performance requiremens for slew
    recovery, filter wheel rotation recovery, and stability during camera exposures. Left: $30^\circ$
    roll slew and recovery.  Right: performance during a typical 8-minute observing window. 
    Polarizing filter wheel
    rotations, and the seven standard observing-sequence exposures (Fig. \ref{fig:observing-sequence}), are visible starting 60s 
    into the 8-minute window.}}
    \label{fig:adcs}
\end{figure}

The attitude determination and control system (ADCS) is an XACT unit
made by Blue Canyon Technologies. The XACT uses twin star trackers for 
fine-pointing attitude determination in celestial coordinates (right ascension and declination).  
In nominal science operations, sun-pointing
is achieved by directly commanding absolute celestial pointing minute to minute, using 
the solar ephemeris.  This strategy permits continuous observation even during the
inevitable ``eclipse season'' in which each spacecraft passes through Earth's shadow once per
orbit, near the unfavorable solstice. Each spacecraft does carry 
Sun sensors, which are used to ensure sun-pointing and spacecraft safety in off-nominal modes.  
XACT also uses a magnetometer and digital geomagnetic model to constrain solar roll
angle when in coarse sun-pointing mode.
The ADCS operates its control loop on a 5 Hz cadence, maintains 16 arcsecond (3$\sigma$) 
stability in fine pointing mode, uses quad reaction wheels, and manages its own angular 
momentum continuously via magnetic torque rods mounted in the spacecraft
interior.  It also includes a GPS receiver which is used to synchronize the on-board clocks 
across the constellation and (on the ground) to generate refined ephemerides and pass
predictions for 
each spacecraft. 
\new{The slew rate is sufficient to execute a 30$^\circ$ roll, 
and recover stability,
in under 60 seconds. Stability requirement during exposures is $\pm 30''$ in pitch (about the 
spacecraft Y axis) and yaw (about the spacecraft Z axis), and $\pm 300''$ in roll (about the X 
axis, which is approximately aligned with the spacecraft-Sun line).  On-orbit performance
exceeds requirements for settling time after slews and after filter wheel rotations, and for 
stability between spacecraft and mechanism moves.  Performance of WFI-1 during a typical 8-minute
observing window is shown in Figure \ref{fig:adcs}.}

The propulsion system is a HYDROS unit, supplied by Tethers Unlimited, Inc., which uses 
water as propellant.  A small water tank feeds an electrolyzer unit, which separates water into hydrogen and oxygen gases; these accumulate in twin plenums
which are vented into a combustion chamber to produce rocket thrust in discrete 
``thrust events'' that impart up to 2~cm/s of $\Delta v$ over approximately 2~s.
Each thrust event requires 1-2~ks of electrolysis time to prepare, and
consumes approximately 0.5~mL of water.  Electrolysis does not 
interfere with science data acquisition.  The rocket nozzle is mounted on the -Z face of each spacecraft.  Total 
$\Delta v$ capability is approximately 20~m/s on each spacecraft, against a calculated 
requirement of 4-6~m/s, driven by estimated orbital perturbation rates from the geoid over a 
two-year mission.

The power system uses deployable solar arrays from Sierra Nevada Corp., together with SwRI 
power-management electronics incorporated into the C\&DH box and a lithium battery from ABSL.  
Peak power trackers match the supply voltage of the solar arrays to the interior 28VDC bus voltage.
Both switched and unswitched power buses are managed by power cards in the C\&DH box.

The thermal control system is a cold-biased passive design with supplementary multi-zone
digitally controlled management heaters and thermostatically switched survival heaters.
Spacecraft subsystems dump heat into the body of the spacecraft. The instruments are thermally
isolated from the spacecraft, and radiate their waste heat to space.  The instruments include CCD
cameras that
are passively cooled by radiation to deep space; this takes advantage of the 
roll
attitude program for WFI, and drives a similar roll program for NFI, both of which maintain the CCD radiator in shade on the \new{+Z} (zenith) side of the spacecraft bus.

The instruments are directly controlled by the C\&DH system, via a SpaceWire interface to the
camera system and electrical pulse interfaces to the instrument mechanisms.  The
instruments were specifically designed to be electrically identical and controlled the same
way, simplifying C\&DH and flight software design by reducing the functional differences between the instrument types to minor differences in on-board configuration tables.

\begin{figure}
    \centering
    \includegraphics[width=4in]{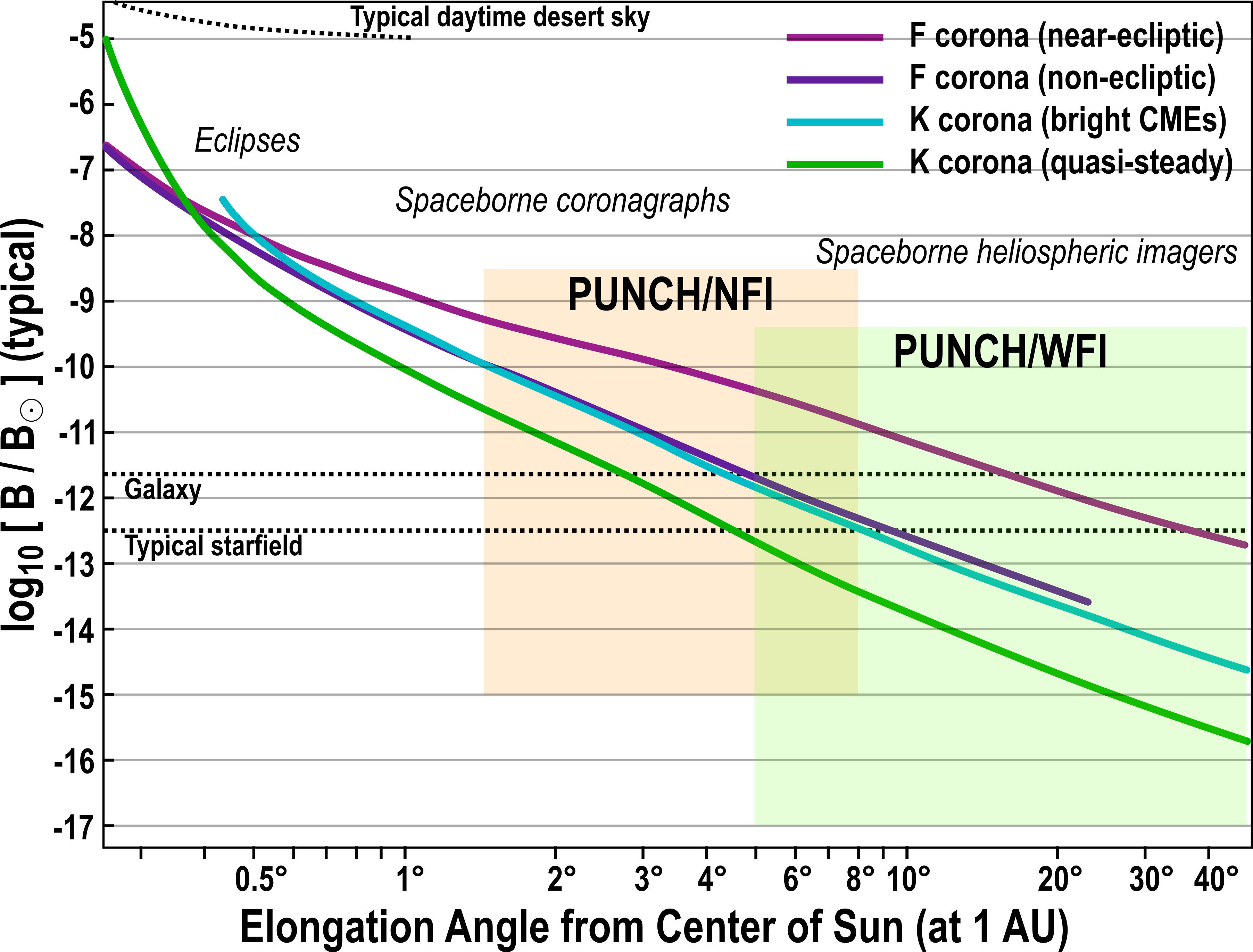}
    \caption{The PUNCH field of view captures an extremely broad dynamic range, spanning nearly seven orders of magnitude between the brightest F corona in the inner field and the quasi-steady K corona in the outer field. This requires two separate instrument types: a Narrow Field Imager (NFI) and a Wide Field Imager (WFI), with significant field of view overlap.}
    \label{fig:brightness}
\end{figure}

\section{PUNCH instruments \label{sec:Instruments}}

PUNCH operates as a single ``virtual instrument'' built from four individual physical 
instruments.  The Narrow Field Imager (NFI; Section \ref{sec:NFI}) and three Wide Field Imagers (WFIs; Section \ref{sec:WFI}) are 
physically matched such that each instrument's instantaneous field of view (IFOV) overlaps
with each of the other three instruments' IFOVs; the wavelength passbands are matched at 450--750~nm; and
the observing sequences are synchronized to make the merged science data as seamless as possible.  

The two different instrument types are needed to accommodate the very
wide dynamic range across the PUNCH field of view (Figure \ref{fig:brightness}).
The instrument types share two important common subsystems: a camera/CCD system (Section \ref{sec:CCD}) 
and a polarizing filter wheel (Section \ref{sec:PFW}).  They are designed to mount interchangeably
on the PUNCH spacecraft, and also share joint electronics and electrically-identical components
for door actuation and thermal control.

\subsection{Narrow Field Imager \label{sec:NFI}}

\begin{figure}
    \centering
    \includegraphics[width=1.0\linewidth]{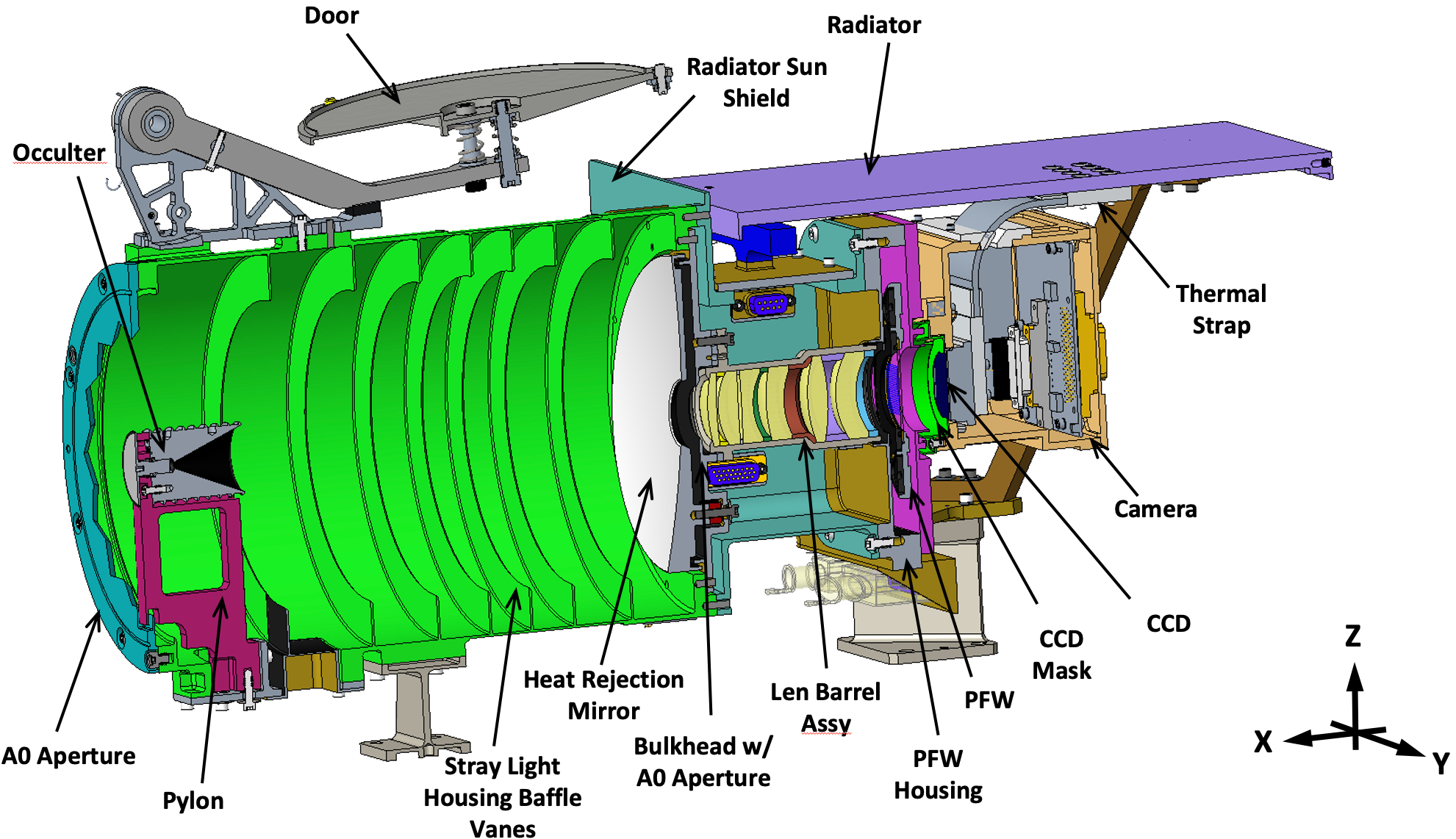}
    \caption{Cutaway drawing shows the structure of the PUNCH/NFI instrument and its major subsystems.}
    \label{fig:nfi}
\end{figure}

\begin{figure}
    \centering
    \includegraphics[width=0.667\linewidth]{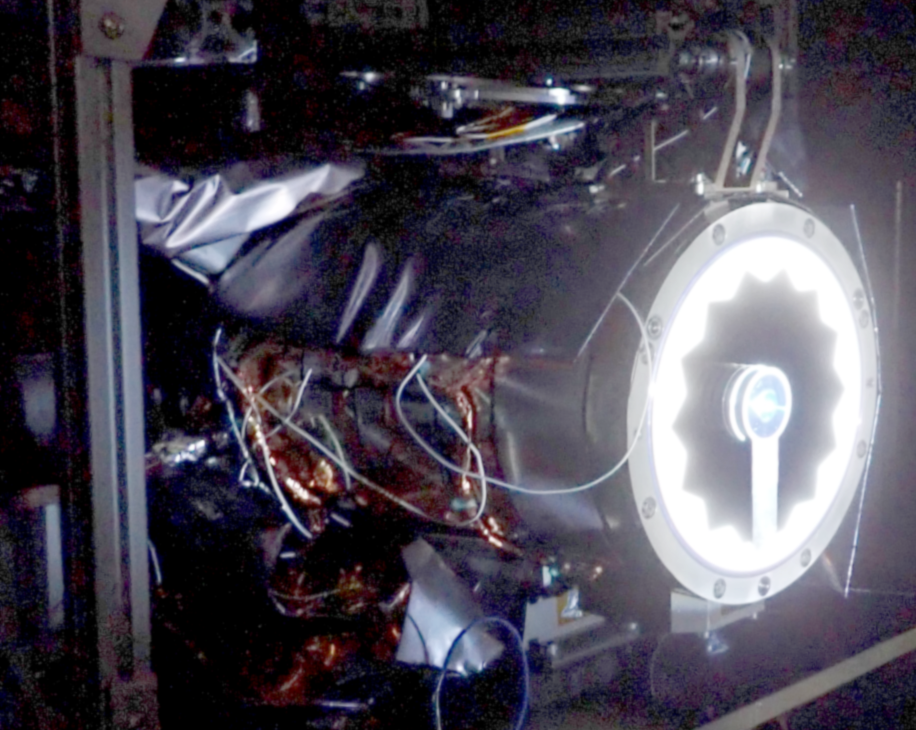}
    \caption{\new{Photograph of NFI during stray light testing in NRL's
    SCOTCH vacuum-testing facility shows the major front end elements
    diagrammed in Figure \ref{fig:nfi}.  The toothed aperture, single
    external occulter, and support pylon are brightly illuminated by 
    a simulated solar beam; the instrument stray light housing baffle, 
    covered in multi-layer insulation, is visible behind them. 
    The door is open in the flight
    configuration at top.} }
    \label{fig:nfi-photo}
\end{figure}

The Narrow Field Imager (NFI) is an externally-occulted compact coronagraph design built by the
U.S. Naval Research Laboratory (NRL). \new{Figure \ref{fig:nfi}. The
instrument is a cutaway diagram of the instrument.  Figure \ref{fig:nfi-photo} is a photograph of the instrument undergoing stray light testing
inside the SCOTCH vacuum chamber at NRL.}  NFI is
described in detail by \citep{colaninno_2025}; the design is summarized here.  NFI builds on a
long history of coronagraph development, beginning with \citet{lyot_1930} and extending 
through OSO-7 \citep{koomen_etal_1975}, Skylab-ATM \citep{macqueen_etal_1974}, Solwind
\citep{Sheeley_etal1980}, \small{SOHO/LASCO} \citep{brueckner_etal_1995},
\small{STEREO/SECCHI} \citep{howard_etal_2008}, and GOES/CCOR \citep{thernisien2021CCOR,dudley2023CCOR}.

\new{In Figure \ref{fig:nfi},} a front-end baffle assembly (green) supports an A0 aperture with a toothed outer edge that minimizes
edge diffraction into the optics, and a pylon and occulter assembly that block direct sunlight
from entering the optics. The occulter is a tapered-ogive\footnote{Readers are reminded that an ``ogive'' is a figure of revolution of a circle, about a non-diameter chord of that circle; or a truncated segment of such a figure.} multi-disk 
design with a 
rear-facing conical
light trap.  The multiple disks force multiple diffraction scatter events, strongly attenuating 
diffracted sunlight.  The conical light trap absorbs scattered light from the rear of the baffle
assembly.
Most sunlight entering the instrument impacts a spherical Heat Rejection Mirror (HRM) at the 
back of 
the baffle assembly, and is focused into the empty space opposite the pylon, and ejected into
space.  The HRM has a central hole that forms the aperture to a lens barrel assembly that 
focuses light through a polarizing filter wheel (Section \ref{sec:PFW}) onto a CCD detector connected to an interface camera module, and operated by an external
Camera Electronics Box (Section \ref{sec:CCD}).  The entire instrument was closed out, at launch,
by a 
ground-resettable one-time-open door.  The
CCD is cooled by thermal conduction through a thermal strap to a radiator aimed at deep space.
The entire instrument structure is supported on the spacecraft by a kinematic mount comprising 
three feet that bolt directly to the \old{top deck of the} spacecraft.

PUNCH/NFI has an optical focal length of 103~mm and is operated at an aperture of f/4.5, yielding a
plate scale of 0.5~arcmin per 15~$\mu$m pixel. The field of view covers nearly all position angles $\alpha$ around the Sun,
extending across solar elongation (apparent radial) angles $\varepsilon$ from
6~R$_\odot$ to 33~R$_\odot$ (1.5$^\circ$--8.3$^\circ$), with the diffraction ``bright ring'' on the
occulter at exactly 1.5$^\circ$ from the Sun itself. Images acquired by the instrument are masked
on-board PUNCH to 5.7~R$_\odot$ to 32~R$_\odot$, to optimize downlink volume.

Each NFI image is acquired as a triple exposure on board PUNCH: three separate camera exposures (digitized to
16 bits) are summed on board to produce 20-bit-deep pixel values, which are further processed on-board
(Section \ref{sec:CCD}) for downlink.  Each polarization sequence comprises three of these 
triple exposures: one summed triple exposure with the polarizing filter wheel in each of its three polarizing
positions. 

\new{During development NFI was tested for solar stray light rejection, at NRL's Solar Coronagraph Optical 
Test Chamber (SCOTCH) facility \citep{korendyke_1993}. SCOTCH is a long-tube baffled vacuum chamber with a solar 
simulator light source, designed to measure stray light while eliminating Rayleigh and Mie scattering 
from the air and dust present in a more typical laboratory environment.  NFI was fully illuminated
by the simulated solar beam, and light levels were measured in image data collected through the 
instrument itself.  Solar stray light in NFI is dominated
by backscatter off the HRM, which illuminates the rear of the occulter.  Doubly scattered sunlight enters
the optics, forming an out-of-focus image of the occulter itself, with effective radiance comparable to that
of the solar F corona.  This anticipated effect is removed in post processing in the science reduction pipeline,
described in Section \ref{sec:SOC}.  Solar stray light from the simulated solar beam was measured at three radial locations; these are tabulated in Table \ref{tab:NFI-stray light}. The stray light was measured with each of three polarizers in the beam, and with a clear (non-polarizing) filter.  The largest of the three values is reported in Table \ref{tab:NFI-stray light}. The units are given as 
$B'_\odot$ rather than $B_\odot$ to acknowledge linear scaling for the measured intensity of the 
simulated solar beam. NFI solar stray light meets design requirements at all locations.
}

\begin{table}[]
    \begin{tabular}{c|c|c}
       Radial location  &  Requirement & Measured stray radiance  \\
    \hline
        $6.6 R_\odot$ & $<41\times 10^{-11} B'_\odot$ & $3.8\times 10^{-11} B'_\odot$  \\
        $15 R_\odot$ & $<13\times 10^{-11} B'_\odot$ & $5.8\times 10^{-11} B'_\odot$  \\
        $27 R_\odot$ & $<3.5\times 10^{-11} B'_\odot$ & $2.0\times 10^{-11} B'_\odot$ 
    \end{tabular}
    \caption{Solar stray light requirements and values measured for NFI at three field-of-view locations, at NRL's SCOTCH facility}
    \label{tab:NFI-stray light}
\end{table}

\begin{figure}
    \centering
    \includegraphics[width=0.75\linewidth]{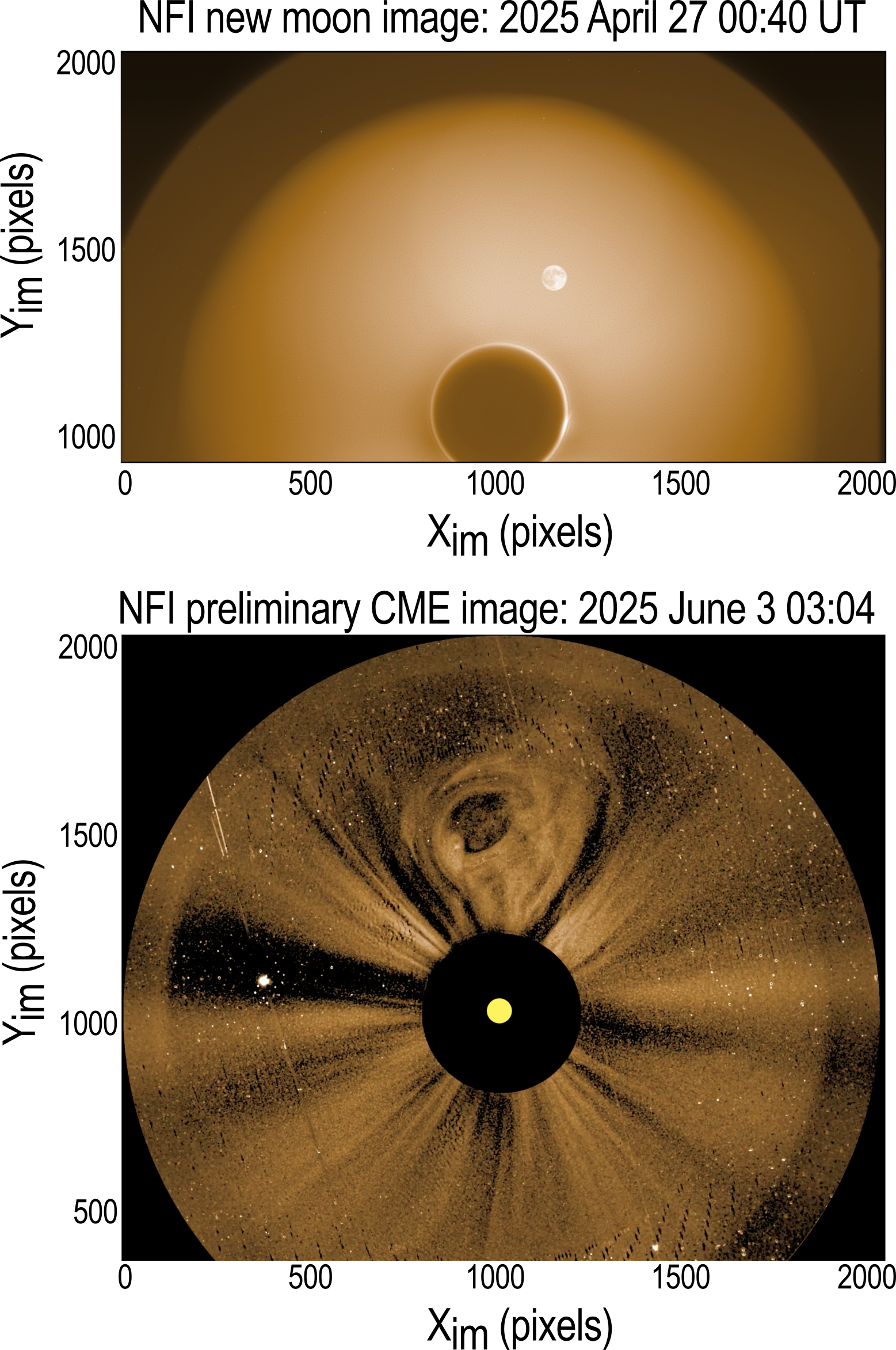}
    \caption{\new{Early example images from NFI demonstrate instrument capability. Top: the new Moon first passed through the NFI FOV on 2025 April 27, providing a scale and brightness reference for the field of view.  The Earth-illuminated Moon is approximately 2x brighter than the fixed +variable stray light pattern.  Bottom: a northbound CME launched on 2025 June 2; preliminary data reduction shows that NFI resolved fine interior details of the CME as it passed through the outer corona. }}
    \label{fig:nfi-demo}
\end{figure}

During early operations, total stray light in the NFI instrument was found
to be greater, and far more variable, than \new{the solar stray light performance
in Table \ref{tab:NFI-stray light}.
This has been attributed to variable Earthshine illuminating the 
interior of the baffle tube, and thereby scattering onto the rear of
the occulter; the effect was not characterized during the SCOTCH testing campaign in the 
pre-launch phase. The additional variable stray light is up to $5\times$ brighter than the solar stray 
light, though more typically $0.2\times$ to $2\times$ as bright; and is confined to the zone (inside of approximately $22 R_\odot$) covered by the defocused image of the occulter.} 
The process to scrub this variable stray 
light from 
the data is the subject of a rapid development effort early in the
mission; it is described in more detail by \citet{hughes_etal_2025}.

NFI sample data, collected during commissioning, included images of the new moon and of
several bright CMEs. Example images are in Figure \ref{fig:nfi-demo}, and reveal the
instrument's high sensitivity and resolution even with preliminary data processing.

\begin{figure}
    \centering
    \includegraphics[width=0.75\linewidth]{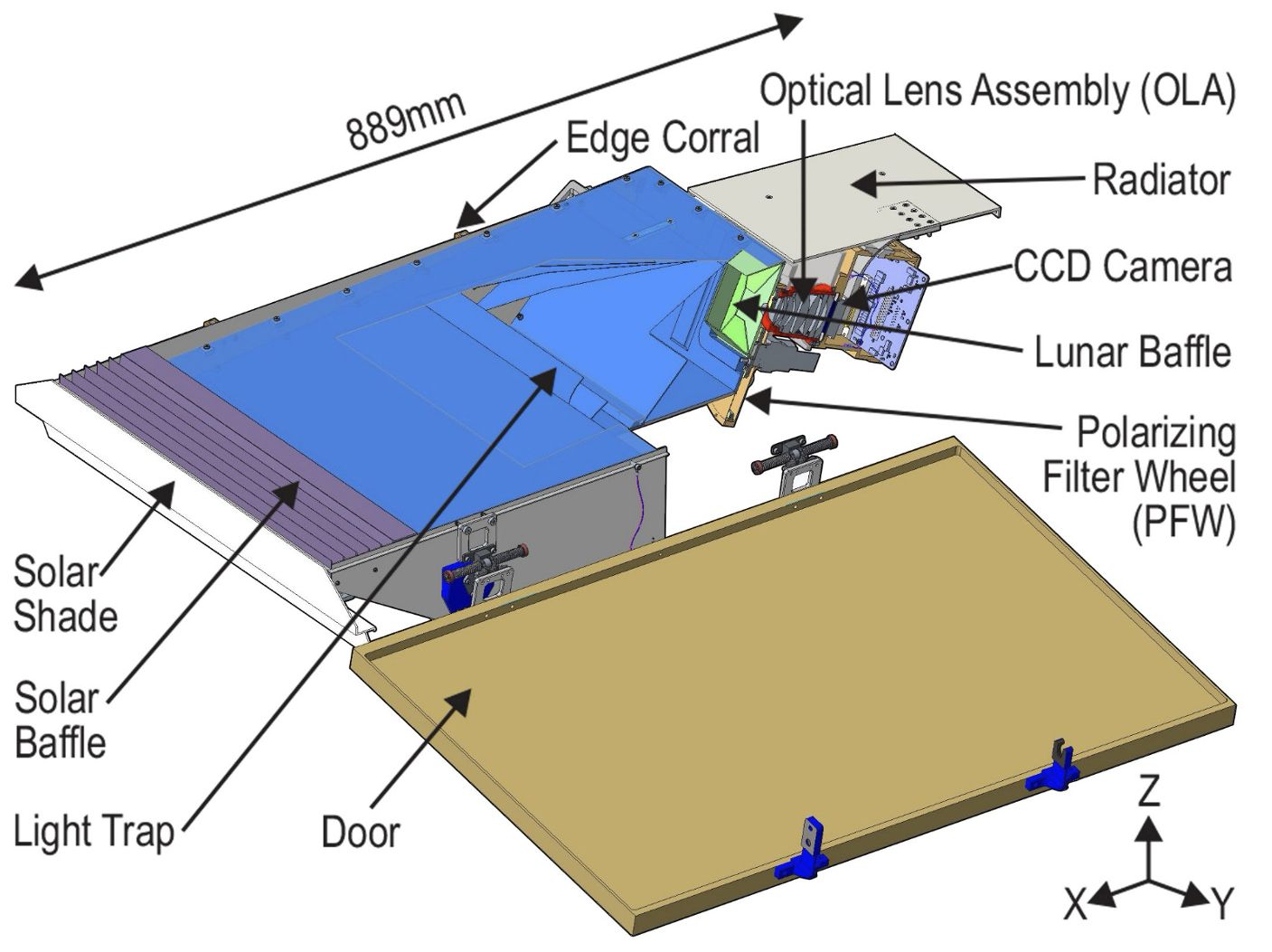}
    \caption{Cutaway drawing shows the structure of the PUNCH/WFI instrument and its major subsystems.}
    \label{fig:wfi}
\end{figure}

\begin{figure}
    \centering
    \includegraphics[width=0.9\linewidth]{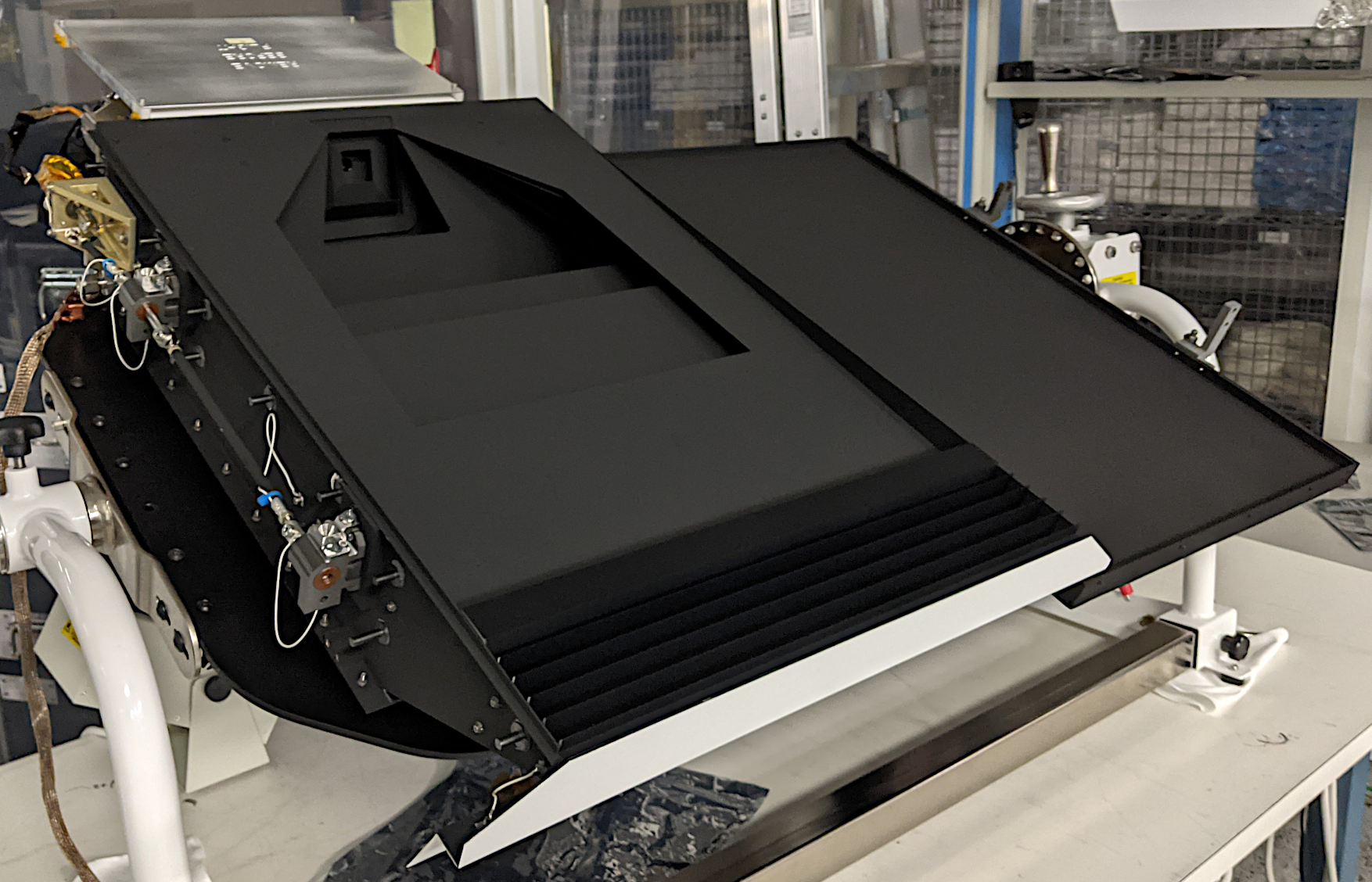}
    \caption{\new{Photograph of WFI-2 during final instrument 
    integration shows the major elements diagrammed in Figure 
    \ref{fig:wfi}. From front to back, the major features are 
    the solar shade, coated in white thermal paint; the multi-vaned solar
    baffle, with its Acktar optical black coating; the deep light trap
    vanes, with Aeroglaze Z307 black coating; the square-geometry lunar
    baffle; and the optical aperture.  The CCD radiator is protected by a
    protective shield.  At left is the door release mechanism.
    The door is in its deployed flight configuration.}
   }
    \label{fig:wfi-photo}
\end{figure}

\subsection{Wide Field Imager \label{sec:WFI}}

The Wide Field Imager (WFI) is a corral-baffle heliospheric imager design; three were 
built for PUNCH at Southwest 
Research Institute (SwRI). \new{Figure \ref{fig:wfi} is a diagram of WFI. Figure \ref{fig:wfi-photo} is a photograph of one of the instruments, under development.} WFI is described
in detail by \citep{laurent_2025}; the design is summarized here.  The  is analogous to an 
externally-occulted coronagraph, in linear geometry (i.e. the baffle structure is 
approximately symmetric via extrusion along the spacecraft Y axis, rather than via revolution 
about the spacecraft X axis). Heliospheric imaging was conceived in the late 20th century using 
the three photometers of the spinning Helios mission \citep{jackson_leinert_1985}, and developed through 
Coriolis/SMEI \citep{eyles_etal_2003, jackson_etal_2004}, STEREO/HI 
\citep{howard_etal_2008,eyles_etal_2009}, SolO/SOLOHI \citep{howard_etal_2020}, and PSP/WISPR 
\citep{Vourlidas_etal2016}. The more modern designs include a full optical imaging system, a solar 
baffle, and a deeply vaned light trap to absorb glint; WFI is similar, with the addition of a 
lunar baffle near the optical aperture, to reduce flooding by moonlight.

The instrument body is a large flat corral fabricated from aluminum.  The corral is surrounded by
a flat lip (dark grey on $\pm$Y sides) that defines an instrument shadow plane, hiding the Earth itself from the main baffle
structure.  The body supports a leading solar
shade (white) that diverts the bulk of the solar beam away from the baffle and body, to avoid thermal 
flexure.  The multi-vane solar baffle (periwinkle) intercepts sunlight that would otherwise enter the corral. It is
machined from a single piece of aluminum and coated with optical black from Acktar Corporation.
The baffle edges follow a cylindrical envelope that is analogous to the ogive shape of the NFI 
occulter, and enforce seven separate scattering events for any photon diffracted from the main
beam around the baffle into the optics, following the applied scattering theory outlined by \citet{Deforest_etal2025}.
Light that passes over
the solar baffle crosses into the corral area, crossing over deep ``light trap'' vanes (blue) that are 
designed to absorb glint, moonlight, or other back-illumination of the instrument.  At the rear of the 
light trap is a small lunar baffle (green) in square geometry. The lunar baffle is a conventional two-bounce
design to reduce the field of regard, mitigating the effects of moonlight in the gibbous and full phases
(when the Moon is outside the field of view but also bright enough to wash out WFI images).  An
aperture
at the rear of the lunar baffle coincides with an external pupil of the optical design.  The external 
pupil reduces stray light and provides room (optical ``eye relief'') for a polarizing filter wheel (PFW; Section \ref{sec:PFW}) in front of the
optics.  The image is projected onto a CCD (Section \ref{sec:CCD}) and transferred to the spacecraft for 
processing and download.

PUNCH/WFI has an optical focal length of 35.2mm, operated at f/3.2 (11mm aperture), yielding a plate scale of
1.47 arcmin per 15~$\mu$m pixel and an optical field of view (FOV) that is a circle truncated by the ``horizon'' of the solar vane.
In nominal observing attitude, the Sun is on the vertical centerline of the image, below the horizon; the field 
of view extends along that centerline across elongation angles $\varepsilon$ from 20~R$_\odot$ 
to 180~R$_\odot$ (5$^\circ$--45$^\circ$).  The nominal FOV is an approximate square 40$^\circ$ on a side, 
truncated by a circle 50$^\circ$ in diameter, centered on the instrument boresight; the actual 
useful FOV is over 55$^\circ$ in diameter.  Because WFI operates in linear 
geometry, the inner edge of the field of view may be adjusted by shifting the spacecraft attitude. To take advantage of the as-built stray light performance and field of view, all three WFIs are operated with the inner edge of the field of view close to 12~R$_\odot$ 
(3$^\circ$) from the Sun itself.  Images from WFI are masked to exclude areas below the horizon and outside the
observed useful field of view, to optimize downlink volume.

WFI images are acquired in single exposures on board PUNCH: a single 51-second (polarized) or 26-second (unpolarized)
image is read out at 16 bit dynamic range.  Single-exposure acquisition improves signal-to-noise ratio (SNR) in the outer portion of
the field of view, where image-independent camera read noise dominates the noise profile of the camera; the cost is that objects with magnitude 2.0 or less (brighter) saturate the image. There are on average 3.3 saturated objects in each WFI's field of view
throughout the mission.  Accounting for bleeding artifacts, and the destreaking and PSF correction steps of the data processing pipeline (Section \ref{sec:SOC}), 
these saturated objects invalidate an average of approximately 0.2\% of the total WFI field of view from each frame.  Because the instruments rotate every 8 minutes, the location of these artifacts changes rapidly so that the effect of any one artifact is minimized.
The invalid pixels are marked in the calibrated data products.

\new{One WFI was tested for solar stray light rejection, at NRL's SCOTCH facility.
A simulated solar beam was directed at the WFI baffle, as in the NFI test, and the instrument itself 
was used to measure the radiance of stray light; details are given by \citet{laurent_2025}.  The 
WFI stray light rejection performance exceeded considerably the stray light rejection performance 
of the SCOTCH facility, and WFI images of the SCOTCH interior were dominated by interior features 
rather than by instrumental stray light.  Typical dark chamber-interior image points measured a local 
radiance near $1\times 10^{-11} B'_{\odot}$ (compared to $B'_{\odot}$, the average radiance of the 
simulated solar beam in the SCOTCH chamber), and differential hit-miss beam measurements 
afforded sensitivity as low 
as a few $\times 10^{-14} B'_{\odot}$; this was not sufficiently sensitive to measure the 
required FOV-average
solar stray light radiance attenuation of $10^{-16} B'_{\odot}$ directly.  
Instead, we used the measured
lens stray-light attenuation of $3\times 10^{-3}$, a geometric factor for the apparent size of the
solar vane edge ``bright line'' in the field of view, and differential measurement of the bright line.  This yielded calculated overall solar stray light levels of a few $\times 10^{-18} B'_\odot$.  Given the
wide margin on stray light rejection, subsequent WFIs were not end-to-end tested for  
stray light performance.  In-flight stray light performance was found to meet requirements with wide margin, as expected.
}

\new{WFI is subject to variable stray light from Earth, near both solstices and especially during the ``eclipse season'' surrounding the unfavorable (winter) solstice. During the eclipse season, the Earth is in front of WFI, approaching the actual instrument field of view, during passage over the north
pole; and stray light increases by up to three orders of 
magnitude near the baffle in the corners of the image.  This contaminates some images each orbit.  
These periods were considered during mission
design and are omitted from the mission total duration count. Details are given in \citet{laurent_2025}.}

\begin{figure}
    \centering
    \includegraphics[width=0.8\linewidth]{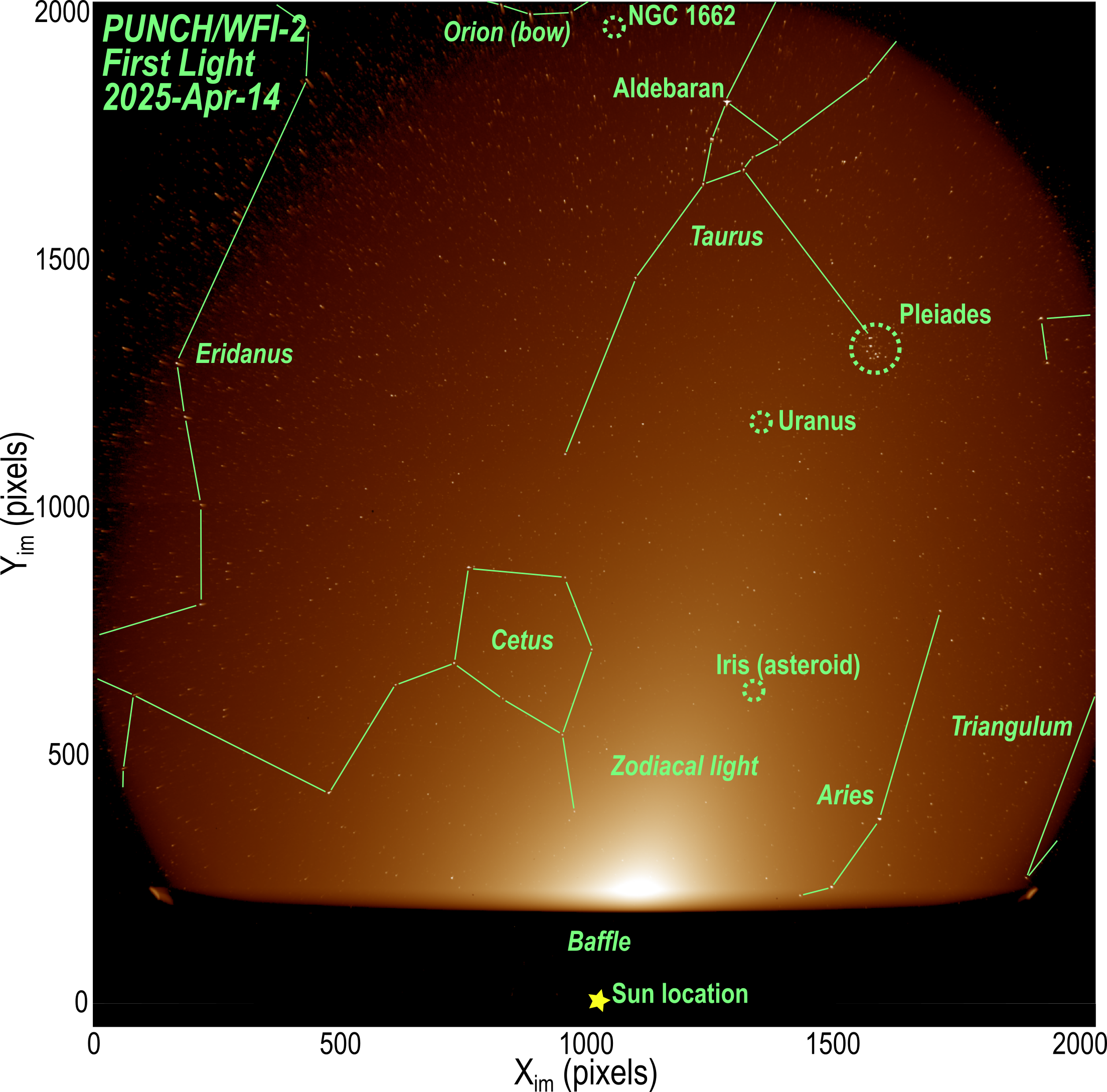}
    \caption{\new{First-light image from WFI shows the very large field of view of each WFI camera, 
    spanning over $40^\circ$ of sky. The dominant source of brightness is the zodiacal light (extended F 
    corona). Several relevant constellations and objects have been marked.  Asteroid Iris (magnitude 
    10.2) is clearly visible in the data, indicating good photometric sensitivity.}}
    \label{fig:wfi-demo}
\end{figure}

\new{Figure \ref{fig:wfi-demo} shows first light from the WFI-2 instrument, demonstrating low stray light
performance and field of view. The limiting magnitude is roughly 11 in a single image: asteroid Iris,
with magnitude 10.2, is clearly visible in the original data.  The brightest object visible to WFI 
without saturation is Saiph ($\kappa$ Orionis, magnitude 2.1), which registers a peak brightness of 
up to 65,000 digitizer counts (of 65,535 possible) when optimally aligned with the WFI-2 pixel grid.}

\subsection{Camera \& onboard data handling} \label{sec:CCD}

The focal planes are 2048$\times$2048 charge-coupled device (CCD) photodetector arrays, made by 
Teledyne E2V, with 2048$\times$4096 total 15-$\mu$m square pixels. The CCDs are operated in 
quasi-frame-transfer mode: one 2048$\times$2048 block of pixels is masked with a metallic cover to 
form a ``storage'' area, and the remaining 2048$\times$2048 block is left bare (with anti-reflection
coating) to form an ``active'' area. The CCDs are thinned and back-illuminated for quantum
efficiency and radiation tolerance; and in operation are cooled to below -50~C. During image
acquisition, the CCDs are reset multiple times to clear
residual charge, then allowed to accumulate photons during the exposure time.  At the end
of the exposure, the latent image is transferred to the storage area for slower digitization 
and readout.  Image transfer takes 125~ms; readout requires 3~s.  This mode of operation yields 
extremely stable exposure times, with better than $10^{-4}$ relative precision, at the cost of 
requiring each image to be destreaked in post-processing.  The CCDs are each operated and read 
out by a Camera Electronics Box (CEB) from RAL Space.  The CEB digitizes the image data to
16 bit depth at a gain value of 4.9 electrons per digitizer count (320k electrons at saturation), 
then transfers image data over a digital link to the C\&DH box for on-board processing and transfer to the ground.  

The C\&DH
processing includes a switchable compression pipeline that enables square-root coding and lossless 
image compression.

Camera pixels are square-root coded for compression: the 
digitizer value $P_{xy}$ of each pixel is multiplied by an integer scale factor $\alpha$ to 
a total of 
up to 32 bits width,
then the integer square root of that value, $(\alpha P_{xy})^{0.5}$, is recorded instead of
$P_{xy}$. The square root operation reduces the pixel values back to 16 bit integers, which 
typically occupy 
fewer bits of actual dynamic range.  The purpose of the square root operation is to match digital
transition
step size in the final data product, to the variable level of quantum shot noise across the dynamic range
\citep{deforest_etal2022a}.  This
in turn enables further compression of the remaining bits of each image, by removing useless 
high-entropy noise while
preserving useful and lower-entropy data.  In flight, $\alpha$ was selected to be 16, or $2^4$.  WFI 
images, single
16-bit-deep camera frames, thus retain
only 10 bits of information (roughly 1,000 unique values) across the range of the digitizer.  
NFI raw images, being sums of three camera frames, begin as 18-bit integer values 
occupying 17.5 bits of dynamic range, and conclude with 10.75 bits of dynamic range (roughly 1700 unique 
values).  This setting preserves a
small amount of photon noise to dither photometric values, so that digital steps do not dominate 
the images.  This 
dynamic-range ``toeroom'' (room to expand at the bottom) is at least 
0.7~(RMS) digitizer counts at all brightness levels, and 1.5~(RMS) digitizer counts
across most of the dynamic range.  The smaller values occur for dark pixels, where the camera digitizer 
``pedestal'' (approximately 400 counts per exposure) is significant compared to the photon signal; 
and the larger values occur at pixel values above roughly 2,000 counts (or 6,000 counts for NFI 
sums) where 
photon counting
noise is the most important element of uncertainty in each pixel value.  

After square-root coding, image values are compressed on-board with the 
JPEG-LS lossless image-compression algorithm, 
applied to values that occupy up to 16 bits of dynamic range. Non-square-rooted PUNCH 
raw images are observed to 
compress by a factor of approximately 2.7 with this algorithm. WFI images that have been 
square-root-coded to a depth
of 10 pixels are observed to compress by a further factor of 3.5, 
for an overall compression factor of 5.6. Masking out
unused pixels (to avoid downlinking read noise and photon noise from those areas) 
improves overall compression slightly,
to an overall factor of 6.0 (1.33 MB/image).

Compressed image data are stored in an on-board flash RAM unit and downlinked during ground passes.

\subsection{Polarizing filter wheel} \label{sec:PFW}

Each PUNCH imager uses a polarizing filter wheel (PFW) to measure the
linear polarization of light across the image plane.  Each 
polarizing filter wheel intersects the main optical beam and has five positions, 
containing: 
three linear polarizers oriented at -60$^\circ$, 0$^\circ$, and +60$^\circ$ relative to the instrument
``horizontal'' axis (aligned with spacecraft $\pm$Y); a non-polarizing ``clear'' optical
flat; and a ``dark'' opaque metal sheet.  Three polarizers were chosen following historical usage
\citep[e.g.,][]{howard_etal_2008}; the nuances of the tri-polarizer formulation are described by
\citet{deforest_etal2022b}. Each filter wheel is actuated by a stepper motor, commanded directly by the
C\&DH electronics; its
position is reported by redundant parallel resolvers and downlinked in the engineering/housekeeping 
telemetry channel.

The NFI filter wheel is mounted between the lens barrel assembly and the camera.  The WFI filter wheel
is mounted in front of the optical lens assembly, in the empty space between the leading lens and the 
external pupil. The position in WFI was chosen to better regulate polarization over the very wide field
of view and fast optical beam. The WFI filter wheels are mounted in the reverse direction to the NFI filter wheel: NFI mounts
the motor on the +X (sunward) side of the filter wheel, while WFI mounts it on the -X (camera) side of
the filter wheel.  This means that the polarizer positions are reversed between the two instruments, a 
nuance that is managed on the ground by the Science Operations Center (SOC) pipeline during data product
assembly.

\subsection{Student Thermal Energetic Activity Module}

The PUNCH/NFI spacecraft hosts a student-contributed do-no-harm class instrument, the Student 
Thermal Energetic 
Activity Module (STEAM). STEAM was conceived as a dual band X-ray full-sun spectrometer, implemented
using solid state detectors and off-the-shelf electronics. The STEAM project is described by 
\citet{hanson_2024}.  STEAM comprises two integrated solid-state X-ray spectrometers (off-the-shelf X123
units) supplied
by Amptek, together with control electronics designed, assembled, and integrated by 
undergraduate students at the University of Colorado.  STEAM was an educational success, helping provide
hands-on experience for over 50 students across its five-year development program.  The STEAM instrument
suffered an anomaly during Observatory level thermal vacuum testing, and is permanently switched off.

\section{Data products and Ground Pipeline\label{sec:SOC}}

PUNCH has an 
open data policy: all scientific data are released to everyone simultaneously, with no restrictions 
on use. 
Scientists and
others making use of the data are \textit{requested} to acknowledge the PUNCH mission 
with the statement ``PUNCH is a heliophysics mission to study the corona, solar wind, and space 
weather as an integrated system, and is part of NASA’s Explorers program (Contract 80GSFC14C0014)'' 
and to report their usage to the Project Scientist (S.E.G.) for
impact tracking.

PUNCH data are archived and distributed via NASA's Solar Data Analysis Center 
\citep[SDAC;][]{Ireland_etal_2022}, and may be retrieved directly from the SDAC file system or 
via the Virtual Solar Observatory \citep[VSO;][]{Hill_etal_2009}.  During mission operations, 
current information and links to relevant analysis software, documentation, and team
contact information may be 
found at the PUNCH mission website ``https://punch.space.swri.edu''.

PUNCH data products are generated in the Science Operations Center (SOC), which is described in detail by 
\citet{hughes_etal_2025} and summarized here. Data are produced as FITS files \citep{Wells_etal1981} in four principal processing Levels, 
numbered from 0 through 3.  Background subtraction is the driving challenge for PUNCH data reduction, and drives precision of all levels of data
reduction. 

Level 0 data are direct from each spacecraft, unpacked from the downlinked 
CCSDS packet stream and delivered as integer files in digitizer units.  
Level 1 data are photometrically and astrometrically calibrated but remain associated with individual downlinked images.  
Level 2 data are resampled into a common coordinate system and
merged across spacecraft into mosaic images, as if from a single instrument. 
Level 3 data are 
background-subtracted and ready for scientific analysis of the K corona and solar wind.  

In addition to the primary numbered data levels, the SOC produces both QuickLook (``Level L'') 
images, in conventional display formats; and low-latency QuickPUNCH (``Level Q'') data suitable for forecasting, as FITS files.  
Level Q products, and the utility of PUNCH for
space weather forecasting, are described in detail by \citet{seaton_etal_2025}.

Data products are identified by their level number, a three-character code, and a timestamp; and the PUNCH data
repositories are organized as a hierarchical data structure by those identifiers in that order.  In the following subsections
we describe the numbered reduction levels and the principal data product codes of interest to the community.  

PUNCH FITS files have more structure than simple images. FITS supports multiple header-data units (HDUs) per file.  The 
zeroth (HDU 0) contains information about tile compression used in subsequent images in each file.  The first (HDU 1) 
contains the image data, which may be two-dimensional or, in polarized data products, three-dimensional.  The second
(HDU 2) contains an uncertainty array, representing the uncertainty in the primary data product. 

Detailed contents of the data and metadata, and mission-supplied analysis tools to read and manipulate them, are described by
\citet{hughes_etal_2025} and by the online mission documentation.

The L3 data products are designed to be the primary data used to address the science
topics outlined in Section \ref{sec:Science} and close the mission's science objectives.
However, data products at L0, L1, and L2 are also archived and distributed to support 
reproducibility of PUNCH data analysis, and to enable additional science (Section \ref{sec:additional-science}).

\subsection{Level 0 data}

\begin{table}[tbh]
    \begin{tabular}{c|l}
        Code & Definition \\
        \hline
        CR\textit{n} & Clear exposure (no polarizer) \\
        PM\textit{n} & Linearly polarized Minus (-60$^\circ$) exposure \\
        PZ\textit{n} & Linearly polarized Zero (0$^\circ$; horizontal) exposure\\
        PP\textit{n} & Linearly polarized Plus (+60$^\circ$) exposure\\
        DK\textit{n} & Dark exposure (for calibration)\\
        PX\textit{n} & Nonstandard or unknown PFW position (calibration or error)\\
    \end{tabular}
    \caption{PUNCH primary Level 0 \& Level 1 product codes. All these codes represent 2048$\times$2048 pixel single image files, organized by polarizer position and Observatory.}
    \label{tab:L0codes}
\end{table}

PUNCH Level 0 data files are produced rapidly, typically within less than an hour of receipt at the 
SOC.  Each data file has a 
two-letter code indicating the type of
exposure, followed by a single digit indicating which PUNCH Observatory
produced the file (Table \ref{tab:L0codes}).  
Data from the WFI Observatories are 
numbered 1-3, and data from PUNCH-NFI00 is numbered 4.  The codes CR, PM, PZ,
and PP correspond, respectively, to 
clear, Minus, Zero, and Plus polarizer positions (Section 
\ref{sec:PFW}) and are produced by the 
normal science campaign.  The other codes are relevant to calibration 
activities.

Level 0 pixels have integer
values from the camera digitizer and are neither calibrated nor flat-fielded in
any way, so (for example) they include
digitizer pedestal, exposure streaking, and all other camera and optical artifacts. They are typically 
square-root compressed: if the header field ``ISSQRT'' is boolean-true, then the values in the file represent the 
scaled square root of the value from the digitizer; in that case, camera pixel values $P_{cam}(x,y)$ may be calculated
by squaring each image pixel value then dividing by the header field ``SCALE''.  The images have fixed geometry relative to the instrument, with CCD ``columns'' running in the horizontal direction in each image.  The cameras use dual digitizers so that
two different pedestal values and gain curves apply in the top and bottom half of
each image.

The images are masked on-board each spacecraft, to eliminate unused pixels outside the FOV 
of the camera, reducing downlink volume.  NFI and all three WFIs include
slightly more active pixels than required.  Unused/masked pixels have value 
0, which is not a possible value for a valid pixel because of the 
digitizer pedestal of roughly 400 counts.

Level 0 headers are assembled from engineering data and science metadata from the spacecraft. 
Pointing information is stored using the World Coordinate System (WCS); at this stage, it is ``roughed-in'' to few-pixel precision based on a priori 
knowledge of the instrument geometry and the spacecraft-reported pointing information.  No 
uncertainty layer is present at Level 0.

\subsection{Level 1 data\label{sec:L1}}

PUNCH Level 1 data files are also produced rapidly, within hours of receipt by the SOC.  They 
remain organized by spacecraft, and retain the one-to-one
relationship between in-space camera images and ground data products.  Each data file has
a two-letter code indicating the type of exposure, followed by a single
digit indicating which PUNCH Observatory produced the file.  The main data 
products use the same codes as Level 0 (Table \ref{tab:L0codes}). 

Level 1 pixels have floating-point values that are intended to capture 
photometric values on the sky, but are still individual exposures from 
particular Observatories through particular polarizers.  The 
Level 0 to Level 1 steps include: improved square-root decoding 
\citep{deforest_etal2022a}; destreaking to account
for quasi-frame-transfer streaks in camera operation; 
quartic-polynomial flat-field correction including both detector effects
and optical vignetting; cosmic ray and saturated-pixel detection and removal; 
pointing and distortion-function 
correction using the starfield; and point-spread function (PSF) 
regularization \citep{hughes_etal_2025}.  The 
pixel values are calibrated in $B_\odot$ units (using the mean 
photospheric radiance of $7.58\times10^6$~W~m$^{-2}$~SR$^{-1}$, adjusted to the 
passband of PUNCH). 

Stray light in the instruments is removed at Level 1. 
Stray light is removed from each of the polarized
(or unpolarized) channels (M,Z,P,C) separately, in four separate data streams
per instrument.

Solar stray light from the instruments is quasi-stationary in the instrument
reference frame (changing month to month due to eccentricity of Earth's orbit) and is removed 
at Level 1, using the minimum-value-over-time method with analytic correction for the 
statistical bias introduced by minimum-value sampling.  Because of the eccentricity of Earth's orbit, the Sun's apparent size (and therefore total 
intensity) change slowly over the course of the year, and therefore a posteriori instrument stray light models are 
regenerated every few days and linearly extrapolated to the time of current data.  These stray light models necessarily 
include quasi-stationary, circularly symmetric elements of the F and K corona, in addition to instrumental stray light.

The NFI instrument was observed during commissioning to be subject to variable stray light 
patterns from earthshine illuminating the interior of the instrument.  The patterns are dominated
by earthlight scattering off an interior tube baffle vane and illuminating the rear of the 
occulter.  Glint from the circular rim at the rear of the occulter then enters the A1 aperture,
producing large and variable 
stray light patterns that cannot be treated as quasi-stationary.  
This effect is removed on the basis of spatial structure and spatiotemporal variability, using
linear transformations and sparse matrix inversion.  The
NFI-specific removal algorithm and module were
a subject of rapid development early in the operational phase of the mission, and are detailed by
\citet{hughes_etal_2025}. 

PSF regularization is tuned to produce an absolutely uniform PSF on the celestial
sphere: the target PSF in the instrument plane is set to match the measured distortion function of
each instrument, so that (for example) deprojecting any given stellar
neighborhood to a locally-flat projection of the celestial sphere will yield
the same PSF regardless of position in the L1 image.  This means that careful analysis will
show slightly different PSF values across each instrument's focal plane, reflecting the 
optical projection of the celestial sphere onto the flat CCD.

Level 1 images are not resampled from the original camera pixels, but World 
Coordinate
System \citep[WCS;][]{Calabretta_Greisen_2002} coordinates, including distortion
parameters, are updated using the starfield in each image; alignment precision is 
under 10 arcsec RMS, well below the pixel size.

HDU2 of the L1 FITS files is populated with images containing photometric uncertainty information
for the corresponding HDU1 pixels, based on a priori photon noise estimation with correction
to account for subsequent steps.  This uncertainty value is also propagated and updated through
subsequent
processing levels.

\subsection{Level 2 data\label{sec:L2}}

PUNCH Level 2 data files are typically produced within 1-3 days of receipt of the first data
at the SOC; this is limited by the need to accumulate all relevant data from all four 
spacecraft, for each output image.  The data are merged across spacecraft and across polarizers, and
rectified to orient projected solar north with up on the image plane.  They retain the one-to-one 
temporal association between data product acquisition and individual data products, but 
lose the identity of individual camera frames or spacecraft.  For example, one polarized
acquisition sequence on-orbit results in twelve individual images (three from each of four
spacecraft) that are merged into a single polarized mosaic. Each data file has a three-letter 
code indicating the type of data product (Table \ref{tab:L2codes}).  

The mosaics are reprojected
into the Sun-centered azimuthal equidistant projection, so that radius on the image plane
is proportional to the solar elongation angle $\varepsilon$ (Appendix \ref{sec:geometry}). 
The mosaic pixel scale was chosen as a compromise between representing the $90^\circ$ field at full
resolution and the need to retain data products at a manageable size.  The mosaics support a spatial
resolution of approximately 3 arcmin, at a plate scale of 1.3 arcmin/pixel.

Because NFI offers higher resolution (approximately 1.5 arcmin),
a separate full-resolution NFI data product is offered at $2048\times2048$ resolution.
NFI full-resolution L2 products are resampled into the Sun-centered gnomonic (``tangent plane'') 
projection often called ``solar observing coordinates'', so that
radius on the image plane is proportional to tan($\varepsilon$).  The difference in projection is to
improve compatibility between NFI data and the legacy data from SOHO/LASCO \citep{brueckner_etal_1995},
while also working in a useful standard projection for the wide PUNCH FOV.  The difference between the
two projections amounts to a 1-NFI-pixel offset at 16 $R_\odot$, or up to 7 NFI pixels (two mosaic pixels)
at the outer edge of the NFI field of view; this is trivial for many applications but significant for 
PUNCH, which must co-align starfield images to roughly 10 arcsec (RMS) to achieve starfield 
subtraction.

The polarized
data products are typically 3-D ``cubes'' of size $w\times h\times 3$, with $w$ being the image width, $h$ being the image height, and the third dimension running across MZP\footnote{for Minus, Zero, Plus [60$^\circ$] polarizer angle relative to a reference angle} virtual-polarizer
triplet images (Appendix \ref{sec:photometry}), while the unpolarized/clear 
data products are typically 2-D $w\times h$ images. The MZP virtual polarizer angles are
measured relative to the system in which solar north is up; the system is described by \citet{deforest_etal2022b} and developed in 
more detail by 
\citet{patel_etal_2025}.

Level 2 PUNCH data include uncertainty information in HDU 2 of the FITS file.  The uncertainty layer is
used both to track signal to noise ratio (SNR) and also to mark bad pixels, including cosmic ray hits, known detector defects, and 
objects such as the high altitude aurora and other satellites.

\begin{table}[tbh]
    \centering
    \begin{tabular}{c|c|l}
       Code & Shape & Description  \\
       \hline
        PTM & $4096\times 4096\times 3$ & Polarized trefoil science mosaic (MZP) \\
        PNN & $2048\times 2048\times 3$ & Polarized NFI full-resolution image (MZP) \\
        CTM & $4096\times 4096$ & Clear (no polarizer) trefoil science mosaic \\
        CNN & $2048\times 2048$ & Clear (no polarizer) NFI full-resolution image
    \end{tabular}
    \caption{PUNCH Level 2 data product codes}
    \label{tab:L2codes}
\end{table}

\subsection{Level 3 data\label{sec:L3}}

PUNCH Level 3 data files comprise scientific image files and extracted solar-wind velocity maps.
The image files are background-subtracted.  Latency is typically 14 days from receipt of the 
data by the SOC; this is limited by the need to accumulate long-baseline datasets for the
background subtraction.  Image products are named with a leading ``P'' (Polarized) or ``C'' (Clear).
These images are produced in full-cadence and low-noise forms.  The full-cadence products are marked with
either ``T'' (for Trefoil) or ``N'' (for NFI), and are produced at the full observing cadence for that 
type of image:  4 minutes for P types and 8 minutes for C types. Low noise products are named with ``A''
(for Average) and consist of 32-minute averages (across four 8-minute observing holds each).  Velocity
maps are produced via the delayed-autocorrelation method \citep{Attie_etal_2025} and are named ``VAM''.

\begin{table}[tbh]
    \centering
    \begin{tabular}{c|c|l}
    Code & Shape & Description \\
    \hline
    PTM & $4096\times 4096\times 2$ & Polarized K-only trefoil science mosaic (B, $^{\perp}$pB)   \\
    PNN & $2048\times 2048\times 2$ & Polarized K-only NFI full-resolution image (B, $^{\perp}$pB) \\
    PAM & $4096\times 4096\times 2$ & Polarized K-only averaged (low-noise) mosaic (B, $^{\perp}$pB) \\
    PAN & $2048\times 2048\times 2$ & Polarized K-only averaged (low-noise) NFI image (B, $^{\perp}$pB)\\
    CTM & $4096\times 4096$ & Clear (no polarizer) K-only trefoil mosaic \\
    CNN & $2048\times 2048$ & Clear (no polarizer) K-only NFI full-resolution image \\
    CAM & $4096\times 4096$ & Clear (no polarizer) K-only averaged (low-noise) mosaic \\
    CAN & $2048\times 2048$ & Clear (no polarizer) K-only averaged (low-noise) NFI image\\
    VAM & $720\times 8$ & Velocity map at 8 altitudes, derived from mosaic correlation\\
    \end{tabular}
    \caption{PUNCH Level 3 data product codes}
    \label{tab:my_label}
\end{table}

Figure \ref{fig:brightness} illustrates the fundamental challenge of
background subtraction for all heliospheric imagers including PUNCH: the F
corona (zodiacal light) and starfield dominate over the signal by
2-3 orders of magnitude in the far field.  PUNCH uses the same
six-step process for background subtraction as was developed for STEREO/HI
\citep{deforest_etal_2011,deforest_etal_2016}, with the advantage of
additional lessons learned from that mission, and also the challenge of using spliced
mosaic images rather than single exposures from one optical system.  The process is summarized here and detailed by \citet{hughes_etal_2025}.

PUNCH background subtraction requires separate treatment of three steady backgrounds.  
They are all separated on the basis of stationarity in a particular reference frame.  

The first background is steady solar stray light, which is removed at Level 1 (and described in Section \ref{sec:L1}).  The other two are removed at Level 3.

The second background is the F corona (zodiacal light), which is quasi-stationary in the heliocentric
reference frame and is 
removed on the basis of a running-minimum fit, corrected for variable noise level effects, in the Level 2 data.  Each pixel location in the mosaic
frame is fit with a running-minimum polynomial background estimator; these polynomials are conditioned to
produce daily F corona models, which are interpolated to provide an F background for each data product, and
subtracted in the L3 pipeline.  This method makes use of the observed approximate symmetry of the heliospheric dust cloud; it is expected that any dynamic features such as a dust enhancement near Venus \citep[e.g.,][]{jones_etal_2017,stenborg_etal_2021a} or local dust depletions \citep{stenborg_etal_2021b} will be recognizable as their dynamics follow Keplerian orbital time scales of months, rather than solar wind time scales of days, and may be removed on that basis. 

The third background is the stellar background, which is stationary in the celestial reference frame.  L2 data
products, with the F corona removed, are resampled to celestial coordinates for an interval, nominally
$\pm$~2~weeks, around each
day, and a minimum value fit is applied to produce a daily celestial background -- which is then interpolated
and resampled to match each L2 mosaic, then subtracted from the data themselves.  This removes stationary celestial objects, is 
enabled by photometric adaptive-filter resampling \citep{deforest_2004}, and requires careful treatment of the PSF as each resampling 
operation slightly blurs the data.  Non-stationary celestial objects include the major planets and the Moon, which are identified 
by pixel saturation and marked using the uncertainty layer; and comets and asteroids.  Asteroids at or below 12th
magnitude, of which there are roughly $10^4$ \citep{Bottke_2020}, are below the noise floor in a
single PUNCH/WFI image and do not significantly affect PUNCH science.  Considered as a long-exposure
detecting telescope for asteroids or comets, PUNCH has a limiting magnitude of approximately 18 
when 60 days of images are combined coherently. 

The very brightest celestial objects -- stars, the Moon, and bright planets -- saturate the CCD and invalidate 
pixels both in the vicinity of the PSF-correction neighborhood around the object, and also over
the locus of associated CCD charge-bleeding (which produces saturated streak artifacts along 
the readout direction); this affects approximately 0.02\% to 0.5\% of pixels in each mosaic 
depending on the time of year and lunar phase.  The WFI exposure time is adjusted to balance
between preserving the faint end of the dynamic range, and saturating the brightest objects 
in the sky. Therefore  
objects of magnitude 
2.0 or brighter saturate the detector in the nominal observing sequence.  There are roughly 65
celestial objects
(including the Moon and the bright planets) that saturate WFI images; the star Saiph 
($\kappa$ Orionis, Orion's right knee directly ``below'' Betelgeuse,
visual magnitude 2.09) is the brightest celestial object that does not saturate WFI clear 
exposures.  Saturated streaks from bright objects rotate with the spacecraft and therefore do
not impact any particular piece of dark sky for more than one 8-minute observing window at a
time.

Following \citet{deforest_etal_2011}, a final step of Fourier filtering is applied
to clean the L3 data via motion filtering.  The motion-filtering step for PUNCH is
sufficiently broad to pass SIRs and other caustic structures that may move
$3\times$--$5\times$ slower than solar wind itself.

PUNCH background subtraction is complicated by the fact that every image is polarized, as are all of 
the backgrounds: starfield, F corona, and instrumental stray light. The three polarization channels 
are treated with individual background processes and models
(eight in all: four for the mosaics and four for the NFI full-resolution products). Using
virtual polarizer channels, rather than pB or Stokes parameters, allows re-using baseline-estimation methods already 
adopted for unpolarized brightness images: each virtual polarizer channel represents an actual 
filtered radiance, from which a background may be estimated using conventional minimum-value
methods.  Each step of background subtraction is resolved not only in spatial coordinates (in the 
instrument, solar, and celestial frames), but also in polarization angle, by mathematical
transformation of the virtual polarizer triplets.  The method is described in detail by \citet{hughes_etal_2025}.

\section{Additional science\label{sec:additional-science}}

PUNCH was specifically designed to meet the requirements in Section \ref{sec:Requirements} and
thereby fulfill the science described in Section \ref{sec:Science}. PUNCH observations are 
also suitable for addressing additional science topics in astrophysics, planetary science, and
geospace science.  Here we discuss the relevance of PUNCH observations to several topics 
unrelated to the PUNCH primary science objectives: stellar polarimetry,  
asteroid observations (particularly NEOs and possible vulcanoids), interplanetary dust 
observations, and geospace observations of high-altitude aurora.

\subsection{Stellar polarimetric observations\label{sec:stellar-polarimetry}}

As a byproduct of Level 3 background subtraction, PUNCH data include a comprehensive polarized map of 
the sky at all ecliptic latitudes out to $\pm45^\circ$ from the ecliptic, with $< 0.1\%$ polarimetric
precision.  Stars are observed to be polarized \citep{hiltner_presence_1949,hall_observations_1949} and
these observations and the physics are nicely reviewed by \citet{clarke_stellar_2010} and by \citet{trippe_polarization_2014}.  
The effect is
troublesome for the primary PUNCH science as it requires precise polarimetric
characterization of the entire visible celestial sphere. These polarimetric data contain useful
information for understanding stars and the galactic magnetic field, supplementing the many 
important telescopic surveys that exist \citep[e.g.,][]{mathewson_polarization_1970,hsu_standard_1982,heiles_2000,fossati_etal_2007,versteeg_interstellar_2023,angarita_interstellar_2023}.

The bulk of the stellar linear polarization signal is not due to intrinsic polarization of emitted
starlight, but to aspherical grains interacting with the light as it propagates: either in 
interstellar space \citep{versteeg_interstellar_2023} or (less commonly) in 
local dust clouds
\citep{poeckert_intrinsic_1979}.  The polarization of starlight by interstellar dust is interesting in part because 
it is a measure of the direction and strength of the interstellar magnetic field \citep{versteeg_interstellar_2023}, 
and of turbulence in the interstellar
medium \citep{angarita_interstellar_2024}. 

Compared to contemporary surveys, PUNCH is both more comprehensive and more limited by spatial
resolution. At the nominal 3 arcmin resolution of mosaic data, crowding places the limiting apparent
magnitude at roughly 14: although PUNCH can detect fainter stars photometrically, there are only roughly
$10^7$ resolution elements in the entire 9-steradian effective field of view, matching the number of 
magnitude 14 objects in the same field. PUNCH can therefore isolate most stars only at 
about magnitude 12 or brighter, although it is capable of ``crowded-field polarimetry'' of much fainter objects.

\subsection{Asteroid and comet observations\label{sec:asteroids}}

\begin{figure}
    \centering
    \includegraphics[width=0.75\linewidth]{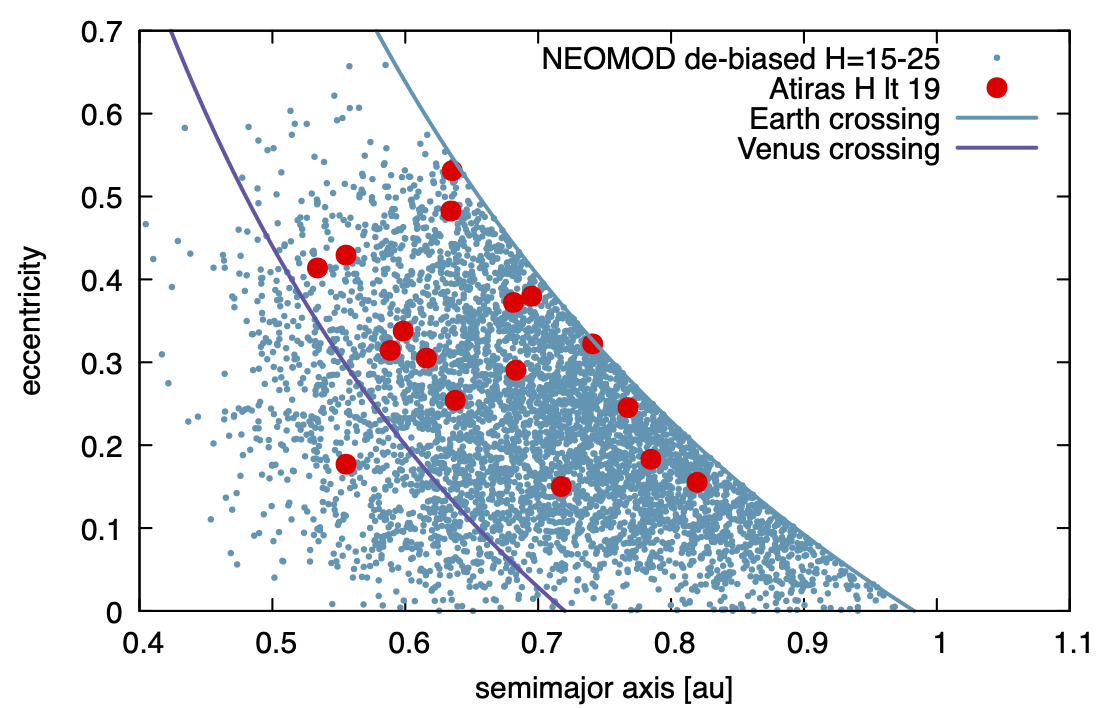}
    \caption{PUNCH should be able to detect many Earth-crossing or 
    sub-Earth objects in solar orbit, as seen in this scatterplot of 
    orbital characteristics of predicted and observed objects in the 
    solar system.  Blue points indicate objects in the predicted 
    population, with orbits entirely within 1~AU from the Sun 
    \citep[NEOMOD solar system model;][]{Deienno_2025Icar}, which 
    should be detectable by PUNCH. Red circles indicate known objects 
    in this region with absolute magnitude brighter than 19. The blue 
    line indicates orbital properties that lead to an Earth-crossing orbit, 
    and the purple line is the same for Venus. The only red dot to the 
    left of the Venus-crossing line is the only known asteroid whose 
    orbit is entirely inside that of Venus \citep{Bolin_2022}.}
    \label{fig:asteroids}
\end{figure}

Recent asteroid surveys from ground-based telescopes have identified a handful of asteroids on orbits 
entirely within the orbit of Earth or Venus (e.g. \citet{Bolin_2022,Sheppard_2022}). These are notable 
discoveries because the nature of these asteroids' orbits only permit their observation from the Earth 
when they have a very low solar elongation, which is impossible for some telescopes or only accessible 
for short time windows each night for others. However, models for the population of near-Earth objects 
predict a sizable population of objects in this region of the solar system (Figure \ref{fig:asteroids}), 
including 5-10 as of yet undiscovered asteroids larger than 1 km \citep{Nesvorny_2023,Deienno_2025Icar}.

\new{PUNCH observes continuously in a part of the sky that is historically difficult: areas with low solar elongation. Many of the objects predicted to populate this region are always 
in the PUNCH FOV.} The ability for PUNCH to discover, track and characterize 
asteroids and comets depends on the trade between its relatively small imaging aperture, its large 
field of view and its rapid and constant imaging cadence. There is precedent for observing asteroids 
and comets with SOHO and STEREO \citep{Knight_2010,Li_Jewitt_2013}, and the larger field of view, 
smaller pixels and lack of on-board image stacking may permit PUNCH to be a more effective tool 
for characterization and for discovering new objects. 

\subsection{Solar system dust observations\label{sec:dust}}

PUNCH is well placed to make significant contributions toward
understanding the
zodiacal dust cloud (ZDC).  Recent coronagraph observations from the STEREO and PSP missions have shown that the photospheric light scattered by the ZDC exhibits small variations that need further definition.
This scattered light is referred to as the ``zodiacal light'' or 
``F corona'' \citep{Schuerman_Weinberg_1983}. With its complete $360^\circ$ coverage out to $45^\circ$ 
from the Sun, PUNCH is able to measure the brightness and polarization of 
the F-corona with greatly increased sensitivity and coverage compared to prior observations.

Measurements of the polarization of zodiacal light are scarce for the regions observed by PUNCH.  
\citet{koutchmy_lamy1985} reported on the status of contemporary and
earlier 
observations of F-corona polarization. They revealed a 
discrepancy in contemporary observations of the high corona between 
7~$R_\odot$ and 15~$R_\odot$. No polarization measurements had ever 
been reported between elongations of roughly 4$^\circ$ and 45$^\circ$ 
although very wide angles were covered by the Teide observatory 
\citep{dumont_sanchez1976}.  No newer survey measurements have covered
this missing area; the polarization profile of the inner zodical cloud
(throughout the PUNCH field of view) remains unknown.
 
The orbits of the dust in the ZDC are affected by the gravitational
influences exerted by the solar system planets.  The center of mass 
of the solar system (the barycenter) is in constant motion around the 
Sun, varying in longitude, latitude and radius from the Sun. The barycenter's
distance variation, in particular, creates an asymmetry in the  
distribution of the inner ZDC, and a corresponding asymmetry in the appearance of the zodiacal light; this asymmetry has been observed \citep{stenborg_etal_2024} with the PSP/WISPR instrument.  With its 360$^\circ$ coverage, PUNCH will be able to determine whether any north-south asymmetry exists, and the extent and evolution of the east-west asymmetry.

Finally, with its increased sensitivity, PUNCH will likely be able to detect subtle brightness increases due to, for example, asteroids or comets entering the inner heliosphere as well the impact of interstellar dust increasing the brightness in a range of ecliptic longitudes centered at $259^\circ$, due to the solar system traveling through the galaxy.  The range of longitudes could be as high as $180^\circ$ (full-width), but the increase would be highest at the central longitude.  Additional searches include finding the dust depletion zone, which begins at 35~$R_\odot$ and peaks at 25~$R_\odot$ \citep{stenborg_etal_2021b}, or even if there are additional sublimation zones beyond 35~$R_\odot$; and characterizing
and understanding dust depletions behind CMEs as have been seen with WISPR \citep{stenborg_etal_2023}, or associated with other plasma structures such as SIRs.

\subsection{Aurora and airglow observations\label{sec:aurora}}

PUNCH may contribute significantly to geospace studies through novel observations of upper atmospheric luminous phenomena, notably the aurora and airglow.  

At Earth, the aurora is a persistent phenomenon, and a strong diagnostic of space weather near Earth. As energetic electrons and protons interact with the neutral atmosphere, ionization and excitation processes occur. The basic mechanism is well understood. Electrons precipitate along magnetic field lines into the ionosphere, where they collide with atmospheric particles, exciting them and releasing photons as ``electron aurora'' \citep[e.g.,][]{Rees1963}. Protons also undergo charge-exchange collisions with atmospheric particles, creating excited hydrogen atoms that emit characteristic ``proton aurora'' \citep[e.g.,][]{mcilwain1960}. 

Most auroral emission occurs from 100 to 400~km altitude; the specific altitude serves as a proxy for the energy of precipitating particles and provides key insights into magnetosphere-ionosphere coupling \citep[e.g.,][]{kaila1989}.  Discrete green-line aurora typically occurs below 150km, and red-line aurora produced by atomic oxygen is seen up to 400km. The Solar Mass Ejection Imager \citep[SMEI;][]{eyles_etal_2003} serendipitously detected luminous phenomena at altitudes $\ge 840$~km over the auroral zones and polar caps \citep{mizuno_etal_2005}. To date, the SMEI result remains the only report of this ``high-altitude aurora'', which has never been systematically characterized or investigated in detail. As a result, its energy source, particle population, and underlying physical mechanism are poorly understood.  
Informed by the SMEI result, PUNCH was designed specifically to mitigate the effects of 
high-altitude aurora for heliospheric imaging; but Level 1 data include hypothetical effects of high-altitude aurora and other high-altitude luminous phenomena including airglow, affording a unique opportunity to investigate these effects systematically and understand a novel and as-yet unexplored phenomenon of the magnetosphere. 

Other optical emissions in the upper atmosphere, collectively known as airglow, are produced by chemical reactions between solar radiation and atmospheric constituents \citep[e.g.,][]{silverman1970}. Unlike aurora, airglow is continuously present and occurs at discrete wavelengths throughout the spectrum. PUNCH observes airglow emissions from a novel vantage point, offering a new probe of the morphology, potential dynamics, and altitude of the quasi-steady airglow emission.

\section{Conclusions\label{sec:conclusions}}

PUNCH addresses large-scale and meso-scale science of the solar wind and corona, with a 
larger field of view and deeper sensitivity than have been possible with prior missions.  The mission
builds on capabilities proven through earlier iterations of similar instruments
and observing concepts; these capabilities are supported by a novel mission design that enables a science of
unification, viewing the solar corona and inner heliosphere as a single unified system, within
its celestial context on the sky. 

The unique challenges of wide-angle heliospheric imaging drive a unified view of
the mission itself, considering the post-processing pipeline as a critical 
element of the instrument suite.  The rigor required in merging and calibrating
the PUNCH images well enough to support its primary science make PUNCH a useful
observatory for 
astronomical, planetary, and geophysical topics well beyond its primary 
mission science.

\acknowledgements

PUNCH is the result of a very large scientific and technical community working together over a decade, 
combining outstanding talent across many fields and heroic eﬀort through many types of adversity. The 
hundreds of individuals, whose dedication and willing support brought PUNCH to fruition, span four 
continents and dozens of institutions. PUNCH builds on the major achievements of prior missions, 
including PSP/WISPR, STEREO/SECCHI, Coriolis/SMEI, SOHO/LASCO, and the Helios three-channel 
photometer. 

PUNCH owes much of its overall design and scientific concept to early eﬀorts by T.A. Howard, 
whose work was instrumental in co-developing and exploiting methods 
for deep background-subtraction of heliospheric images; in predicting and promoting the 
possibility of wide-field 3-D analysis via polarization; and in helping demonstrate the feasibility 
of heliospheric imaging from below the Van Allen belts.

We particularly acknowledge Tele Vue Optics and remember their founder, Al Nagler (1935-2025),
 for designing and supplying the optics for the Wide Field Imagers, working well 
outside Tele Vue’s “comfort zone” of eyepiece design, and stepping up to produce a 
very compact, achromatic, space-qualified system with very low distortion, several 
important stray-light reducing features, and excellent focus over a very wide field of view. 
PUNCH/WFI is but one of Al Nagler's many contributions to the astronomy, spaceflight, 
and scientific communities, and he is both sorely missed and fondly remembered.

We also gratefully acknowledge C. Eyles for early consultation on heliospheric imager design based on his 
experience with SMEI; the SwRI Internal Research program for much early support that led to 
the successful PUNCH proposal; J. Burch for consultation and institutional support
throughout the project; and J. Andrews for invaluable advice on project management.

J.B., S.F., S.E.G., and A.M. acknowledge support of the National Center for At-
mospheric Research, a major facility sponsored by the National Science Foundation
under Cooperative Agreement No. 1852977. D.J.M acknowledges support from the
IMAP mission as a part of NASA’s STP mission line (80GSFC19C0027).

PUNCH is a heliophysics mission to study the corona, solar wind, and space weather as an integrated
system, and is part of NASA’s Explorers program (Contract 80GSFC14C0014).

\appendix

\begin{table}
    \centering
    \begin{tabular}{c|c}
             \multicolumn{2}{c}{~}\\
         \multicolumn{2}{c}{\textbf{Coordinate axes and distances}}\\
         \hline
         {RTN coords} & Radial/Tangential/Normal coordinates \citep{burlaga_1984} \\
         {X axis} & Direction from Sun center toward observer \\
         {Y axis} & Cross product of X and solar north (projected solar west) \\
         {Z axis} & Perpendicular to X and Y (projected solar north) \\
         {Plane of Sky} & YZ plane \\
         \hline
         {r} & 3-D distance from Sun center to a scattering point \\
         {r$_{pos}$} & 2-D distance from Sun center to a line of sight, in the YZ plane \\
        {r$_{obs}$} & Distance from Sun center to Observer (along X axis) \\
         \hline
         {s} & Distance toward the observer from the Thomson surface \\
         {$\ell$} & Distance from the observer, along a given line of sight \\
         \hline
         {X}$_{im},~{Y}_{im}$ & Image-plane coords: $X_{im}$ horizontal, $Y_{im}$ vertical, origin: lower left \\
         \hline
         $r_\odot$ & Actual solar radius (696,340 km) \\
         \hline
         \multicolumn{2}{c}{\textbf{Relevant angles}}\\
         \hline
         $R_\odot$ & Convenience unit: $0.25^\circ$ (approx. apparent solar radius from 1 AU) \\
         \hline
         $\varepsilon$ & Elongation angle from Sun: apparent distance from Sun center \\
         $\alpha$ & Solar ``azimuth'' (position angle), CCW about X from Z \\
         \hline
         $\chi$ & Scattering angle ($\chi=0$ implies no scatter) \\
         $\chi'$ & Supplement to $\chi$ (interior to the scattering triangle) \\
         $\tau$ & Complement to $\chi$ (Thomson angle as in Fig. \ref{fig:appendix-diagram}) \\
         \hline
         $\xi$ & Out-of-sky-plane angle ($\xi = \varepsilon + \tau$) \\
         $\psi$ & Complement to $\xi$ (forms scattering triangle, with $\chi'$ and $\varepsilon$)\\
         \hline
         \multicolumn{2}{c}{\textbf{Photometric quantities}}\\
         \hline
         $B_\odot$ & Mean solar photospheric radiance (units of W~m$^{-2}$~sr$^{-2}$)\\
         $I_\odot$ & Total solar intensity (units of W~m$^{-2}$)\\
         \hline
          $I$ & Stokes parameter (total intensity or radiance; clarify when used) \\
          $Q$ & Stokes linear-polarization parameter (vert. - horiz.) \\
          $U$ & Stokes linear-polarization parameter (diagonal) \\
         \hline
         $tB$ & ``total brightness'': Stokes $I$ radiance (also $B$, deprecated) \\
         \hline
         $B_R, B_T$ & Radiance through a radially or tangentially oriented polarizer \\
         $B_M, B_Z, B_P$ & Radiance through a polarizer at $-60^\circ$, $0^\circ$, or $+60^\circ$ orientation \\
         $B_C$ & Radiance through no polarizer (``clear''); similar to $tB$ \\
         \hline
         $pB$ & ``tangential polarized brightness'': $B_T-B_R$ (also $^{\perp}pB$)\\
         $pB'$ & ``alt polarized brightness'': Stokes-like parameter, conjugate to $pB$\\
         \hline
         $L$ & total linear polarized brightness $\sqrt{Q^2+U^2}$ (also $^{\circ}pB$) \\
         \hline
         $p$ & ``Degree of [tangential linear] polarization'' $(pB / tB)$ \\
         $PR$ & ``Polarization ratio'' $B_R/B_T= (1-p)/(1+p)$ \\
    \end{tabular}
    \caption{Glossary of PUNCH nomenclature for geometric and photometric analyses}
    \label{tab:defs}
\end{table}

\begin{figure}[tb]
    \centering
    \includegraphics[width=0.9\linewidth]{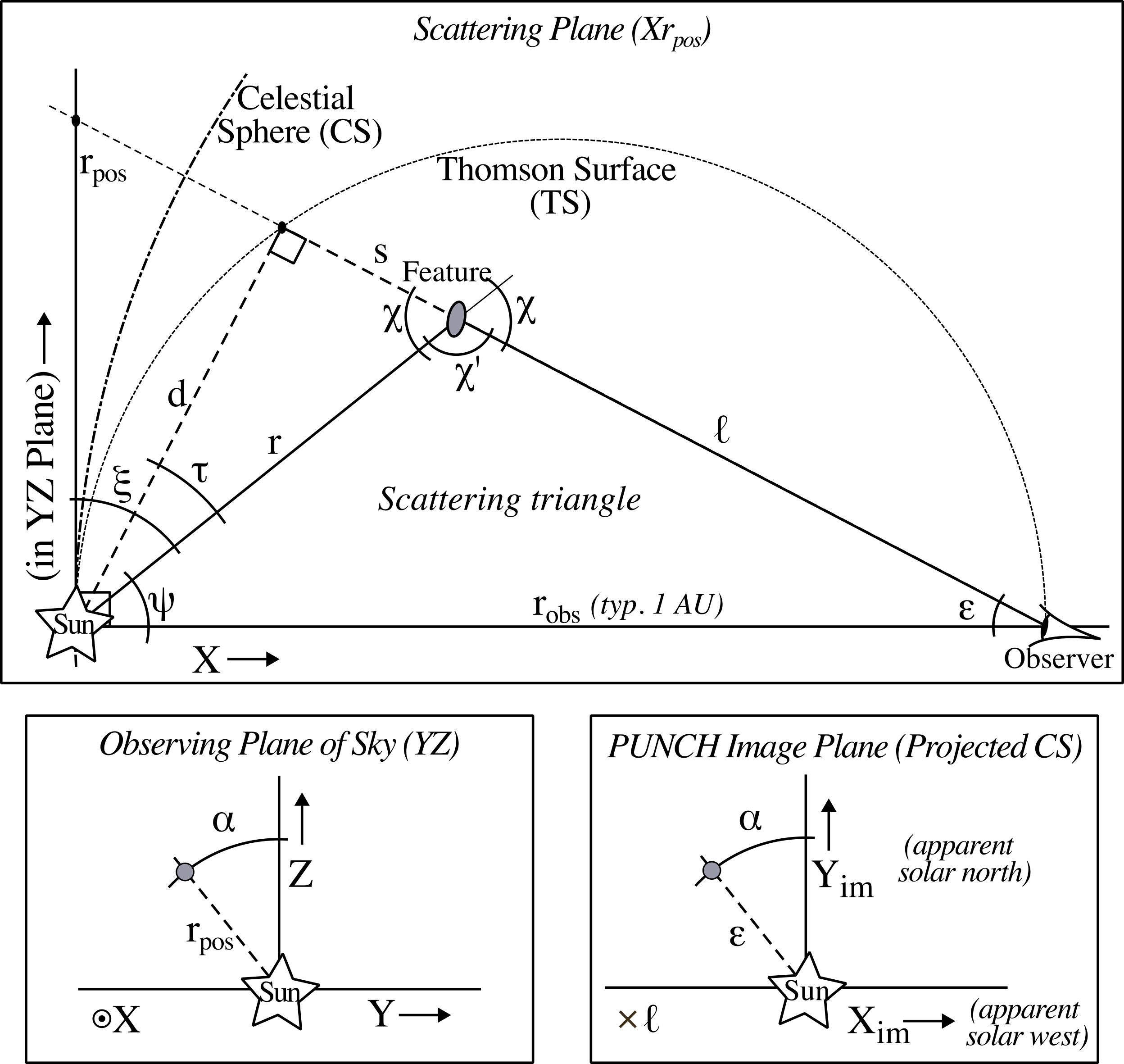}
    \caption{Diagrams and standardized symbols for PUNCH observations show many of the relevant quantities and constructions used for PUNCH analysis. PUNCH mosaic images are delivered as 
    azimuthal equidistant (AZP) projections of the celestial sphere (CS), such that the Sun is
    at the coordinate origin (at the center of the pixel grid), apparent solar north (projected Z) is in the Y$_{\mathrm{im}}$ direction, and coordinate radius in the 
    (X$_{\mathrm{im}}$,Y$_{\mathrm{im}}$) plane is equal to the elongation angle $\varepsilon$.}
    \label{fig:appendix-diagram}
\end{figure}

\section{Nomenclature: Geometric and Photometric Quantities\label{sec:appendix-nomenclature}}

Analysis of PUNCH polarimetric wide-field images makes use of 
different
geometry than has been typical for solar remote sensing missions; and polarimetric 
imaging of distributed objects
requires a specialized vocabulary.  In the interest of coordinating 
across analyses of PUNCH data, we 
define here a ``PUNCH standard'' nomenclature for geometric (Section \ref{sec:geometry})
and photometric/polarimetric (Section \ref{sec:photometry}) terms thought to be of interest
for all PUNCH analyses.  Table \ref{tab:defs} 
is a glossary of these terms.

\subsection{Geometric definitions\label{sec:geometry}}

Distances, coordinates, and angles relevant to PUNCH observations are drawn in Figure \ref{fig:appendix-diagram}.  

The lower right panel of Figure \ref{fig:appendix-diagram} shows the image plane used for 
PUNCH mosaic image products.  For normal L2 and higher science data products from 
PUNCH, the $X_{im}$ and $Y_{im}$ axes of the image plane correspond to
observer's ``solar west'' and ``solar north'' respectively.  Mosaic products are in 
the azimuthal equidistant projection, so that radial distance from the center of the image 
is directly proportional to the elongation angle $\varepsilon$, also called the ``apparent
distance'' between the feature the Sun.  Likewise, ``azimuth'' ($\alpha$ angle) on the image,
measured counterclockwise from solar north (the $Y_{im}$ direction), corresponds to solar
position angle on the sky.  (Full-resolution NFI images are delivered in
the gnomonic projection, to better match the projection used by prior instruments
such as SOHO/LASCO.  For this projection, radius on the image plane is proportional to 
tan($\varepsilon$) rather than $\varepsilon$ itself.)

The top panel of Figure \ref{fig:appendix-diagram} shows a cross-section 
of the 
three-dimensional geometry of the system.  The ``scattering triangle'' formed by a particular
small feature, 
the Sun, and PUNCH itself defines a plane, the ``scattering plane'', in three dimensions;
this plane projects to a radial line in the image plane (the dashed line in the 
lower right panel of Figure \ref{fig:appendix-diagram}). 

For analysis in Cartesian coordinates, we adopt the 
familiar RTN Cartesian system defined by \citet{burlaga_1984}
In this system the 
X axis extends from 
the Sun to the observer; the Y axis extends in the direction of the cross product of
solar north with X (``solar west''); and the Z axis extends in the mutually perpendicular direction (``projected solar north''), to form a right-handed coordinate system.
In RTN coordinates the YZ plane, shown in the 
lower left panel of Figure \ref{fig:appendix-diagram}, is the ``plane of the sky'', also 
called the ``tangent plane''.  Note that, because PUNCH is located at Earth, RTN
coordinates applied to PUNCH data match the heliographic-RTN (HGRTN) coordinates
defined by \citet{franz_harper_2002}. In cases where multiple spacecraft or vantage
points may be used, we recommend using X$_{HG}$, etc. to denote HGRTN coordinates.
Also note that RTN is similar to the heliocentric-cartesian (HC) coordinates defined
by \citet{thompson_2006}, but permuted: X$_{HC}$ and Y$_{HC}$ loosely correspond to
the image coordinates X$_{im}$ and Y$_{im}$ on a rectified solar image, while the 
corresponding RTN coordinates are Y and Z.  We adopt RTN despite the permutation, both
to highlight that PUNCH images are different in nature from a narrow-field 
sky plane projection,
and also to enable easier comparison with the large body of existing solar-wind work
conducted in RTN coordinates.

L2 and higher image data products from PUNCH are produced in specific projections.
Historically, coronagraph images \citep[e.g. from SOHO/LASCO][]{brueckner_etal_1995} have 
been created in an approximate
``gnomonic'' (tangent-plane) projection. PUNCH/NFI data products are produced in this 
projection to match the historical data sets.  Gnomonic projection leaves 
much to be desired when extended to wide fields of view, and therefore PUNCH mosaic images 
are delivered in the azimuthal equidistant projection shown in the lower right panel of Figure 
\ref{fig:appendix-diagram}.  In azimuthal equidistant projection, distance from the center
of the image is proportional to the elongation angle $\varepsilon$ rather than (as in gnomonic
projection) its tangent.

The two projections differ slightly but significantly in the inner part of the PUNCH 
field of view.  In particular, 
the NFI FOV apparent radius (8 degrees from Sun center) is approximately 7 pixels larger in 
NFI data products 
than it would be with a comparably scaled azimuthal equidistant projection.  

The ``Thomson Surface'' (TS) is an important figure for analysis of Thomson scattered light
\citep{vourlidas_howard_2006}.
In the point-Sun approximation \citep{howard_deforest_2012}, the TS is the locus on which 
the scattering angle 
$\chi$ happens to be $90^\circ$, and therefore
the intersection of a given line of sight with the TS is the location of closest
approach of that particular line of sight to the Sun.  The TS
is the sphere
with a diameter extending from the Sun to the observer, following the theorem attributed to 
Thales of Babylonia \citep{laertius_50}.

Direct 2-D image analyses of the corona and heliosphere have typically treated features 
as if they're projected onto the tangent plane: 
because $r_{obs}$ is known and $\varepsilon$ is measured by direct location in the image plane,
the related distance $r_{pos}=r_{obs}\tan\varepsilon$ is simple to calculate.  Alternatively,
in wide field studies \citep[e.g. with STEREO/SECCHI,][]{howard_etal_2008}, features are
sometimes considered to be on or near the Thomson surface 
\citep[e.g.][]{vourlidas_howard_2006, deforest_etal_2013}, placing them at the 
line-of-sight impact parameter distance $d=r_{obs}\sin\varepsilon$ from the Sun.

PUNCH is designed for 3-D analysis using polarimetry, which makes use of additional named
angles and distances in the scattering plane.  The desired measurement for 3-D location is
$\ell$, the distance from the observer to the feature\footnote{$\ell$, denoted 
``\textbackslash ell'' in \LaTeX{} math mode,
has also been called $z$ by
some authors, a usage we deprecate because it is similar to the name of a standard Cartesian
axis.}. 
Rather than measuring $\ell$ directly, polarimetry yields an estimate of the scattering 
angle $\chi$; or equivalently its complement, the Thomson angle $\tau$ 
(Appendix~\ref{sec:appendix-polarization}). That angle gives 
the distance $s$ of a feature from the Thomson surface via 
the relation $s=r_{obs}\sin\varepsilon\sin\tau$; this, in turn, yields the feature's
distance $\ell = r_{obs}\cos\varepsilon - s$ from the observer .  

Some authors have preferred to use either the 
solar angle $\psi$ or its complement, the
out-of-sky-plane angle $\xi$, for direct comparison to 2-D sky-plane analyses; we
observe that $\xi=\tau+\varepsilon$.
These and other  
trigonometric relationships may be directly inferred by inspection of 
Figure \ref{fig:appendix-diagram}. 

\subsection{Photometric and polarimetric definitions\label{sec:photometry}}

PUNCH measures distributed objects (features in the corona and solar wind).  The fundamental
measurement is photon fluence on each pixel in the instrument focal plane, through 
the filtering and geometry of the optics and the switchable polarizers in the instrument. 

Digitizer output number from each exposure is approximately proportional to the number of
visible-light photons incident on 
each CCD pixel during an exposure interval; after background subtraction and spectral 
calibration this
is equivalent to a visible energy $E_{vis}$ deposited on each pixel during each
exposure; dividing by the exposure time yields a power $P_{vis}$ (in W or, more 
appropriately, in femtowatts or fW) deposited on each 
pixel by Thomson-scattered light during data collection. Dividing by the effective area 
of the instrument yields an absolute intensity $I_{vis}$ (in 
W~m$^{-2}$) at that pixel.  Dividing by the solid angle subtended by each 
pixel yields the radiance or ``brightness'' $B_{vis}$ (in W~m$^{-2}$~sr$^{-1}$), which is 
a photometric invariant in linear ray optics.  The intensity is an extensive property and
therefore is summed or integrated across features that include more than a single pixel,
while the radiance
is an intensive property and therefore is averaged across features that include more than a 
single pixel. Because PUNCH's primary science target is scattered sunlight, the natural 
unit for radiance is $B_{\odot}$, the mean solar photospheric radiance (i.e. the solar 
intensity $I_\odot$ divided by the subtended solid angle of the Sun
$\Omega_\odot$\footnote{Note that $\Omega_\odot$ is also used in other contexts to denote 
the solar average rotation rate; for that reason we typically use instead the expression
$\pi(r/r_\odot)^2$, which is valid for features throughout the PUNCH field of view, as well as for the familiar case where $r=1$AU.}). 
Delivering radiances in $B_{\odot}$ removes the spectral profile of the instrument
and the Sun itself, and the effects of solar limb darkening, from subsequent calculations
of polarimetry and feature density. 

PUNCH is a polarimetric instrument.  Level 3 data products are delivered in dual 
radiance channels $tB$ (``total brightness'') and $pB$ (``polarized brightness'').
$tB$ is preferred nomenclature for an unpolarized total brightness data product, while the 
more ambiguous $B$ may be used for generic radiance quantities but is deprecated by PUNCH.
PUNCH produces two kinds of $tB$: ``clear'' images, taken with no polarizing filter
in the beam, at 8 minute cadence; and ``three-polarizer tB'' images, delivered as part
of a ($tB$, $pB$) image pair, at 4 minute cadence. 

Polarimetry may be (and has been) described in multiple ways by multiple authors; the 
fundamentals of the Stokes parameter formulation are covered in standard textbooks
\citep[e.g.][]{hecht_zajac_1974,bornwolf1999}.  The polarization signal expected from
the far-field K corona, and the PUNCH approach to measuring it, are  
developed in detail by \citet{deforest_etal_2013} and \citet{deforest_etal2022b},
respectively, and summarized here. 

The Stokes system of polarimetry defines an overall unpolarized parameter $I$ and 
three polarization 
parameters $Q$, $U$, and $V$.  Each parameter name has been used to describe 
intensity or radiance of a beam of light (or other wave), in different contexts.  
$I$ represents the total radiant 
energy in a beam, while $Q$, $U$, and $V$ represent difference signals between 
the amount of radiant energy passing through conjugate 
polarizers (and hence these parameters do not represent actual radiant energy transfer
but rather the character of the energy flux represented by $I$).
$Q$ is the difference between 
brightnesses through a vertical and horizontal polarizer, $Q \equiv B_{|}-B_{-}$; and $U$ is
the difference between brightnesses through two diagonal polarizers, 
$U\equiv B_{\backslash}-B_{/}$.  By convention, these orientations are defined relative to
a fixed laboratory or instrument.  Stokes $V$ is the difference between brightnesses 
observed through two opposite circular polarizers and is not relevant to linear 
polarization analyses.  The Stokes parameters bear the relationship $I^2~\ge~Q^2+U^2+V^2$,
because violating that relationship would require negative brightness values through 
one or more polarizers.  Treated as unit vectors in a 3-space, $I$, $Q$, and $U$ form 
a complete basis for describing linear polarization.

Thomson scattering polarizes light perpendicular to the scattering plane (Figure 
\ref{fig:appendix-diagram}), by projection of the electric field from the incident ray into
the plane perpendicular to the scattered ray.  This means that the radiance $B_R$ viewed
through a radially-aligned linear polarizer is less than the radiance $B_T$ viewed through a 
tangentially-aligned polarizer (i.e. one oriented $90^\circ$ to the solar-radial direction).
The quantity $B_T-B_R$ (``tangential polarized brightness'') is similar to the Stokes 
parameter $Q$, with the notable 
difference that $Q$ is defined in the fixed reference frame of a laboratory, while the 
orientation of the polarizers for $B_T$ and $B_R$ varies around the image plane.  We refer
to $B_T-B_R$ as $pB$ and identify it with the $^{\perp}pB$ defined by 
\citet{deforest_etal2022b}.  The appendix to that article, incidentally, includes 
a discussion of the (confusing) history of $pB$ nomenclature, including the contrast between
$^{\perp}pB$ and the related but distinct quantity $^{\circ}pB~\equiv~\sqrt{Q^2+U^2}$
(``total linear polarized brightness'').  Both quantitites have been called $pB$ in the 
historical literature by different (or sometimes even the same) authors;
to avoid confusion, we encourage authors to define $pB$ explicitly in every article that
uses the term.  
The quantity $^{\circ}pB$ has also been called $L$ \citep{rachmeler_etal_2013}, a usage
we encourage for its distinctness from $pB$.

By analogy to the Stokes parameters, there exists a parameter $pB'$ that is 
conjugate to $pB$ and describes
linear polarization 
diagonal to the radial direction on the image plane. $pB'$ is both predicted 
and observed to be zero in coronal Thomson-scattered light 
\citep[e.g.,][]{Filippov_etal_1994}.  It is 
omitted from PUNCH L3 published data products, although the data reduction pipeline 
can produce $pB'$ image products as a check on data quality.

PUNCH does not measure $B_R$ and $B_T$ directly.  Instead, each instrument measures 
polarization triplets: sequences of three exposures taken through linear polarizers
oriented $60^\circ$ apart.  The theory has been known for some time
\citep{fesenkov_1935,ohman_1947} and is recapitulated, with an eye toward modern 
application, by \citet{deforest_etal2022b}. We 
refer to the polarization-triplet system as MZP, 
and to each brightness value in a triplet as $B_M$, $B_Z$, or $B_P$ respectively. PUNCH 
collects MZP triplets in each instrument's reference frame on-orbit, then mathematically
transforms them to MZP in rectified solar coordinates.  This requires accounting not
only for spacecraft rotation angle, but also esoterica such
as projection effects that change the effective angle of each polarizer across the image
plane \citep{patel_etal_2025}.  

3-D image analysis makes use of the ``degree of [tangential linear] polarization'' $p$, 
which
we define as $p \equiv pB/tB$, and 
the related
``polarization ratio'' $B_R/B_T$, which provide indicators of the scattering angle 
$\chi$ and are described further in Appendix \ref{sec:appendix-polarization}.

We recommend notating the Stokes-like polarization parameters with a single prefix
letter as in $pB$ or $tB$, and notating real or computed radiance values (which are 
positive-definite) with a subscript, as in $B_T$, $B_R$, $B_M$, $B_Z$, or $B_P$.  
Radiances acquired with no polarizer in the beam may be labeled $B_C$ (for ``Clear'') when
it is necessary to distinguish them from the (possibly computed) $tB$.

\section{Polarization\label{sec:appendix-polarization}}

PUNCH polarimetry is specifically designed to locate objects in three dimensions, via the
physics of Thomson scattering.  The theory has been developed to varying degrees of 
detail by many authors, from \citet{minnaert_1930} and \citet{vandehulst_1950}, through
\citet{billings1966book}, to modern works including \citet{mierla_etal_2009} and
\citet{howard_tappin_2009}. It has been demonstrated by several authors including \citet{deforest_etal_2017} and \citet{mierla_2022}.  PUNCH observes features that are quite far from the Sun and 
therefore polarimetric analysis may be simplified considerably via the 
point-sun approximation
\citep{deforest_etal_2013}, which is important given the complexity of large 
distributed objects such as CMEs \citep{howard_etal_2013}.  
In this Appendix we very briefly discuss this elementary theory of polarimetry to locate
features, and refer to both prior articles that lay a more comprehensive foundation and
other articles in this collection, which explore the details of
distributed object polarimetry in the context of PUNCH.

PUNCH observes sunlight that is Thomson scattered by free electrons in coronal
features. Thomson scattering polarizes light through projection of the inbound
electric field into the perpendicular plane of the outbound rays. The electric field
component perpendicular to the scattering plane (Figure \ref{fig:appendix-diagram}) remains 
unchanged by 
these projection effects, while
the electric field component in the scattering plane is reduced by a factor of $\cos \chi$.  The
brightness
$B_R$ from a small feature, observed through a radially-aligned polarizer, is therefore
reduced by a factor of $(\cos \chi)^2$; while the brightness $B_T$, through a polarizer
perpendicular to the radial, is independent of $\chi$.  Calculating the polarization
ratio $PR\equiv B_R/B_T$ of a small bright feature cancels out any density or geometric
considerations,  so that $\chi$ may be determined via 
\begin{equation}
    \chi=\arccos\left(\sqrt{PR}\right) = \arccos\left(\sqrt{\frac{1-p}{1+p}}\right)\,,
\end{equation}
where $p\equiv pB/tB$ is the degree of linear polarization from Thomson scattering.
Both the $\arccos$ and the square root have sign ambiguities, but the positive branch of 
the square root is forced by the geometry of the scattering plane, leaving only the sign
ambiguity of the $\arccos$ function.  That ambiguity leads to a split between a ``real'' and
a ``ghost'' location along the line of sight, for a feature seen in any one observation.  
The two locations may be distinguished by their behavior over time in multiple image sets:
small ejecta from the Sun are known to follow radial trajectories, and the ghost location
follows a clearly nonphysical trajectory \citep{deforest_etal_2013,gibson_etal_2025}.  

\begin{figure}
    \centering
    \includegraphics[width=0.5\linewidth]{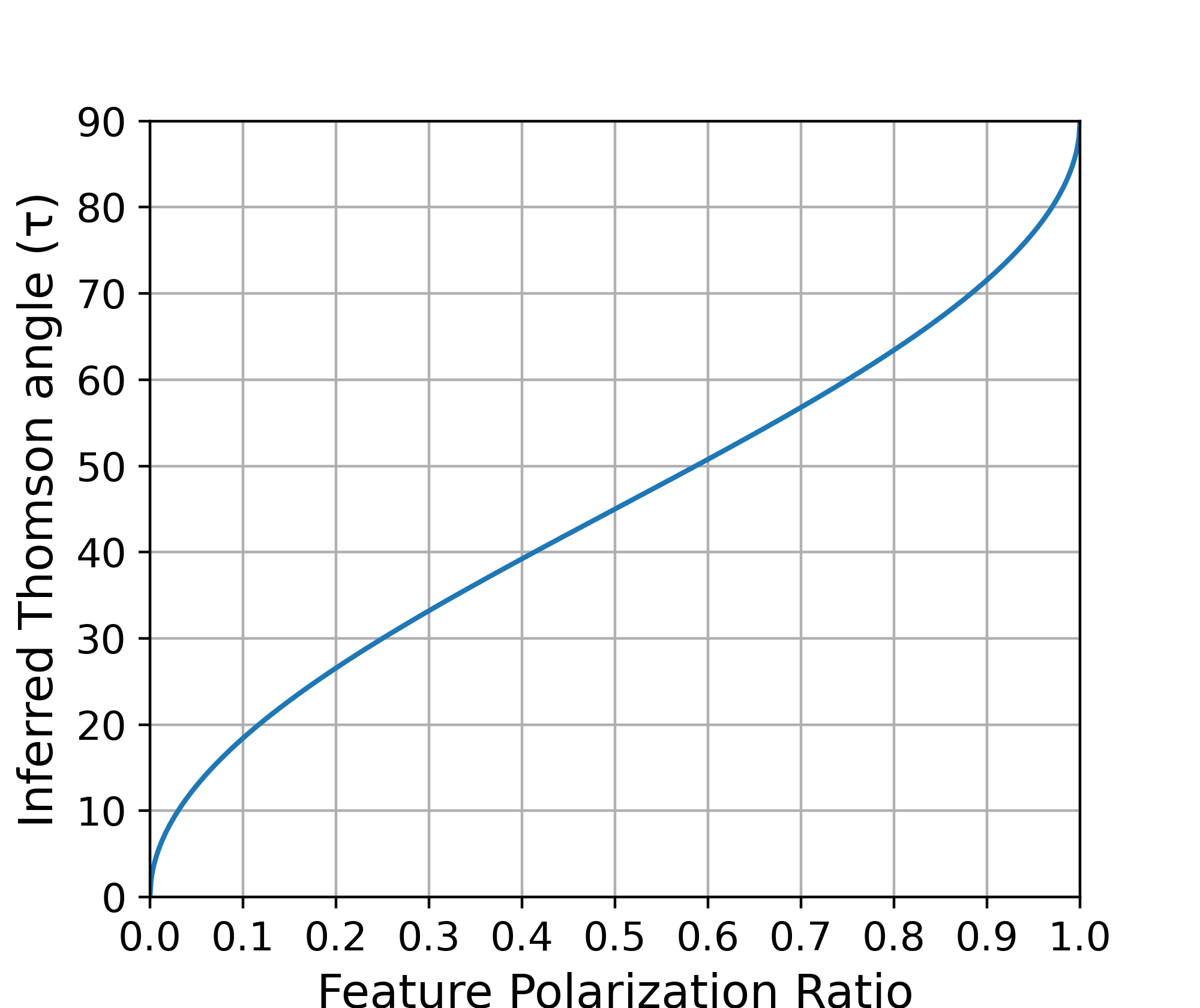}
    \caption{The polarization ratio $PR\equiv B_R/B_T$ (Table \ref{tab:defs}) determines 
    scattering angle $\chi$ or, equivalently, its complement $\tau$ (Figure 
    \ref{fig:appendix-diagram}) for isolated, background-subtracted bright features.  
    For features at intermediate $\tau$ (or $\chi$) angles between roughly 
    20$^\circ$-70$^\circ$, the slope 
    is approximately $0.6^\circ$ per percent in $B_R/B_T$, so although a feature with
    polarization ratio of 0.3 is at approximately $\tau=\pm32^\circ$, a 5\% uncertainty 
    in that feature's $PR$ corresponds to a roughly 3$^\circ$ uncertainty in 
    $\tau$.
    }
    \label{fig:location-by-polarization}
\end{figure}

Figure \ref{fig:location-by-polarization} illustrates simple 3D inversion, applicable to 
compact features that are well isolated and fully background subtracted.  Reading off the 
polarization ratio yields an estimate of the $\tau$ angle (complement of the scattering 
angle $\chi$).  From that and the elongation $\varepsilon$ of the feature, one may infer the
out-of-sky-plane angle $\xi$, as shown in Figure \ref{fig:appendix-diagram}.  This 
compact-feature case, for features at least several $r_\odot$ from the Sun itself,
has been developed analytically by several authors including
\citet{deforest_etal_2013}.  This compact-feature approximation may be used even for portions
of large features (whose size is comparable to the distance from the observer or the Sun), 
because the 2-D
leading edge of any large feature must necessarily be spatially compact along the line of
sight \citep{howard_etal_2013}.  However, all coronal features exist against the background
of myriad other K corona (or solar wind) features along each line of sight, and that ``K background'' must be 
considered
and removed in addition to the backgrounds from stray light, F corona, and stellar 
backgrounds described in Section \ref{sec:L3} \citep[e.g.,][]{vandehulst_1950, deforest_etal_2017}. 

Both distributed-object effects and background effects are important to many of the
objects observed by PUNCH; polarimetric theory for these 
measurements is developed in some detail elsewhere in this collection: by
\citet{gibson_etal_2025} for general solar wind observations, by
\citet{malanushenko_etal_2025} in the context of coronal mass ejections, and by 
\citet{dekoning_etal_2025} in the context of stream interaction regions. 
\bibliographystyle{spr-mp-sola}
\bibliography{punch-bib}  


\end{article} 

\end{document}